\documentclass[superscriptaddress,twocolumn,floatfix,10pt,longbibliography]{revtex4-1}

\usepackage[a4paper, left=.6in, right=.6in, top=.6in, bottom=.7in]{geometry} 

\usepackage{times} 

\usepackage[colorlinks=true,citecolor=blue,linkcolor=blue,urlcolor=blue]{hyperref}

\usepackage[utf8]{inputenc}
\usepackage{graphicx}
\usepackage{amsmath}
\usepackage{amssymb}

\newcommand{\romb}{\text{\protect\rotatebox[origin=c]{45}{$\square$}}}

\usepackage{dsfont}
\usepackage{mathrsfs}
\usepackage{bbm}
\usepackage{relsize} % for \mathsmaller

\usepackage{braket}
\usepackage{float}
\usepackage{afterpage}

\usepackage[capitalise]{cleveref}
\crefname{equation}{Eq.\!}{Eqs.\!}
\crefname{figure}{Fig.\!}{Figs.\!}
\crefname{chapter}{Chap.\!}{Chaps.\!}
\crefname{section}{Sec.\!}{Secs.\!}
\crefname{appendix}{App.\!}{Apps.\!}

\newcommand{\bs}[1]{{\boldsymbol{#1}}}
\newcommand{\ba}[1]{{\bar{#1}}}

\newcommand{\T}{\mathcal{{T}}}
\newcommand{\K}{\mathcal{{K}}}
\renewcommand{\S}{\mathcal{{S}}}
\renewcommand{\P}{\mathcal{{P}}}

\newcommand{\sT}{{\mathsmaller\T}}
  
\newcommand{\DD}{\mathbb{D}}
\newcommand{\UU}{\mathbb{U}}
\newcommand{\RR}{\mathbb{R}}
\newcommand{\NN}{\mathbb{N}}
\newcommand{\LL}{\mathbb{L}}

\newcommand{\ax}{\alpha}

\newcommand{\asym}{\beta}

\newcommand{\mf}[1]{{\mathfrak{#1}}}

\begin{document}

\title{Nonlocal discrete continuity and invariant currents in \\ locally symmetric effective Schr\"odinger arrays}

\author{C.~V. Morfonios}
\affiliation{Zentrum f\"ur Optische Quantentechnologien, Universit\"{a}t Hamburg, 22761 Hamburg, Germany}

\author{P.~A. Kalozoumis}
\affiliation{Department of Physics, University of Athens, 15771 Athens, Greece}

\author{F.~K. Diakonos}
\affiliation{Department of Physics, University of Athens, 15771 Athens, Greece}

\author{P. Schmelcher}
\affiliation{Zentrum f\"ur Optische Quantentechnologien, Universit\"{a}t Hamburg, 22761 Hamburg, Germany}
\affiliation{The Hamburg Centre for Ultrafast Imaging, Universit\"{a}t Hamburg, 22761 Hamburg, Germany}

\begin{abstract}
\begin{center}
 Journal reference: \href{https://doi.org/10.1016/j.aop.2017.07.019} {{Ann. Phys.}\ \textbf{385},\ {623} ({2017})}
\end{center}

We develop a formalism relating nonlocal current continuity to spatial symmetries of subparts in discrete Schr\"odinger systems.
Breaking of such local symmetries hereby generates sources or sinks for the associated nonlocal currents.
The framework is applied to locally inversion-(time-) and translation-(time-) symmetric one-dimensional photonic waveguide arrays with Hermitian or non-Hermitian effective tight-binding Hamiltonians.
For stationary states the nonlocal currents become translationally invariant within symmetric domains, exposing different types of local symmetry.
They are further employed to derive a mapping between wave amplitudes of symmetry-related sites, generalizing also the global Bloch and parity mapping to local symmetry in discrete systems. 
In scattering setups, perfectly transmitting states are characterized by aligned invariant currents in attached symmetry domains, whose vanishing signifies a correspondingly symmetric density. 
For periodically driven arrays, the invariance of the nonlocal currents is retained on period average for quasi-energy eigenstates.
The proposed theory of symmetry-induced continuity and local invariants may contribute to the understanding of wave structure and response in systems with localized spatial order.
\end{abstract}

% \begin{keyword}
% Discrete symmetries \sep Current-density continuity \sep Waveguide arrays \sep Photonics \sep PT-symmetric quantum mechanics \sep Discrete Schr\"odinger equation
% \end{keyword}

% PACS codes here, in the form: \PACS code \sep code
% 03.65.-w, % Quantum mechanics
% 42.25.Bs, % Wave propagation, transmission and absorption   
% 42.82.Et, % Waveguides, couplers, and arrays  
% 78.67.Pt} % Optical properties of multilayers / of photonic structures
      
\maketitle

\section{Introduction}

Symmetry under spatial transformations is a simple yet fundamental concept underlying the description of most systems in contemporary physics as the origin of conserved quantities.
In quantum theory and wave mechanics, symmetry-induced conservation laws are extended from continuous to discrete transformations in terms of commutation of the corresponding operators with the Hamiltonian, which in turn dictates the spatial structure of stationary states.
Ubiquitous paradigms are reflection ($\P$) or finite translation ($\K$) symmetry imposing definite parity or Bloch momentum on Hamiltonian eigenstates, which then provide the basis for understanding phenomena such as selection rules of level transitions in atoms \cite{Zettili2009____QuantumMechanics} or the formation of band structure in crystals \cite{Jiles1994____IntroductionElectronic}.
In intricate lattice systems of intense current interest such as topological insulators
\cite{Hasan2010_RMP_82_3045_TextitcolloquiumTopological}, Weyl semimetals \cite{Burkov2011_PRB_84_235126_TopologicalNodal}, and Lieb lattices \cite{Vicencio2015_PRL_114_245503_ObservationLocalized}, state localization and transport properties are also predominantly traced back to broken or unbroken spatial symmetries in combination with the operation of time reversal ($\T$).
The significance of broken spatiotemporal symmetries has further been addressed for driven quantum systems \cite{Grossmann1991_PRL_67_516_CoherentDestruction,Lehmann2003_JCP_118_3283_RectificationLaser-induced} in relation to directional transport  \cite{Denisov2007_PRA_75_063424_PeriodicallyDriven}.

Since symmetries are naturally addressed with respect to commutation with the global system Hamiltonian, they are usually regarded dichotomously as broken or unbroken.
Nevertheless, in a composite system the Hamiltonian elements may retain a certain spatial symmetry \textit{locally} in a subdomain of configuration space although it is broken \textit{globally};
we refer to this case as a `local symmetry' (LS).
Although there is clearly a remnant of the considered symmetry in the corresponding subsystem, it is not evident how to track this information in the eigenstates of the Hamiltonian since it generally does not commute with the LS operation.
At the same time, LSs are actually inherently paramount to a variety of systems such as aperiodic lattices with long-range order \cite{Macia2006_RPP_69_397_RoleAperiodic,Albuquerque2003_PR_376_225_TheoryElementary,Lahini2009_PRL_103_013901_ObservationLocalization,Morfonios2014_ND_78_71_LocalSymmetry}, partially disordered media
\cite{Wochner2009_PNAS_106_11511_X-rayCross,Lahini2008_PRL_100_013906_AndersonLocalization,Mustapha2013_JPCM_25_105401_UseSymmetry}, and single complex molecules \cite{Pascal2001_JPCA_105_9040_ConciseSet,Domagaa2008_JAC_41_1140_OptimalLocal}. 
A typical example are also generic solid state nanostructures \cite{Ferry1997____TransportNanostructures}, where unavoidable impurities and defects (usually modeled as $\P$-symmetric) separate the host crystal into finite $\K$-symmetric parts.
LS may further be present by design in artificial devices like, e.\,g., multilayered photonic setups \cite{Zhukovsky2010_PRA_81_053808_PerfectTransmission,Peng2002_APL_80_3063_Symmetry-inducedPerfect}, acoustic waveguides \cite{Hladky-hennion2013_JAP_113_154901_AcousticWave,Theocharis2014_NJP_16_093017_LimitsSlow}, or magnonic systems \cite{Hsueh2011_JOSAB_28_2584_FeaturesPerfect}, due to restrictions of finiteness or functionality.
In fact, even global spatial symmetry is generally rendered local as soon as a system's immediate environment is included in the description.

The spatial structure of stationary states in continuous, one-dimensional (1D), locally symmetric scattering potentials was recently pinpointed within the framework of symmetry-adapted nonlocal currents (NLCs) \cite{Kalozoumis2013_PRA_87_032113_LocalSymmetries,Kalozoumis2014_PRL_113_050403_InvariantsBroken,Zampetakis2016_JPAMT_49_195304_InvariantCurrent}:
At any real frequency, two distinct NLCs are spatially constant within subdomains of local $\P$ or $\K$ symmetry.
They provide a mapping of field amplitudes between LS related points, thus generalizing the mapping through parity or Bloch factors for global symmetry \cite{Kalozoumis2014_PRL_113_050403_InvariantsBroken}.
The NLCs were further linked to perfectly transmitting states \cite{Kalozoumis2013_PRA_88_033857_LocalSymmetries} and proposed as a natural order parameter for spontaneous symmetry breaking in non-Hermitian $\P\T$-symmetric systems \cite{Kalozoumis2014_PRA_90_043809_SystematicPathway,Kalozoumis2016_PRA_93_063831_Mathcalpt-symmetryBreaking}, as well as observed experimentally in lossy acoustic setups \cite{Kalozoumis2015_PRB_92_014303_InvariantCurrents}.

The stationary translational invariance of the NLCs suggests that they are governed by a continuity equation incorporating the associated symmetry transform.
In the case of global space reflection in continuous systems, the conservation of a symmetry-adapted two-point Schr\"odinger current has been shown under the combination of $\P$ with time reversal $\T$ \cite{Bagchi2001_MPLA_16_2047_GeneralizedContinuity,Japaridze2002_JPAMG_35_1709_SpaceState,Jr2015_JPAMT_48_155304_BilocalPicture}, and can be derived alternatively via gauge invariance of a nonlocal Lagrangian \cite{Sinha2015_PRE_91_042908_SymmetriesExact}.
The underlying global $\P\T$ symmetry has gained increased attention since it admits real discrete spectrum for non-Hermitian systems \cite{Bender1998_PRL_80_5243_RealSpectra,Bender2007_RPP_70_947_MakingSense} in parametric regions where density loss and gain are balanced \cite{Weigert2004_CJP_54_1139_PhysicalInterpretation}, and has also been intensively investigated in wave scattering \cite{Cannata2007_AP_322_397_Scattering-symmetric,Chong2011_PRL_106_093902_Mathcalpmathcalt-symBreaking,Schomerus2013_PTRSL_371_20120194_ScatteringTheory,Ambichl2013_PRX_3_041030_BreakingPt,Garmon2015_PRA_92_022125_BoundStates}.
General non-Hermitian Hamiltonians are of particular interest in discrete models, where the coupling between sites constitutes an additional degree of freedom for design and control \cite{Longhi2016_PRA_93_022102_Non-HermitianTight-binding,Makris2016_IJSTQ_22_42_ConstantIntensity}.
A reliable platform of applicability is provided by photonic waveguide arrays, which are well described by an effective discrete Schr\"odinger equation with tight-binding Hamiltonian \cite{Eisenberg2000_PRL_85_1863_DiffractionManagement,Schomerus2013_PTRSL_371_20120194_ScatteringTheory}.

In discrete models, already the local current becomes a two-point quantity, naturally defined to flow on the `link' between two  lattice sites \cite{Baranger1991_PRB_44_10637_ClassicalQuantum,Boykin2010_EJP_31_1077_CurrentDensity}.
The question is then raised of how to formulate discrete conservation laws for nonlocal quantities entailing different links.
At the same time, discrete arrays are ideal for implementing global or local spatial symmetries, by adjusting the onsite Hamiltonian elements or the inter-site hoppings.
As an example, in a certain class of discrete structures with `hidden' symmetries in the state space \cite{Teimourpour2014_PRA_90_053817_LightTransport}, conserved quantities from global operators commuting with the Hamiltonian can be identified.
Partial $\P\T$ symmetry along a given coordinates has also been implemented for coupled oscillator systems \cite{Beygi2015_PRA_91_062101_CoupledOscillator}, in the form of dimer arrays featuring synchronous Bloch-Zener oscillations \cite{Bender2015_PRA_92_041803_Wave-packetSelf-imaging}, or in synthetic lattices with solitonic excitations \cite{Wimmer2015_NC_6_7782_ObservationOptical}.
In view of the above, connecting discrete nonlocal link currents to general symmetry transformations restricted to lattice subdomains would constitute a framework to establish a relation between field state properties and LS.

In the present work, we combine the concept of local symmetry transformations with that of nonlocal current continuity for discrete---generally non-Hermitian---Schr\"odinger models.
We thereby identify the breaking of LSs as sources for the NLCs flowing along transformation related links.
In this way, the proposed framework unifies the treatment of Hermitian and non-Hermitian setups in the context of LS and its breaking, for arbitrary imposed boundary conditions.
Further, the NLCs are given in operator form, making the extension to multiple links per site straightforward.
With photonic waveguide arrays in mind, we apply the framework to 1D tight-binding Hermitian and non-Hermitian lattice systems with local $\S$ or $\S\T$ symmetry, respectively, with $\S = \P$ (inversion) or $\K$ (finite translation).
The evolution of the associated LS-adapted nonlocal charge is then given in terms of the NLCs at LS domain boundaries.
Together with a defined dual NLC, mapping relations are derived which relate the amplitudes of general states at symmetry-transformed sites and connect the NLCs with the local current.
It is then demonstrated how domainwise constant NLCs of stationary states reveal different types of LS encoded in arbitrarily irregular eigenstates, generalized also to periodically driven arrays.
Finally, the density profile of perfectly transmitting scattering states is shown to be characterized by NLCs in terms of its LS.

The paper is organized as follows.
In \cref{sec:transform} we introduce the concept of LSs and the corresponding local site permutations.
Section \ref{sec:discrete_continuity} is devoted to the discrete nonlocal current-density continuity and its relation to LSs.
In \cref{sec:stat_invariants_mappings} we address the properties of NLCs in stationary states, derive symmetry-induced amplitude mapping relations and the connection to the local current, and generalize the NLC to driven arrays.
In \cref{sec:bound_eigenmodes} we discuss the cases of complete, overlapping, and gapped LS, and provide examples of invariant NLCs in non-Hermitian and driven finite arrays.
In \cref{sec:scattering} we consider the NLCs in scattering states, derive their relation to perfect transmission and illustrate their invariance in locally symmetric non-Hermitian scatterers.
Section \ref{sec:conclusions} concludes with future perspectives.

%%%%% fig: setup_hamiltonian %%%%%%%%%%%%%%%%%%%%%%%%%%%%%%%%%%%%%%%%%%%%%%%%%%%%%%%%%%%%%%%%%%
\begin{figure}[t!] 
\centering
\includegraphics[width=.83\columnwidth]{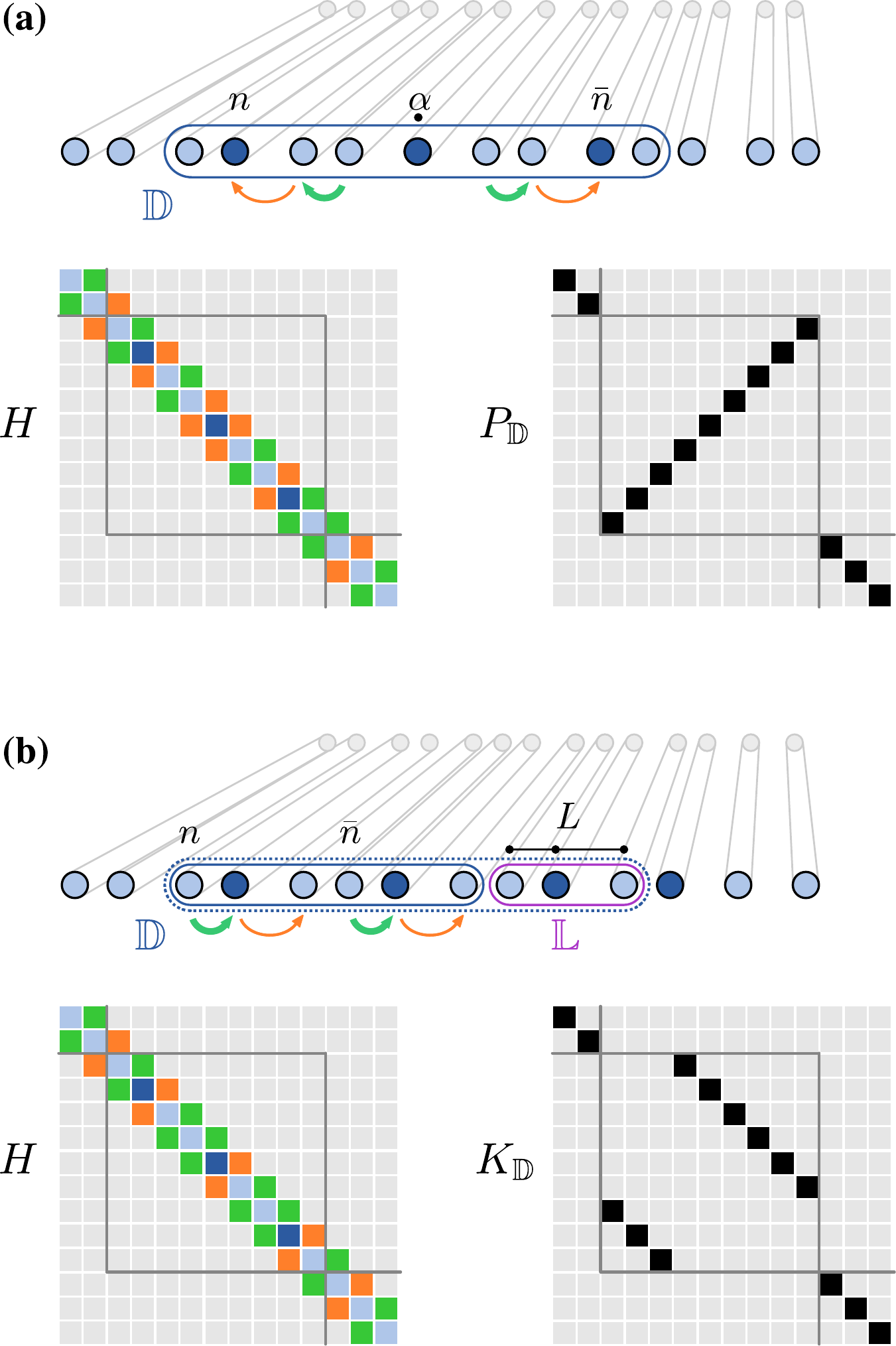}
\caption{Sketch of a 1D finite lattice in the form of a photonic \hyphenation{wave-gui-de} waveguide array which is \textbf{(a)} locally inversion ($\P$) symmetric about a center $\ax$ or \textbf{(b)} translation ($\K$) symmetric with period $L$, for sites $n$ within a domain $\DD$ (solid blue line), together with the corresponding close-coupling Hamiltonian matrix $H$ and local transformation (site permutation) matrix $P_\DD$ or $K_\DD$, respectively.
$P_\DD$ maps $\DD$ to itself, while $K_\DD$ performs a cyclic shift permutation by $L$ sites on $\UU = \DD \cup \LL$, where $\LL$ (solid purple line) is the rightmost period in $\UU$ (dotted blue line).
}
\label{fig:setup}
\end{figure}
%%%%%%%%%%%%%%%%%%%%%%%%%%%%%%%%%%%%%%%%%%%%%%%%%%%%%%%%%%%%%%%%%%%%%%%%%%%%%%%%%%%%%%%%%%%%%%%

\section{Local symmetry operations}
\label{sec:transform}

To define the concept and explicit form of a LS operation in a 1D discrete model, we start by considering a single-particle, time-independent Hamiltonian $\hat{H}$ represented on $N$ localized site excitations $\ket{n}$, $n = 1,2,\dots,N $, with onsite potential elements $v_n = H_{nn} = \braket{n|\hat{H}|n}$ and hopping elements $h_{n,n'} = H_{nn'} = \braket{n|\hat{H}|n'}$ from site $n'$ to $n \neq n'$.
The temporal evolution of a quantum state
\begin{equation} \label{eq:psi}
 \ket{\psi} = \sum_{n=1}^N \psi_n \ket{n}
\end{equation}
is then governed by the Schr\"odinger equation (SE)
\begin{equation} \label{eq:Schr}
 i\partial_t \ket{\psi} = \hat{H} \ket{\psi}
\end{equation}
where we set $\hbar = 1$.
Such a discrete model is widely used to describe coherent electron transport in 1D lattice potentials like the ones induced in quantum dot arrays, but also as a finite difference approximation to the continuous SE by choosing appropriate (typically equal) hopping values.
In particular, \cref{eq:Schr} effectively also describes light propagation in photonic waveguide arrays within the paraxial approximation \cite{Eisenberg2000_PRL_85_1863_DiffractionManagement,Schomerus2013_PTRSL_371_20120194_ScatteringTheory} for small lateral wave vector components.
1D versions of such a photonic array are schematically depicted in \cref{fig:setup}.
Time $t$ is here represented by the longitudinal spatial coordinate $z$ along the waveguides, and energy $E$ is accordingly represented by the light propagation constant along $z$.
Onsite potentials $v_n$ correspond to the refractive indices of waveguides, while the hoppings $h_{n,n'}$ are proportional to the overlap of the light field localized in waveguides $n,n'$ and thereby mapped to their distance. 
Remarkable agreement with experimental results is achieved already in the tight-binding approximation \cite{Szameit2007_OE_15_1579_ControlDirectional} with $h_{n,n'} = 0$ for $|n-n'|>1$, which we will focus on in the following.

Given the above framework for a discrete 1D Schr\"odinger model, we consider a bijective mapping $\S_\DD: \DD \rightarrow \bar{\DD}$ given by 
\begin{equation} \label{eq:site_mapping}
 \S_\DD: n \to \bar{n} = \S_\DD(n) = \begin{cases} \P_\DD(n) &= 2\ax - n \\ \K_\DD(n) &= L + n  \end{cases},
\end{equation}
which performs a local inversion ($\P$) through a center $\ax$ or translation ($\K$) by a length $L$ of sites $n \in \DD$ to sites $\bar{n} \in \bar{\DD}$, where the domain $\DD \subseteq \NN$ is generally a subset of the total set $\NN$ of all sites.
Suppose now that the discrete system described by the Hamiltonian elements $H_{mn}$ remains invariant under this site mapping, that is,
\begin{equation} \label{eq:Hmn_local_Ssym}
H_{\bar{m}\bar{n}} = H_{mn} ~~~\forall~ m,n \in \DD,
\end{equation}
so that $\S_\DD$ is a symmetry transformation of the system.
We then say that the system possesses a \textit{local} symmetry in the union $\UU \equiv \DD \cup \bar{\DD}$.
A LS is thus contrasted with a \textit{global} symmetry of the system for which $\DD = \NN$.
Locally $\P$- and $\K$-symmetric photonic arrays are depicted in \cref{fig:setup}\,(a) and (b), respectively \cite{note}.
$\P_\DD$ maps a domain $\DD$ to itself, so that $\DD = \bar{\DD} = \UU$, while $\K_\DD$ shifts $\DD$ by $L$ sites, so that $\UU = \DD \cup \mathbb{L}$ where $\mathbb{L}$ are the rightmost $L$ sites in $\UU$ (for $L \leqslant D$); see \cref{fig:setup}. 

With the (symmetry) mapping $\S_\DD$ transforming the site indices $n$, we correspondingly define an operator 
\begin{equation} \label{eq:Sigma_operator}
 \hat{\varSigma}_\DD = \sum_{n=1}^N \hat{\sigma}_{n;\DD}, ~~~ \hat{\sigma}_{n;\DD} = \ket{n}\bra{\S_\DD(n)} = \ket{n}\bra{\bar{n}} 
\end{equation}
which replaces the wave amplitude $\psi_n$ with $\psi_{\bar{n}}$ for the basis ket $\ket{n}$ when acting on a state $\ket{\psi}$ (cf. \cref{eq:psi}),
\begin{equation} \label{eq:Sigma_psi}
 \hat{\varSigma}_\DD \ket{\psi} = \sum_{n=1}^N \psi_{\bar{n}} \ket{n} \equiv \ket{\bar{\psi}},
\end{equation}
since the basis $\{\ket{n}\}$ is assumed orthonormal ($\braket{m|n} = \delta_{mn}$), so that we can symbolically write the action as an amplitude mapping $\hat{\varSigma}_\DD: \psi_n \rightarrow \psi_{\bar{n}}$, with $\hat{\varSigma}_\DD = \hat{P}_\DD, \hat{K}_\DD$,
in analogy with the local \textit{site} mapping in \cref{eq:site_mapping} for $\S_\DD = \P_\DD,\K_\DD$.
The associated matrix ${{\varSigma}}_\DD$ then has elements $[{{\varSigma}}_\DD]_{mn} = \braket{m|\hat{\varSigma}|n} = \delta_{\bar{m}n}$, as depicted in \cref{fig:setup}: 
$P_\DD$ performs a (partial) mirror permutation of the sites within $\DD$, while $K_\DD$ shifts sites within $\DD$ by $L$---we choose $[{{\varSigma}}_\DD]_{mn} = \delta_{\bar{m}-D,n}$ for $m \in \mathbb{L}$ so that the last $L$ sites in $\UU$ are mapped to the first $L$ ones.
The permutation matrix thus acts on the whole array but permutes only sites amplitudes within the domain $\UU$ where the LS of the system resides while leaving sites outside $\UU$ unchanged.
Note that multiple symmetry domains $\DD_d$ ($d=1,2,\dots$) may in general be present, with corresponding centers $\ax_d$ or periods $L_d$.
The total symmetry domain is then the (generally disconnected) set $\DD = \bigcup_d \DD_d$, with the mapping for each separate $\DD_d$ entering \cref{eq:site_mapping}.
Different types of LS will be addressed in \cref{sec:symmetry_types}.

Symmetry of a system under a transformation is usually associated with the commutation of the corresponding symmetry operator with the Hamiltonian.
In the case of global symmetry, that is, $\DD = \NN$ in the present setting (with $N = \infty$ implied for translation symmetry), the Hamiltonian indeed commutes with the matrix ${{\varSigma}}_\DD$ and the energy eigenstates are also eigenstates of the symmetry operation with the corresponding eigenvalues:
Under inversion or translation the wave amplitudes are mapped to $\psi_{\bar{n}} = \lambda \psi_n$ with the parity $\lambda = \pm 1$ or the Bloch phase $\lambda = e^{ikL}$ ($k$ being the quasimomentum) as prefactors, respectively.
In the case of a LS, although the \textit{system} remains invariant under the site permutation, the associated operator generally does not commute with the Hamiltonian,
\begin{equation} \label{eq:noncommutation}
 [\hat{\varSigma}_\DD,\hat{H}] \neq 0.
\end{equation}
The reason is the change in coupling of the domain $\DD$ to its surroundings:
The end sites of $\DD$ in a 1D array are not coupled to the same sites after the local permutation, and so the transformed Hamiltonian $\bar{H} \equiv \varSigma^{\dagger}H\varSigma$ (recall that $\varSigma^{-1} = \varSigma^{\dagger}$ for any permutation matrix) generally contains altered nonzero hopping elements $\bar{H}_{mn} \neq H_{mn}$ for $m \in \DD$ and $n \notin \DD$; cf. \cref{eq:Hmn_local_Ssym}.
As outlined in \cref{app:time_reversal}, for a non-Hermitian $\hat{H}$ it becomes relevant to combine the spatial transformation with time reversal $\T$, and the corresponding local $\S_\DD\T$ symmetry is expressed by $H_{\bar{m}\bar{n}} = H_{mn}^*$ for $m,n \in \DD$, while generally $[\hat{\varSigma}_\DD\hat{T},\hat{H}] \neq 0$.

We here raise question if the presence of symmetries in the system are in some way imprinted on the structure of its states, even if those symmetries are not global but of finite extent.
The aim is thus to find a characterization of the field amplitude configuration in terms of a quantity which follows the underlying LSs of the system.
To do so, in the following section we turn to the notion of nonlocal current-density continuity in its discrete form and adapt it to the LS framework.

\section{Discrete nonlocal current-density continuity}
\label{sec:discrete_continuity}

\begin{figure}[t!]
\centering
\includegraphics[width=.9\columnwidth]{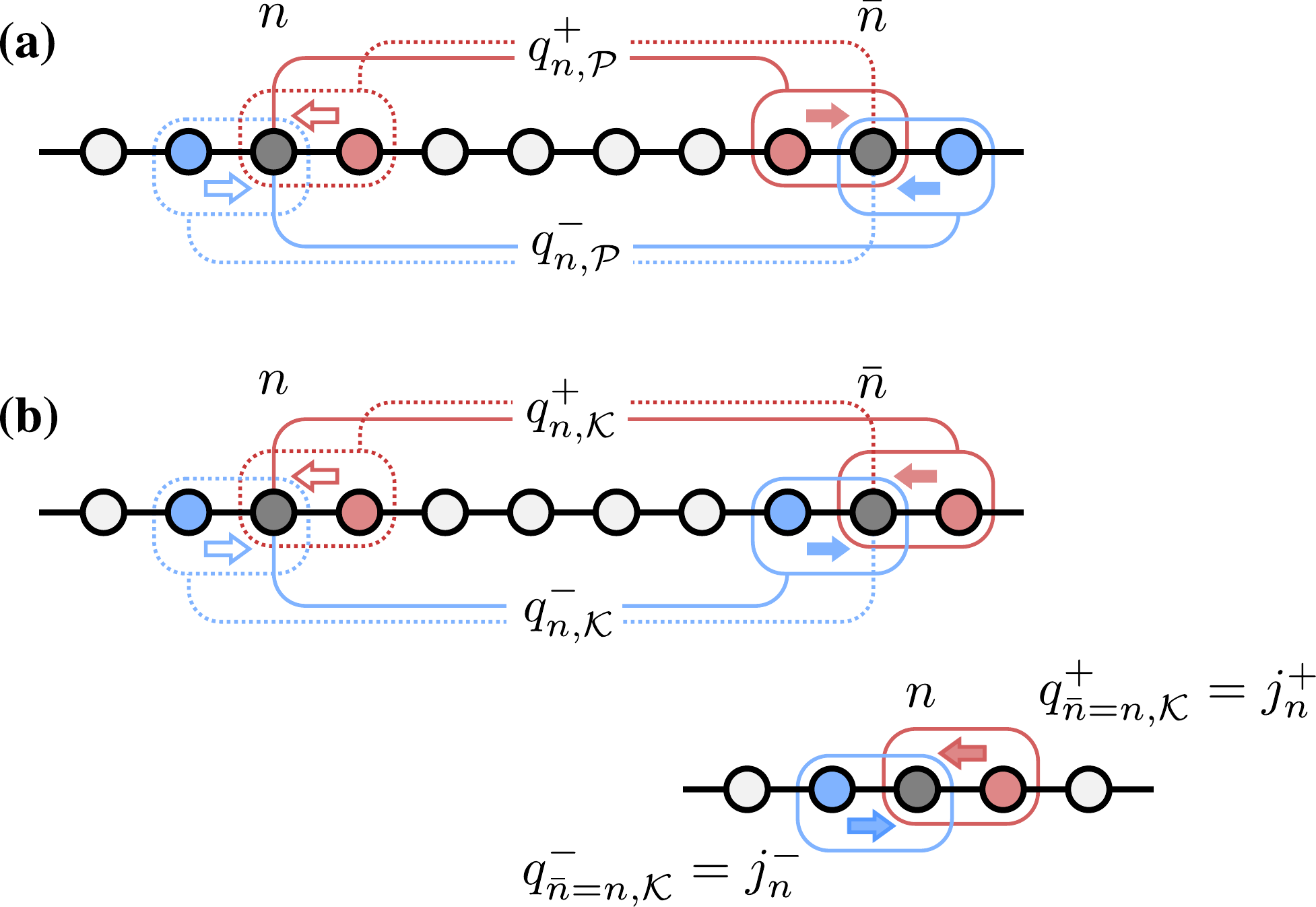}
\caption{Nonlocal current flow on pairs of $\S$-related links for \textbf{(a)} inversion $\S = \P$ and \textbf{(b)} translation $\S = \K$ transform. 
Solid (dotted) lines indicate the term $\psi_n^* h_{\S(n),\S(n \pm 1)} \psi_{\S(n \pm 1)}$ ($\psi_{n \pm 1}^* h_{n,n \pm 1}^* \psi_{\S(n)}$) entering each upper ($+$) and lower ($-$) NLC $q_{n,\S}^\pm$ ($\S = \P,\K$) on the links into a site pair $n,\bar{n}$; see \cref{eq:q_n+-}.
Inset: The translation NLC reproduces the local link current $j_n^\pm$ for $\bar{n} = n$.
}
\label{fig:nonlocal_flow}
\end{figure}

With respect to the spatial transformation $\S_\DD$, we generalize the local discrete current-density continuity, outlined in \cref{app:continuity_local}, to a nonlocal form by replacing the local density operator $\hat{\rho}_n$ with $\hat{\sigma}_n = \ket{n}\bra{\bar{n}}$ from \cref{eq:Sigma_operator} (we drop the subscript $\DD$ from now on). 
The latter is now identified with a nonlocal density operator corresponding to the off-diagonal part of the site-represented density operator $\hat{\rho} = \ket{\psi}\bra{\psi}$ associated with the transformation $\S$, so that
\begin{equation} \label{eq:nonlocal_density}
 \sigma_n \equiv \braket{\psi|\hat{\sigma}_n|\psi} = \braket{\bar{n}|\hat{\rho}|n} = \psi_n^* \psi_{\bar{n}}
\end{equation}
constitutes a symmetry-adapted nonlocal density in state $\psi$
Its sum over $n$ yields the total `nonlocal charge' 
\begin{equation} \label{eq:nonlocal_charge}
 \varSigma_\psi = \braket{\psi | \hat{\varSigma} | \psi} = \sum_n \sigma_n,
\end{equation}
also known as `quasipower' \cite{Sarma2014_PRE_89_052918_ContinuousDiscrete} in globally $\P\T$-symmetric photonic systems, in analogy to the usual charge (i.\,e. quantum probability), or power in photonics, $I_\psi = \braket{\psi | \hat{I} | \psi} = \sum_n \rho_n$ ($\hat{I}$ being the identity operator).
Using the SE, the temporal evolution of $\sigma_n$ is given by 
\begin{equation} \label{eq:continuity_nonlocal}
 \partial_t \sigma_n = q_n - i (v_{\bar{n}} - v_n^*) \sigma_n = q_n + \asym_n \sigma_n,
\end{equation}
with $\asym_n \equiv (v_{\bar{n}} - v_n^*)/i$ characterizing the onsite asymmetry, and where $q_n = q_n^+ + q_n^-$ assigned to site $n$ is the sum of the \textit{nonlocal} currents defined by
\begin{align} \label{eq:q_n+-}
iq_n^\pm &\equiv iq_{n,\S}^\pm \equiv iq_{n,n \pm 1; \S} =\nonumber \\
&\psi_n^* h_{\S(n),\S(n \pm 1)} \psi_{\S(n \pm 1)} - \psi_{n \pm 1}^* h_{n,n \pm 1}^* \psi_{\S(n)}
\end{align}
for a local transformation $\S = \S_\DD$ given in \cref{eq:site_mapping}, with $\S(n \pm 1) = \bar{n} \mp 1$ and $\bar{n} \pm 1$ for $\S = \P$ (inversion) and $\K$ (translation), respectively.
Considering Hermitian hoppings, $h_{m,n} = h_{n,m}^*$, these four-site currents $q_{n,\P}^\pm$ and $q_{n,\K}^\pm$ can be written as the expectation values 
\begin{align} \label{eq:q_n_+-_expval}
&{q}_{n,\S}^\pm = \braket{\psi|\hat{q}_{n,\S}^\pm|\psi} ~~~~ (\S = \P,\K)
\end{align}
of the corresponding nonlocal current operators
\begin{subequations} \label{eq:q_n_+-_operator}
\begin{align} 
 \hat{q}_{n,\P}^\pm &= \frac{1}{i} (\hat{\sigma}_{n,\P}\hat{H}^\pm - \hat{H}^\pm\hat{\sigma}_{n,\P}), \label{eq:q_n_+-_P_operator} \\ 
 \hat{q}_{n,\K}^\pm &= \frac{1}{i} (\hat{\sigma}_{n,\K}\hat{H}^\pm - \hat{H}^\mp\hat{\sigma}_{n,\K}), \label{eq:q_n_+-_K_operator}
\end{align}
\end{subequations}
in analogy with \cref{eq:j_n_+-_operator}, with the transformation type $\P$ or $\K$ entering $\hat{\sigma}$ here explicitly indicated, $\hat{H}^\pm$ being the upper/lower hopping operator (see \cref{app:continuity_local}).
Outside the domain $\UU$ of the mapping where $\bar{n} = n$, the translation current reproduces the usual current, $\hat{q}_{n=\bar{n},\K}^\pm = \hat{j}_n^\pm$, and so \cref{eq:continuity_nonlocal} reproduces the local current-density continuity in \cref{eq:continuity_local}.
For both $\P$ and $\K$, the total current operator is given by 
\begin{equation}
 \hat{q}_n = \hat{q}_n^+ + \hat{q}_n^- = \frac{1}{i} [\hat{\sigma}_n,\hat{H} - \hat{V}],
\end{equation}
where $\hat{V} = \hat{H} - \hat{H}^+ - \hat{H}^-$ is the diagonal part of the Hamiltonian.
Note here that, like the usual currents $j_n^\pm$, the $q_n^\pm$ can be seen as link currents, though not flowing along one link but along the two ones corresponding to the hoppings $h_{\S(n),\S(n \pm 1)}$ and $h_{n,n \pm 1}^*$ entering \cref{eq:q_n+-}.
This nonlocal `flow' is illustrated in \cref{fig:nonlocal_flow}.

Using the transformation matrices $\varSigma_\DD = P_\DD, K_\DD$ (depicted in the example of \cref{fig:setup}), the $q_n^\pm = \braket{\psi|\hat{q}_n^\pm|\psi}$ in \cref{eq:q_n+-} for the whole setup $\NN$ can be written as the components of an $N$-entry column vector $\bs{q}^\pm = [q_1^\pm ~q_2^\pm ~ \cdots ~ q_N^\pm]^\top$ given by (cf. \cref{eq:q_n_+-_P_operator,eq:q_n_+-_K_operator})
\begin{subequations} \label{eq:q_n_+-_vector}
\begin{align}
 i\bs{q}^{\pm}_\P &= \text{d}[\bs{\psi}^\dagger] P_\DD H^\pm \bs{\psi} - \text{d}[\bs{\psi}^\dagger H^\pm] P_\DD \bs{\psi}, \label{eq:q_n_+-_P_vector} \\
 i\bs{q}^{\pm}_\K &= \text{d}[\bs{\psi}^\dagger] K_\DD H^\pm \bs{\psi} - \text{d}[\bs{\psi}^\dagger H^\mp] K_\DD \bs{\psi} \label{eq:q_n_+-_K_vector}
\end{align}
\end{subequations}
for local $\P,\K$-transformation, respectively, where $\bs{\psi} = [\psi_1 ~\psi_2 ~ \cdots ~ \psi_N]^\top$ and $\text{d}[\bs{a}]$ denotes the $N \times N$ diagonal matrix with the components of (row or column) vector $\bs{a}$ along its diagonal.
The above expression for the $q_n^\pm$ becomes convenient when evaluating them for a given Hamiltonian and (numerically computed) state amplitudes $\psi_n$, as will be done in the following sections.

%%%%%%%%%%%%%%%%%%%%%%%%%%%%%%%%%%%%%%%%%%%%%%%%%%%%%%%%%%%%%%%%%%%%%%%%%%%%%%%%%%%%%%%%%%%%%%%
%%%%%%%%%%%%%%%%%%%%%%%%%%%%%%%%%%%%%%%%%%%%%%%%%%%%%%%%%%%%%%%%%%%%%%%%%%%%%%%%%%%%%%%%%%%%%%%

For a given site transformation $\S$, the form of the nonlocal continuity equation (\ref{eq:continuity_nonlocal}) is similar to \cref{eq:continuity_local} with the loss/gain function $\gamma_n$ replaced by $\asym_n$ which characterizes the sources for the nonlocal density $\sigma_n$.
For a system with $\S$-symmetric onsite potential, $v_{\bar{n}} = v_n$ or $[\hat{V},\hat{\varSigma}]=0$, we have $\asym_n = \gamma_n$ and the sources (sinks) of $\sigma_n$ are represented by the gain (loss) of $\rho_n$ at individual sites $n$ like in \cref{eq:continuity_local}, and $\hat{q}_n = [\hat{\sigma}_n,\hat{H}]/i$ like in \cref{eq:j_n_tot_operator}.
Those sources and sinks vanish if the potential is real, but if the $v_n$ are not $\S$-symmetric, then in \cref{eq:continuity_nonlocal} there is a contribution to $\asym_n$ from the potential differences between $\S$-related sites.
In other words, locally \textit{broken spatial potential symmetries act as sources} for the associated nonlocal density.
If a system has locally $\S\T$-symmetric non-Hermitian onsite potential, $v_{\bar{n}} = v_n^*$ or $[\hat{V},\hat{\varSigma}\hat{T}]=0$, then $\asym_n = 0$ within the symmetry domain $\DD$ and \cref{eq:continuity_nonlocal} reduces to $\partial_t \sigma_n = q_n$.
This means that the net NLC $q_n$ vanishes within $\DD$ for stationary states, in contrast to $j_n$ which may vary spatially for $\gamma_n \neq 0$.
As shown in \cref{app:bitemporal_nlc}, an alternative nonlocal continuity equation can finally be established for a \textit{bitemporal} nonlocal density and associated current by projecting $\hat{\sigma}\ket{\psi}$ onto the time-reversed state $\T\ket{\psi} = \ket{\psi^\sT}$ instead of $\ket{\psi}$ in \cref{eq:nonlocal_density}.

The relation of \textit{stationary} NLCs to LSs will be elaborated on in the next section.
To see first the impact of the system symmetry on a general (nonstationary) state $\ket{\psi}$ via the NLCs, let us consider the evolution of the nonlocal charge (see \cref{eq:nonlocal_charge}) within a subdomain $\DD \subseteq \NN$, denoted $\varSigma_\psi^\DD$, which obeys (cf. \cref{eq:continuity_nonlocal})
\begin{equation} \label{eq:charge_evolution}
 \partial_t \varSigma_\psi^\DD = Q_\DD + \sum_{n \in \DD} \asym_n \sigma_n,
\end{equation}
where $Q_\DD = \sum_{n \in \DD} q_n$ is the total net NLC within $\DD$.
Now, if the \textit{hoppings} of the system are $\S_\DD\T$-symmetric, $h_{\S(n),\S(n \pm 1)} = h_{n,n \pm 1}^*$, then from the NLC expression (\ref{eq:q_n+-}) we get, in general,
\begin{equation} \label{eq:q_n-1_+_equal_-q_n_-}
 q_{n \mp 1,\S}^\pm = - \eta_{n \mp 1,n}^* q_{n,\S}^\mp ~~~~ (\S = \P,\K)
\end{equation}
where we define the \textit{hopping ratio} 
\begin{equation} \label{eq:eta_mn}
 \eta_{m,n} = \frac{h_{m,n}}{h_{n,m}}
\end{equation}
between the hopping elements in opposite directions on the link $(m,n)$, here with $m = n \pm 1$.
For \textit{equidirectional} hoppings ($\eta_{m,n} = 1$) within $\DD$, $Q_\DD$ thus equals the sum $q_{\partial\DD}$ of currents $q_m^\pm$ on links above/below sites $m$ of the upper/lower boundaries $\partial\DD^\pm$ of $\DD$,
\begin{equation} \label{eq:Q_D}
 Q_\DD = q_{\partial\DD} \equiv \sum_{m \in \partial\DD^\pm} q_m^\pm,
\end{equation}
since the upper and lower NLCs cancel out on each link within $\DD$.
Hence, if also the potential is $\S_\DD\T$-symmetric ($\asym_{n\in\DD} = 0$), then the temporal change $\partial_t \varSigma_\psi^\DD$ in \cref{eq:charge_evolution} is affected only by those boundary currents $q_{\partial\DD}$.

Recall that $\DD$ may be disconnected, as a union of distinct subdomains $\DD_d$ as mentioned above, entailing equally many $q_m^\pm$ pairs at its boundary points.
In the special case of \textit{global} $\P\T$-symmetry in $\DD = \NN$, we have $\asym_n = 0$ but also $q_{\partial\DD} = 0$ since the boundary currents $q_1^-,q_N^+$ vanish for the isolated array (where $h_{1,0} = h_{N+1,N} = 0$ by definition).
Thus $Q_{\mathbb{D=N}} = 0$ in \cref{eq:Q_D} and thereby $\partial_t \varSigma_\psi^\NN = 0$, reproducing the well-known constancy of the nonlocal charge (quasipower) in systems with $\P\T$-symmetric potential \cite{Bagchi2001_MPLA_16_2047_GeneralizedContinuity,Sarma2014_PRE_89_052918_ContinuousDiscrete}.

\section{Stationary state invariants and amplitude symmetry mapping}
\label{sec:stat_invariants_mappings}

Having established the concept of nonlocal discrete continuity in the above operator form and adapted it to symmetry mappings, we will now use the NLCs to show how LSs of the lattice system are encoded into its stationary states.
For an eigenstate 
\begin{equation} \label{eq:mode}
 \ket{\phi} = \sum_n a_n e^{-iE t} \ket{n} = e^{-iE t} \ket{a}
\end{equation}
of the Hamiltonian, $\hat{H} \ket{\phi} = E\ket{\phi}$, with generally complex eigenvalue $E = E^\circ + i\varGamma/2$ ($E^\circ,\varGamma \in \RR$) if the $\hat{H}$ is (effectively) non-Hermitian \cite{Moiseyev2011____Non-hermitianQuantum}, \cref{eq:continuity_nonlocal} yields
\begin{equation} \label{eq:continuity_nonlocal_eigenstate}
 q_n + (\asym_n - \varGamma) \sigma_n = 0.
\end{equation}
Thus, at sites with $\S\T$-symmetric onsite elements ($\asym_n = 0$), a stationary state $\ket{\phi}$ with $\varGamma = 0$ has vanishing sum of NLC amplitudes,
\begin{equation} \label{eq:q_n_vanishing}
 q_{n,\S} = q_{n,\S}^+ + q_{n,\S}^- = 0 ~~~~ (\S = \P,\K)
\end{equation}
following a nonlocal form of a quantum Kirchhoff law.
For $\S\T$-symmetric hoppings $h_{n,n \pm 1}$, the above equation can be combined with \cref{eq:q_n-1_+_equal_-q_n_-} to obtain
\begin{equation} \label{eq:q_n_+-_transfer}
 q_{n,\S}^\pm = \eta_{n,n \mp 1}^* \, q_{n \mp 1,\S}^\pm ~~~~ (\S = \P,\K)
\end{equation}
for consecutive nonlocal link currents.
Thus, in a \textit{connected} domain $\DD$ where the system has $\S\T$-symmetric Hamiltonian elements , obeying \cref{eq:Hmn_local_STsym}, the NLC $q_{n,\S}^+$ ($q_{n,\S}^-$) accumulates the hopping ratios as a prefactor when traversing the domain forwards (backwards).
For complex \textit{Hermitian} hoppings, $\eta_{n,n \mp 1} = e^{2i\varphi_{n,n \mp 1}}$ and the currents accumulate a Peierls-like hopping phase factor along $\DD$.
If the hoppings are $\S\T$-symmetric and equidirectional, $h_{m,n} = h_{n,m}$ (generally complex, as may be the case in photonic arrays \cite{Golshani2014_PRL_113_123903_ImpactLoss} depending on the host medium), the $q_{n,\S}^\pm$ are explicitly given by
\begin{subequations}
\begin{align} \label{eq:q_n_+-_stat_state}
i q_{n,\P}^\pm &= h_{n,n\pm 1} (a_{n}^*a_{\bar{n} \mp 1} - a_{n\pm 1}^*a_{\bar{n}}), \\
i q_{n,\K}^\pm &= h_{n,n\pm 1} (a_{n}^*a_{\bar{n} \pm 1} - a_{n\pm 1}^*a_{\bar{n}}),
\end{align}
\end{subequations}
and are \textit{translationally invariant} (spatially constant) in $\DD$ for a stationary state with real $E$.

To establish a connection between $\S$-transformations and the state structure via NLCs (see \cref{sec:current_mapping} below), it is convenient to introduce a \textit{dual} NLC $\mathring{q}_{n,\S}^\pm$, as described in \cref{app:dual_nlc}. 
For a stationary state, it has the same \textit{spatial} part as the bitemporal NLC in \cref{eq:qTrev_n+-} with $\mathring{q}_{n}^\pm = q_{n}^{\sT\pm} \, e^{-2iEt}$.
Both $q_{n}^{\sT\pm}$ and $\mathring{q}_{n}^\pm$ are invariant in a locally $\S$-symmetric domain for an energy eigenstate, as shown in \cref{app:bitemporal_nlc,app:dual_nlc}, respectively.
We focus on the equal-time NLCs $q_{n}^\pm$ and $\mathring{q}_{n}^\pm$ in the remainder of the paper.

\subsection{Nonlocal invariants in $\S(\T)$-symmetric domains and corresponding stationary states}
\label{sec:stationary_invariants}

Let us now consider the NLCs $q_{n}^\pm$ and $\mathring{q}_{n}^\pm$ in the concrete case of equidirectional hoppings $h_{n,n \pm 1} = h_{n \pm 1,n}$, and distinguish some basic properties as follows:\\

\noindent\textbf{\small (i)} In a stationary state $\ket{\phi}$ with real energy $E$, local $\P_{\DD}\T$-symmetry ($v_{\bar{n}= 2\ax - n} = v^*_n$, $h_{\bar{n},\bar{n}\pm1} = h_{n,n\mp1}^*$) or local $\K_{\DD}\T$-symmetry ($v_{\bar{n}= n+L} = v^*_n$, $h_{\bar{n},\bar{n}\pm1} = h_{n,n\pm1}^*$) yields spatially constant corresponding currents
\begin{equation} \label{eq:q_constant_D}
 q_{n,\S}^\pm = q_{\DD,\S}^\pm ~~~~ (\S = \P,\K)
\end{equation}
for all $n \in \DD$ with $h_{n,n \pm 1} = h_{n \pm 1,n}$.
In particular for $\P_{\DD}\T$ symmetry it can be shown that $q_{\DD,\P}^\pm$ vanishes for a state vector $\bs{\phi}$ whose part $\bs{\phi}_\DD = \{ \phi_{n \in \DD} \}$ is a local $\P\T$ `eigenstate', \textit{and vice versa}, that is,
\begin{equation} \label{eq:q_P_zero}
 \phi_{\bar{n}}^* = \lambda_{\DD,\P\T} \, \phi_n, ~n \in \DD ~~ \Longleftrightarrow ~~ q_{\DD,\P}^\pm = 0,
\end{equation}
with unimodular `eigenvalue', $|\lambda_{\DD,\P\T}| = 1$, which in turn corresponds to locally symmetric density $\rho_n$ within $\DD$. 
In fact, this applies for arbitrary $\P_{\DD}\T$-symmetric hoppings, since no hopping ratio is accumulated in \cref{eq:q_n_+-_transfer} for zero current.
For the (local) $\P\T$ eigenstate above it further holds that
\begin{equation} \label{eq:qtilde_P_j_PTeig}
 \lambda_{\DD,\P\T} \, \mathring{q}_{n,\P}^\pm = - j_n^\pm = j_{\bar{n}}^\mp,
\end{equation}
connecting the dual nonlocal current with the local one.\\

\noindent\textbf{\small (ii)} For local $\P_{\DD}$-symmetry ($v_{\bar{n}= 2\ax - n} = v_n$, $h_{\bar{n},\bar{n}\pm1} = h_{n,n\mp1}$) or local $\K_{\DD}$-symmetry ($v_{\bar{n}= L + n} = v_n$, $h_{\bar{n},\bar{n}\pm1} = h_{n,n\pm1}$), the dual (time-dependent) quantities $\mathring{q}_{n,\S}^\pm$ are spatially constant in $\DD$,
\begin{equation} \label{eq:qtilde_constant_D}
 \mathring{q}_{n,\S}^\pm = \mathring{q}_{\DD,\S}^\pm ~~~~ (\S = \P,\K).
\end{equation}
We underline that those invariants exist also for non-Hermitian $\S_\DD$-symmetric potentials, e.\,g. with (locally) symmetric loss profile or `unbalanced' loss/gain elements.
In analogy to \cref{eq:q_P_zero}, $\mathring{q}_{\DD,\P}^\pm$ vanishes iff the part $\bs{\phi}_\DD$ is a local $\P$ `eigenstate',
\begin{equation} \label{eq:qtilde_P_zero}
 \phi_{\bar{n}} = \lambda_{\DD,\P} \, \phi_n, ~n \in \DD ~~ \Longleftrightarrow ~~ \mathring{q}_{\DD,\P}^\pm = 0,
\end{equation}
now with eigenvalue $\lambda_{\DD,\P} = \pm 1$, and then also
\begin{equation} \label{eq:q_P_j_Peig}
 \lambda_{\DD,\P} \, q_{n,\P}^\pm = j_n^\pm = j_{\bar{n}}^\mp
\end{equation}
holds in $\DD$. \\

\noindent\textbf{\small (iii)} For a general state $\ket{\psi}$, $q_{n}^\pm$ and $\mathring{q}_{n}^\pm$ are $\P_\DD\T$-symmetric and $\P_\DD$-antisymmetric by construction, respectively, 
\begin{equation} \label{eq:q_P_symmetry_by_construction}
 q_{n,\P}^{\pm} = q_{\bar{n},\P}^{\mp *}  ~~~ \text{and} ~~~ \mathring{q}_{n,\P}^{\pm} = -\mathring{q}_{\bar{n},\P}^{\mp}.
\end{equation}
For equidirectional $\P_\DD\T$- or $\P_\DD$-symmetric hoppings we further have that  
\begin{equation} \label{eq:q_P_symmetry}
 q_{\bar{n} \mp 1,\P}^\pm = -q_{n,\P}^{\pm *} ~~~ \text{or} ~~~ \mathring{q}_{\bar{n} \mp 1,\P}^\pm = \mathring{q}_{n,\P}^{\pm}
\end{equation}
within $\DD$, respectively, for \textit{arbitrary} onsite elements $v_n$.
Thus, the $|q_{n,\P}^\pm|$ or $\mathring{q}_{n,\P}^\pm$ are then locally symmetric about the center $\ax$ (that is, about the shifted center $\ax \mp \frac{1}{2}$ if assigned to sites) for any complex potential profile, regardless of its symmetry.
If also the potential is $\P_\DD\T$-symmetric, then \cref{eq:q_constant_D,eq:q_P_symmetry} imply that stationary NLCs are imaginary in $\DD$ for real $E$: $q_{n,\P}^\pm \in i\RR$.\\

\noindent\textbf{\small (iv)} For an eigenstate (Bloch state) $\ket{\phi^k}$ of the finite translation operator $\hat{K}$ with wavefunction $\phi_n^k = e^{ikn} \chi^k_n$, where $\chi^k_n = \chi^k_{n+L}$, the associated NLC equals the local current times the Bloch factor while its dual vanishes,
\begin{equation} \label{eq:q_K_Bloch}
 q_{n,\K}^\pm = j_{n}^\pm  e^{ikL}, ~~~~ \mathring{q}_{n,\K}^{\pm} = 0.
\end{equation}
This applies for energy eigenstates of globally $\K$-symmetric systems of period $L$ with symmetry domain $\DD = \NN$, including finite ring structures (where site $n = N$ is coupled to site $n = 1$).
Recall here that a $\K\T$-symmetric potential, where the $q_{n,\K}^\pm$ above may be spatially constant, has period $L$ only if it is real (whereby it is $\K$-symmetric), in which case also the $j_{n}^\pm$ are conserved.\\

\noindent\textbf{\small (v)} The NLCs for a forward translation $\K^+(n) = \bar{n} = n + L$ in a general state $\ket{\psi}$ are related to those for a backward translation $\K^-(n) = n - L$ from the mapped site $\bar{n}$ by
\begin{equation} \label{eq:q_K_-L}
 q_{n,\K^+}^\pm = -q_{\bar{n},\K^-}^{\mp *}, ~~~
 \mathring{q}_{n,\K^+}^\pm = -\mathring{q}_{\bar{n},\K^-}^{\mp}
\end{equation}
for $\K\T$- and $\K$-symmetric equidirectional hoppings, respectively.
Thus, for a stationary state (with real $E$), although the $q_{n,\K^+}^\pm$ are constant only in the subdomain $\DD$ of a locally $\K\T$-symmetric domain $\UU = \DD \cup \mathbb{L}$ (see \cref{fig:setup}), we can assign $-q_{n,\K^-}^{\mp *} = q_{n-L,\K^+}^\pm = q_{\DD,\K^+}^\pm$ to (links from) sites $n$ in the last period $\mathbb{L}$, and similarly for $\K$ symmetry and the $\mathring{q}_{n,\K}^\pm$.
This way the symmetry is expressed by nonlocal invariants throughout the whole domain $\UU$.

\subsection{Amplitude mapping and current connection}
\label{sec:current_mapping}

\begin{figure}[t!]
\centering
\includegraphics[width=.9\columnwidth]{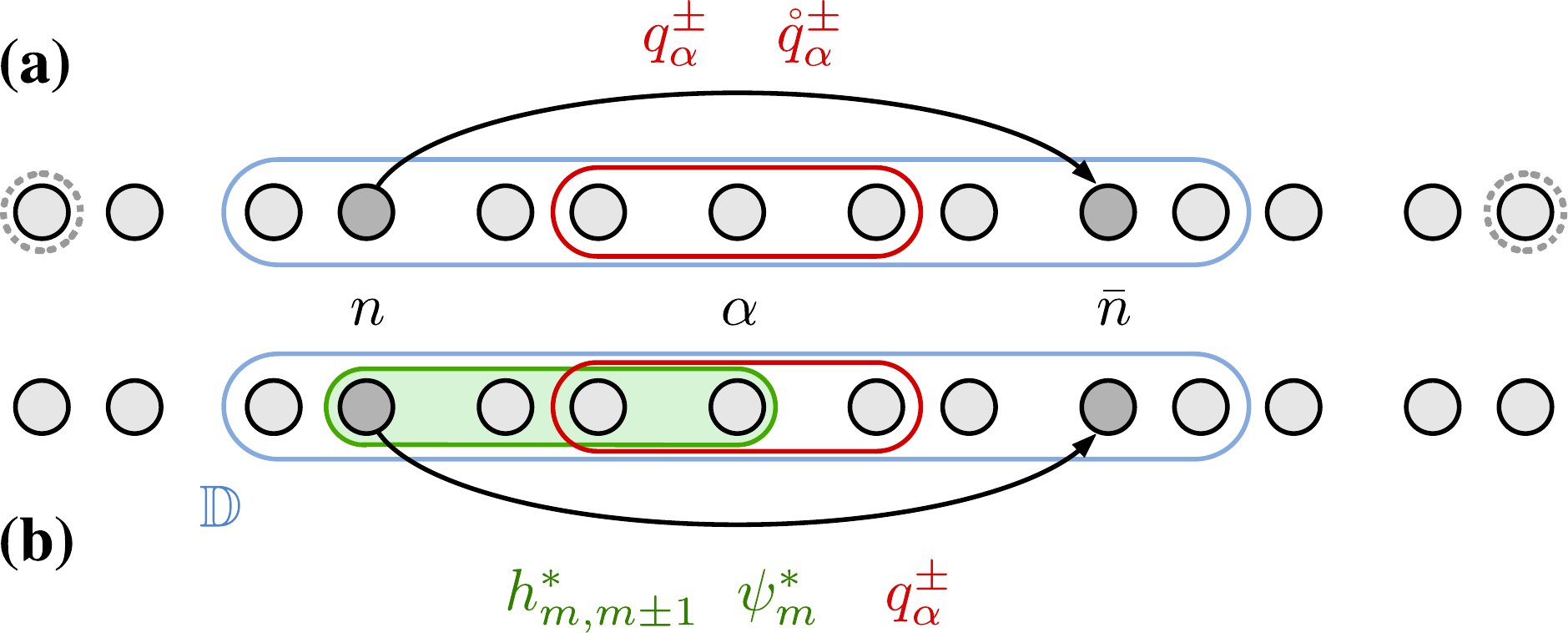}
\caption{Symmetry-induced domain amplitude mapping:
Within a locally $\P$-symmetric subdomain (blue line) of a 1D array, the amplitude at site $n$ is mapped to its image at $\bar{n}$, as indicated by arrows together with the quantities required for the mapping.
\textbf{(a)} Current mapping via \cref{eq:current_mapping_stat} in terms of translation invariant local currents $j_n^\pm$ and NLCs $q_n^\pm$, $\mathring{q}_n^\pm$ evaluated at $n = \ax$ using $a_{\ax}$ and $a_{\ax\pm 1}$ (red line).
Dotted contour on end sites indicate current flow to ensure $j_n^\pm \neq 0$.
\textbf{(b)} Summation mapping via \cref{eq:sum_map_PT} in terms of amplitudes and couplings up to the domain center $\ax$ (green line) and the domain invariant $q^\pm_\ax$.
}
\label{fig:mapping}
\end{figure}

The wave amplitude $\psi_{\S(n)}$ at the transformed site is related to that of the original one $\psi_n$ and its conjugate via the $q_{n,\S}^\pm,\mathring{q}_{n,\S}^\pm,j_n^\pm$ through the general identity 
\begin{equation} \label{eq:current_mapping}
 \psi_{\S(n)} = 
 \frac{1}{j_n^\pm} \left( q_{n,\S}^\pm  \,\psi_n - \mathring{q}_{n,\S}^\pm \,\psi_n^* \right),
\end{equation}
holding for arbitrary $\hat{H}$, where both the upper and lower sign in $\pm$ can be chosen; see derivation in \cref{app:mapping_derivation}.
If the potential $v_n$ is both $\S_\DD \T$- and $\S_\DD$-symmetric (that is, real and $\S_\DD$-symmetric), then both $q_{n,\S}^\pm = q_{\DD,\S}^\pm$ and $\mathring{q}_{n,\S}^\pm = \mathring{q}_{\DD,\S}^\pm$, as well as $j_n^\pm = j_\DD^\pm$, are spatially constant within $\DD$ for a stationary state $\ket{\phi}$.
Hence, \cref{eq:current_mapping} becomes
\begin{equation} \label{eq:current_mapping_stat}
 \phi_{\S(n)} = 
 \frac{q_{\DD,\S}^\pm}{j_\DD^\pm}  \,\phi_n - \frac{\mathring{q}_{\DD,\S}^\pm}{j_\DD^\pm} \,\phi_n^*
\end{equation}
for all $n$ in $\DD$, providing a linear \textit{mapping relation} from $\phi_n,\phi_n^*$ to $\phi_{\S(n)}$ with constant coefficients $q_{\DD,\S}^\pm/j_\DD^\pm$ and $\mathring{q}_{\DD,\S}^\pm/j_\DD^\pm$ which can be evaluated for any link $(n,n \pm 1)$ in $\DD$.
This mapping is illustrated in \cref{fig:mapping}\,(a).
Note that the same stationary mapping (\ref{eq:current_mapping_stat}) would be arrived at by formulating the identity in \cref{eq:current_mapping} using the bitemporal NLC $q_{n,\S}^{\sT\pm}$ instead of $\mathring{q}_{n,\S}^\pm$, as explained in \cref{app:mapping_derivation}.

%%%%% fig: mapping %%%%%%%%%%%%%%%%%%%%%%%%%%%%%%%%%%%%%%%%%%%%%%%%%%%%%%%%%%%%%%%%%%%%%%%%%%%
%%%%%%%%%%%%%%%%%%%%%%%%%%%%%%%%%%%%%%%%%%%%%%%%%%%%%%%%%%%%%%%%%%%%%%%%%%%%%%%%%%%%%%%%%%%%%%

This mapping is clearly not applicable for real spatial components $a_n$ of $\ket{\phi}$ in $\DD$ (or trivially complex through a common prefactor), since the $j_\DD^\pm$ then vanish.
At the same time, real $a_n$ yield equimodular $q_{n,\S}^\pm$ and $\mathring{q}_{n,\S}^\pm$, which in turn coincides quite generally with vanishing local current.
Indeed, a general identity connecting the nonlocal and local currents on links to sites $n$ and $\bar{n} = \S(n)$ is (see \cref{app:mapping_derivation}) 
\begin{equation} \label{eq:current_connection}
 |q_{n,\S}^\pm|^2 - |\mathring{q}_{n,\S}^\pm|^2 = j_n^\pm j_{\S(n)}^{\pm s}
\end{equation}
with the sign $s = -,+$ for $\S = \P,\K$, respectively, implying $j_{\P(n)}^{\mp} = j_{\P(n),\P(n \pm 1)}$ and $j_{\K(n)}^{\pm} = j_{\K(n),\K(n \pm 1)}$.
This \textit{current connection} holds also for $j_n^\pm = 0$, in which case $\mathring{q}_{n,\S}^\pm = e^{2i\theta_n}q_{n,\S}^\pm$, where $\theta_n$ is the phase of the general wavefunction $\psi_n = |\psi_n|e^{i\theta_n}$, as can easily be checked.
For a stationary state we further have that $j_n^\pm j_{\S(n)}^\pm = s(j_n^\pm)^2 \equiv s(j_{\UU}^\pm)^2$ within a Hermitian domain $\UU \ni n,\S(n)$.
Thus, the wave amplitudes in \textit{different} local $\S\T$ or $\S$ symmetry domains $\DD_d$ ($d=1,2,\dots$) in a globally Hermitian system are interrelated by common differences $|q_{\DD_1,\S}^\pm|^2 - |\mathring{q}_{\DD_1,\S}^\pm|^2 = |q_{\DD_2,\S}^\pm|^2 - |\mathring{q}_{\DD_2,\S}^\pm|^2 = \dots = s|j_\NN^\pm|^2$ through \cref{eq:current_connection} for $\S = \P,\K$.

Considering $\S$-symmetry, the mapping relation (\ref{eq:current_mapping_stat}) can be seen as a generalization of the parity and Bloch theorems (which map amplitudes to a symmetry-related site via the associated eigenvalue) from global to local $\P$- and $\K$-symmetry, respectively, where breaking of the global symmetry adds an admixture of the complex conjugate $\phi_n^*$ to the mapping.
Thus, using the NLCs, local symmetry can be used to facilitate the computation of spatial parts of eigenstates (knowing their symmetry-related parts) even in the case where symmetry under a given transformation is broken at domain boundaries.
Taking the continuum limit with lattice constant $\ell \to 0$ (setting all hoppings equal), \cref{eq:current_mapping_stat} reproduces the generalized Bloch and parity theorems of Ref.\,\cite{Kalozoumis2014_PRL_113_050403_InvariantsBroken} for stationary states.
Those are here extended to non-Hermitian Hamiltonians with variable hoppings and are shown to follow from the general identity (\ref{eq:current_mapping}) for arbitrary states.
The usual (global) versions are recovered for $\DD = \NN$; see \cref{app:mapping_derivation}.

\subsection{Summation amplitude mapping}
\label{sec:sum_mapping}

While the current connection (\ref{eq:current_connection}) holds for any $\hat{H}$ and $\ket{\psi}$, the mapping identity (\ref{eq:current_mapping}) is only for nonzero $j_n$, as realized for scattering setups (see \cref{sec:scattering}) or bound setups with sources and/or sinks (see \ref{sec:non-Hermitian}).

As shown in \cref{app:mapping_derivation}, the NLCs can be used to map also amplitudes without using the (potentially vanishing) local current  $j_n$, in the form of a general \textit{summation} mapping; see \cref{eq:sum_map_gen}.
In the case of real amplitudes $a_n$ of a stationary state within an odd-sized domain $\DD$, locally $\P\T$-symmetric around $\ax$, we obtain the mapping
\begin{equation} \label{eq:sum_map_PT}
 a_{\bar{n}} = a_n \left( 1 - i q_\DD^+ \sum_{m=n}^{\ax-1} \frac{1}{a_{m} h_{m,m+1}^* a_{m+1}} \right)
\end{equation}
with constant NLC $q_\DD^+$ in $\DD$ (with the upper sign selected here).
The image amplitudes $a_{\bar{n}}$ can thus be determined from multiple $a_n$'s, as illustrated \cref{fig:mapping}\,(b).
The same can be done for an odd-sized $\P$-symmetric domain using $\mathring{q}_\DD^+$ and $a_{m},h_{m,m+1}$ in \cref{eq:sum_map_PT}.
In other cases (e.\,g. $\P(\T)$ symmetry with even $D$, or $\K(\T)$ symmetry), \cref{eq:sum_map_gen} yields the amplitude mapping provided that the quotient $a_{\bar{n}_0}/a_{n_0}$ can be previously determined.

\subsection{Invariants in periodically modulated systems}
\label{sec:inv_per_mod}

For a time-dependent Hamiltonian $\hat{H}(t)$ there are no stationary eigenstates and so no spatially invariant NLCs in symmetry domains.
In the case, however, of a \textit{periodic} modulation with frequency $\omega$,
\begin{equation}
 \hat{H}(t) = \hat{H}(t + \tau), ~~~~ \tau = 2\pi/\omega,
\end{equation}
the Schr\"odinger equation (\ref{eq:Schr}) admits solutions of the form
\begin{equation} \label{eq:Floquet_state}
 \ket{\psi^{{\mu}}(t)} = e^{-i\epsilon_{{\mu}}t} \ket{\phi^{{\mu}}(t)}, ~~ \ket{\phi^{{\mu}}(t)} = \ket{\phi^{{\mu}}(t + \tau)},
\end{equation}
where the real \textit{quasienergies} $\epsilon_{{\mu}}$ are defined up to multiples of $2\pi$.
Being \textit{quasi}-stationary eigenstates of the modified Hamiltonian operator $\hat{H}_F = \hat{H} - i\partial_t$ with eigenvalues $\epsilon_{{\mu}}$, the periodic Floquet modes $\ket{\phi^{{\mu}}(t)}$ suggest the existence of a type of nonlocal invariants within spatially symmetric domains, in analogy to those for stationary states of a static $\hat{H}$.
Indeed, integrating the nonlocal continuity \cref{eq:continuity_nonlocal} in a Floquet state $\ket{\psi^{{\mu}}}$ over one period $\tau$ for a potential $v_n(t)$ (assumed real for simplicity) which is locally $\S$-symmetric at any $t$, $v_\ba{n}(t) = v_n(t)$, we obtain
\begin{equation}
 \bar{q}_{n;\mu} \equiv \bar{q}_{n;\mu}^+ + \bar{q}_{n;\mu}^+ = \left[ \, \sigma_{n;\mu}(t) \, \right]_0^\tau = 0
\end{equation}
since $\sigma_{n;\mu}(t) = \braket{\psi^{{\mu}}(t)|\hat{\sigma}_{n}|\psi^{{\mu}}(t)}$ is periodic, where 
\begin{equation}
 \bar{q}_{n;\mu}^\pm = \frac{1}{\tau} \int_0^\tau dt \, q_{n;\mu}^\pm(t)
\end{equation}
are the period-averaged NLCs in state $\ket{\psi^{{\mu}}}$.
Just like in the case of static $\hat{H}$, we have $\bar{q}_{n \mp 1;\mu}^\pm = \bar{q}_{n;\mu}^\mp$ for equidirectional (assumed real) $\S$-symmetric hoppings $h_{\S(n),\S(n \pm 1)} = h_{n,n \pm 1}$, so that the $\bar{q}_{n;\mu}^\pm$ are spatially constant in the domain of symmetry $\DD$; cf. \cref{eq:q_n_+-_transfer}.
This generalizes the characterization of the LSs by associated invariants to 1D systems driven periodically with arbitrary strength.
Like their static counterpart, the period invariants $\bar{q}_{n;\mu}^\pm$ also exist independently of boundary conditions; 
they will be illustrated in an example with bound Floquet modes in \cref{sec:inv_per_mod_bound_Floquet}.

\section{Nonlocal currents of bound eigenmodes}
\label{sec:bound_eigenmodes}

Having established the existence of symmetry-induced stationary 1D invariants, we now address their occurrence in bound systems.
We first illustrate the various types of LS in 1D, then show how the NLCs relate to symmetry breaking in globally $\S\T$-symmetric non-Hermitian setups, and finally give an example of invariants in a periodically modulated array.
Lengths, field amplitudes and ${\mathcal{H}}$-elements are measured in units of $\ell$ (lattice constant), $a^\circ$ (amplitude unit) and $\epsilon$ (${\mathcal{H}}$-element unit), respectively, which are in turn set to unity, $\ell = a^\circ = \epsilon = 1$.

%%%%% fig: local_symmetries %%%%%%%%%%%%%%%%%%%%%%%%%%%%%%%%%%%%%%%%%%%%%%%%%%%%%%%%%%%%%%%%%%
\begin{figure*} 
\begin{center}
\includegraphics[width=1\textwidth]{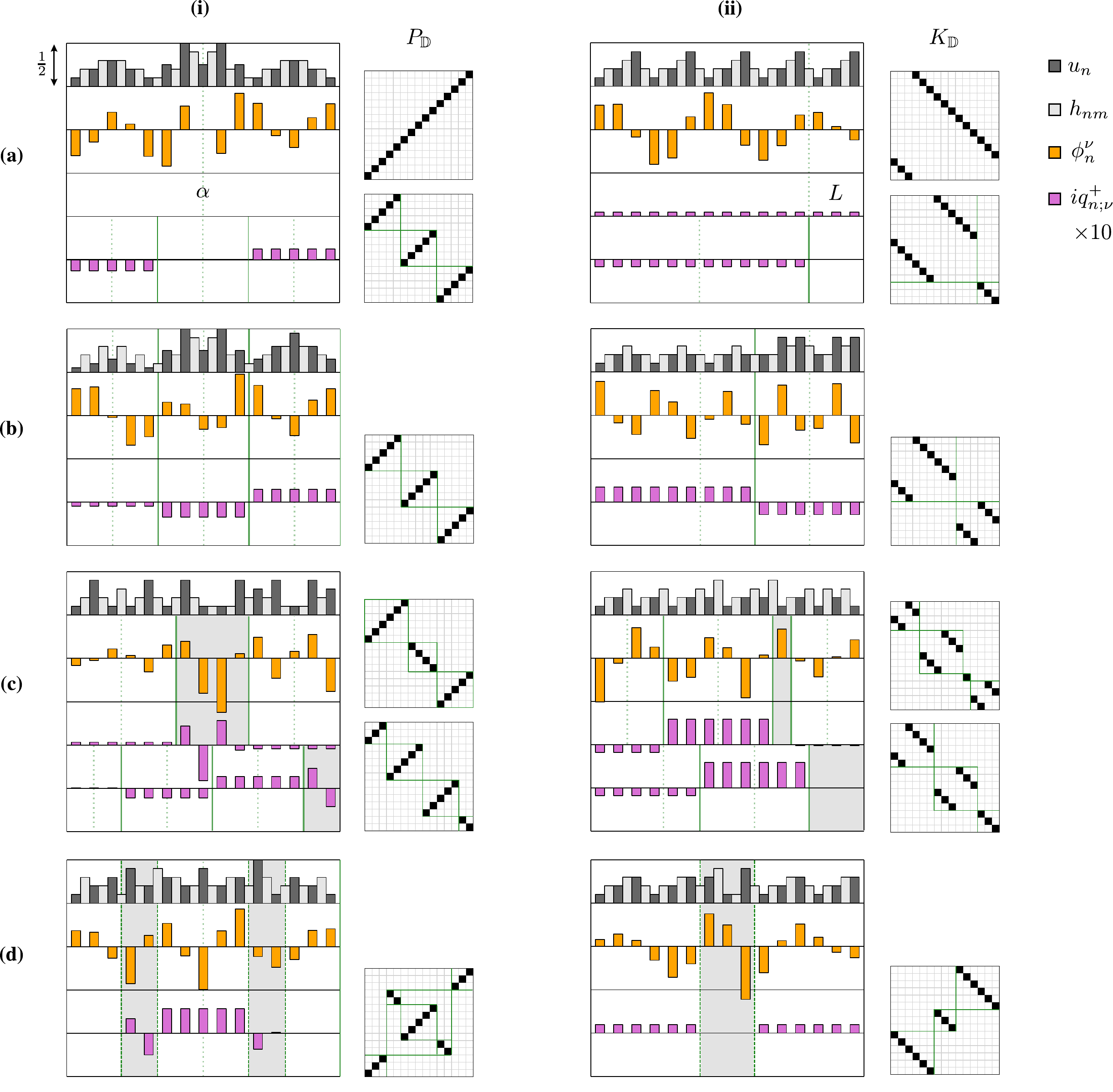}
\caption{Types of local symmetry (LS) and corresponding local permutation matrices for \textbf{(i)} inversion and \textbf{(ii)} translation transforms for finite 1D arrays of $N = 15$ sites.
Plotted quantities (indicated in legend) are: real onsite potential $u_n$ (dark gray), real hoppings $h_{nm}$ between sites (light gray), eigenmode amplitudes $\phi^\nu_n$ (orange), $10i$ times imaginary NLC $q_{n,\nu}^+$ of mode $\nu$ (purple).
\textbf{(a)} Global symmetry within $\UU = \NN$, where $\NN$ covers the complete system,
\textbf{(b)} complete LS (CLS) in non-overlapping subdomains $\DD_d \subset \NN = \bigcup_d \DD_d$,
\textbf{(c)} overlapping LS for two different LS operations, and 
\textbf{(d)} gapped symmetry within $\UU = \NN$.
Eigenmodes (i) $\nu = 8,7,9,7$ and (ii) $\nu = 6,10,8,6$ for (a,b,c,d), respectively, were chosen with visibly distinct $q_{\DD_d}^+$ as a criterion.
In (a,i) and (a,ii) the $q_{n,\S}^+$ for a LS transform are additionally shown.
In order to illustrate their domainwise constancy, the (upper) NLCs are plotted at the sites $n$ below the links $(n,n+1)$ they are defined on.
In (ii), backward invariants $-q_{\bar{n},\K^-}^{\mp *}$ are plotted in the last period of each LS domain; see \cref{eq:q_K_-L}.
Solid vertical lines indicate LS domains, dotted lines the local (i) inversion points $\ax_d$ and (ii) translation lengths $L_d$.
Shaded areas indicate lack of symmetry under the considered transform.
All plotted quantities are dimensionless, with each panel row being of height $1/2$, as indicated on the upper left in (a,i).
}
\label{fig:local_symmetries}
\end{center}
\end{figure*}
%%%%%%%%%%%%%%%%%%%%%%%%%%%%%%%%%%%%%%%%%%%%%%%%%%%%%%%%%%%%%%%%%%%%%%%%%%%%%%%%%%%%%%%%%%%%%%%

\subsection{Types of local symmetry}
\label{sec:symmetry_types}

To illustrate different types of symmetry, we consider an array of $N = 15$ sites with real onsite elements $v_n = u_n$ and hoppings $h_{n,n \pm 1}$, so that the mode amplitudes $a^{\nu}_n$ of an eigenstate $\ket{\phi^{\nu}} = \sum_n \phi^{\nu}_n \ket{n} = \sum_n a^{\nu}_n e^{-iE_\nu t} \ket{n}$ with eigenvalue $E_\nu$ ($\nu = 1,2,\dots,N$) can also be chosen real.
The NLCs 
\begin{equation} \label{eq:q_n_psi_wavepacket}
 q_{n;\psi}^\pm = \sum_\nu |c_\nu|^2 q_{n;\nu}^\pm + \sum_\nu \sum_{\mu \neq \nu}  c_\nu^* c_\mu q_{n;\nu\mu}^\pm
\end{equation}
in a general state (wavepacket) $\ket{\psi} = \sum_\nu c_\nu \ket{\phi^{\nu}}$ are composed of a linear combination of the $q_{n;\nu}^\pm$ of individual modes $\nu$ as well as mixed-mode currents $iq_{n;\nu\mu}^\pm \equiv \phi^{\nu *}_n h_{\S(n),\S(n \pm 1)} \phi^\mu_{\S(n \pm 1)} - \phi^{\nu *}_{n \pm 1} h_{n,n \pm 1}^* \phi^\mu_{\S(n)}$ ($\S = \P,\K$) which oscillate with frequency $E_\nu - E_\mu$.
All general statements of \cref{sec:discrete_continuity,sec:stat_invariants_mappings} about NLCs of general states $\ket{\psi}$ apply to $q_{n;\psi}^\pm$, while, in particular, the stationary currents $q_{n;\nu}^\pm$ are additionally translation invariant in a corresponding LS domain $\DD$.
In the following, we discuss the different types of local $\P$- and $\K$-symmetry in terms of the example arrays shown in \cref{fig:local_symmetries} together with the locally invariant $q_{n;\nu}^\pm$ for selected modes $\ket{\phi^{\nu}}$ as well as the $\P_\DD$- and $\K_\DD$-matrices producing them via \cref{eq:q_n_+-_vector}. \\

\noindent \textit{(a) Global symmetry}  \\
In the case of global symmetry, there is a connected maximal symmetry domain $\UU = \NN$ which extends over the complete setup.
We call a (local) symmetry maximal \cite{Morfonios2014_ND_78_71_LocalSymmetry} if the corresponding domain $\UU = \DD \cup \bar{\DD}$ is the largest one with the given center $\ax$ (for $\P$ symmetry) or period $L$ (for $\K$ symmetry).
In the case of global $\P$ symmetry, shown in \cref{fig:local_symmetries}\,(a,i), the corresponding permutation matrix $P_\DD = P_\NN$ is the usual anti-diagonal unit matrix (upper inset), and the $q_{n;\nu}^\pm$ vanish for any mode $\ket{\phi^\nu}$ due to its definite parity (see \cref{eq:q_P_zero}).
If $\NN$ contains a subdomain $\DD_1$ which is locally $\P_{\DD_1}$-symmetric, then its image $\P(\DD_1)$ under the global inversion  has opposite invariants $q_{\P(\DD_1);\nu}^\pm = - q_{\DD_1;\nu}^\pm$, as seen in the lower panel of \cref{fig:local_symmetries}\,(a,i) together with the corresponding local inversion matrix (lower inset).

We call a finite system `globally' $\K$-symmetric here if it is periodic in $\UU = \DD \cup \LL = \NN$, i.e. consisting of repeated copies of its last period $\LL$, as in \cref{fig:local_symmetries}\,(a,ii).
While the $q_{n;\nu}^\pm$ in $\LL$ differ from those in $\DD$, we assign the backward invariants $-q_{\bar{n},\K^-}^{\mp *}$ (see \cref{eq:q_K_-L}) to $\LL$ in order to illustrate the $\K_\DD$-symmetry.
As seen in the lower panel of \cref{fig:local_symmetries}\,(a,ii), a different period leads to different NLCs. \\

\noindent \textit{(b) Complete local symmetry} \\
In the case of complete LS (CLS), the setup is exactly covered by more than one maximal non-overlapping symmetry domains $\UU_d$ (of size $D_d \geqslant 1$), that is, $\NN = \UU = \bigcup_d \UU_d$ with $\UU_d \cap \UU_{d'} = \varnothing$, as exemplified in \cref{fig:local_symmetries}\,(b).
The corresponding $P_\DD$- and $K_\DD$-matrices are block-diagonal with the blocks being local inversion and translation permutation matrices.
The local order of the Hamiltonian is here completely mapped to the domain invariants $q_{\DD_d;\nu}^\pm$ which thus expose the symmetry information encoded in the states $\ket{\phi^\nu}$.
Note that, for both local $\P$ and $\K$ symmetry, this information is otherwise not at all visible by the mode amplitude profiles alone for an arbitrary CLS setup. 
The concept of CLS may serve to characterize order between the limits of perfect periodicity and uncorrelated disorder in extended systems, alternatively to quasiperiodicity (or deterministic aperiodicity) \cite{Morfonios2014_ND_78_71_LocalSymmetry}. \\

\noindent \textit{(c) Overlapping local symmetry} \\
The case of overlapping LS refers to the existence of \textit{different} symmetry domains of the same setup with common sites, that is, there are (composite) LS domains $\UU = \bigcup_d \UU_d$ and $\UU' = \bigcup_d \UU'_d$ with $\UU_d \neq \UU'_d$ but $\UU_d \cap \UU'_d \neq \varnothing$ for some $\UU_d,\UU'_d$.
Examples of non-CLS setups with overlapping $\P$- and $\K$-symmetric domains are shown in \cref{fig:local_symmetries}\,(c,i) and (c,ii), respectively.
Note that overlapping $\P$-symmetric domains with common center $\ax$ or $\K$-symmetric domains with common period $\LL$ have equal invariant currents $q_{n;\nu}^\pm$, respectively. 
In general, a setup may contain mixed local $\P$ and $\K$ symmetries in different domains which in turn may overlap, and the concept of overlapping LS may be used to analyze order in complex systems at different scales \cite{Morfonios2014_ND_78_71_LocalSymmetry}.
If overlapping LS domains occur in a CLS system, then we speak of different CLS \textit{decompositions} of the system.
As has been shown for continuous models \cite{Kalozoumis2013_PRA_88_033857_LocalSymmetries}, scattering setups with multiple complete local $\P$-symmetry decompositions can enable the design of transparency at multiple incoming frequencies.\\

\noindent \textit{(d) Gapped (local) symmetry} \\
If the Hamiltonian elements in one (or more than one) subdomain(s) of a LS domain are modified so that they break the considered symmetry, then the original LS domain becomes disconnected, or `gapped', as illustrated in \cref{fig:local_symmetries}\,(d).
The $q_{n;\nu}^\pm$ are then spatially constant in the symmetric subparts of $\DD$ outside the gap(s), but generally unequal in different subparts (recall that their constancy pertains from \cref{eq:q_n_+-_transfer} for a connected LS domain).
In the case of gapped $\P$ symmetry, however, the identity (\ref{eq:q_P_symmetry}) yields equal $q_{n;\nu}^\pm$ at symmetry-related subparts for real amplitudes $a_n$, despite intervening gaps, as seen in \cref{fig:local_symmetries}\,(d,i) with equal (vanishing) $q_{n;\nu}^+$ at the left and right ends.
In the case of gapped $\K$ symmetry, gaps induce distinct constant $q_{n;\nu}^\pm$'s in each unperturbed part.
The example in \cref{fig:local_symmetries}\,(d,ii) shows the constancy of $q_{n;\K^+}^+$ (with period $L = 9$) before the gap, equal to the $-q_{n;\K^-}^+$ shown after the gap. 
The blocks of the associated $\varSigma$-matrix corresponding to gaps are diagonal, yielding $q_{n;\nu}^+ = j_{n;\nu}^+ = 0$ in the gaps for $\K$ transformations.
Note that the gaps may as well be locally symmetric under another transformation (e.\,g. $\P$-symmetric about their center), thus yielding a \textit{nested} LS, which can be expressed through locally constant $q_{n;\nu}^\pm$'s by accordingly modifying those permutation matrix blocks.

%%%%% fig: nonherm_modeQ_sites %%%%%%%%%%%%%%%%%%%%%%%%%%%%%%%%%%%%%%%%%%%%%%%%%%%%%%%%%%%%%%%%
\begin{figure}[t!]
\centering
\includegraphics[width=.9\columnwidth]{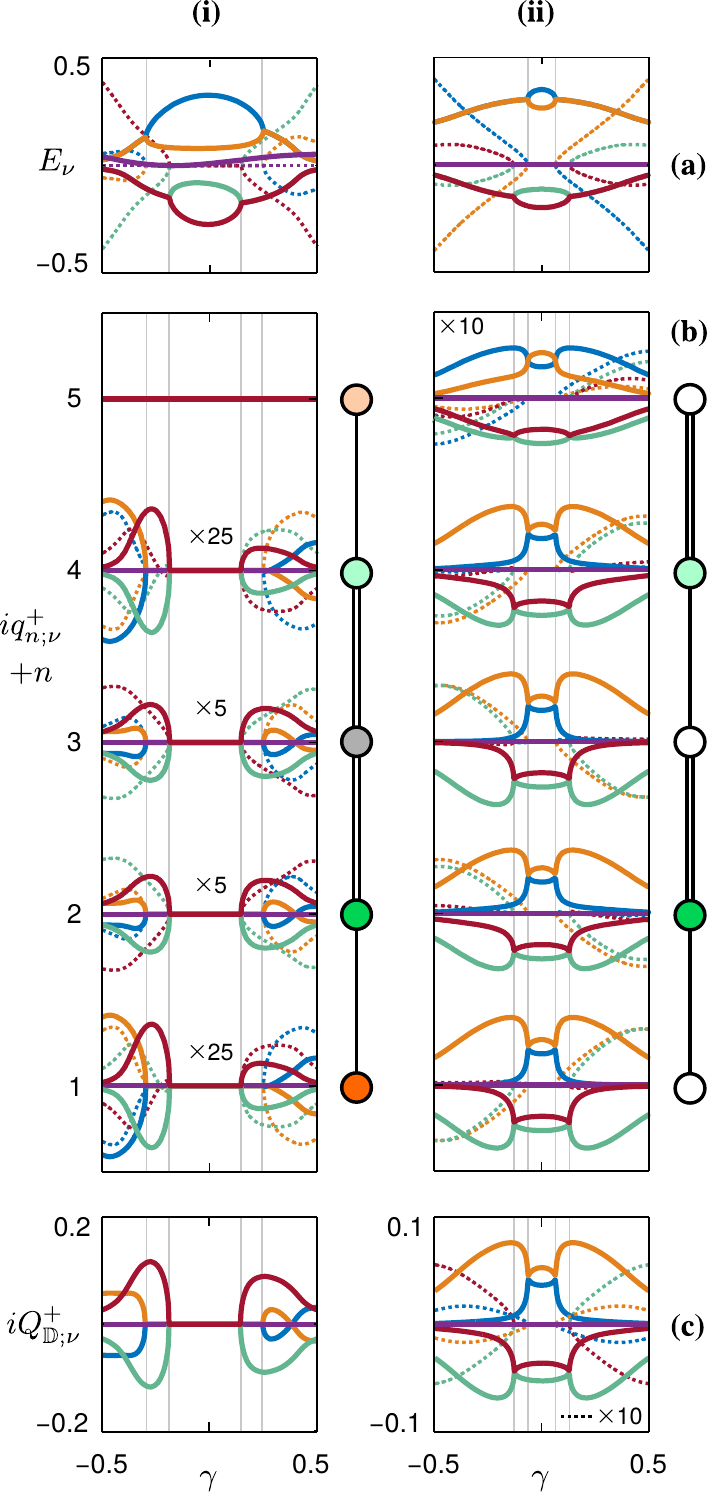}
\caption{Real (solid) and imaginary (dotted) part of \textbf{(a)} spectrum $E_\nu$, \textbf{(b)} spatially resolved upper NLCs $iq_{n;\nu}^+$ (offset by site index $n$), and \textbf{(c)} total upper NLC $iQ_{\DD;\nu}^+$ within $\DD$, in the modes $\nu$ of a \textbf{(i)} $\P\T$-symmetric array with $\{ v_n \} = \{ i\eta, i\gamma, u, -i\gamma, -i\eta \}$, $\{ h_{n,n+1} \} = \{ h, 2h, 2h, h \}$ where $u = \eta = 0.06$ and $h = 0.1$, and a \textbf{(ii)} $\K\T$-symmetric array with $\{ v_n \} = \{ 0, u + i\gamma, 0, u - i\gamma, 0 \}$, $\{ h_{n,n+1} \} = \{ h, 2h, h, 2h \}$ where $u = 0.15$ and $h = 0.1$, for varying loss/gain parameter $\gamma$.
Vertical lines indicate exceptional values of $\gamma$ where transitions from complex to real eigenvalues $E_\nu$ occur, corresponding to transitions from complex to (i) zero $q_{n,\P}^+$ and (ii) imaginary $q_{n,\K}^+$ in the respective modes.
Note that $q_{N,\P}^+ = 0$ identically in (b,i) since it `flows' on the nonexisting links $(0,1)$ and $(N,N+1)$.
By convention, $\DD \equiv \{1,2\}$ in (ii); see text.
}
\label{fig:nonherm_modeQ_sites}
\end{figure}
%%%%%%%%%%%%%%%%%%%%%%%%%%%%%%%%%%%%%%%%%%%%%%%%%%%%%%%%%%%%%%%%%%%%%%%%%%%%%%%%%%%%%%%%%%%%%%%

\subsection{Nonlocal currents in $\P\T$- and $\K\T$-symmetric non-Hermitian arrays}
\label{sec:non-Hermitian}

We now proceed to demonstrate some properties of eigenmode NLCs for minimalistic examples of globally $\P\T$- and $\K\T$-symmetric non-Hermitian arrays, as shown in \cref{fig:nonherm_modeQ_sites}, and in particular their relation to spontaneous symmetry breaking.
The non-Hermiticity is implemented by complex onsite potentials, while the hoppings are taken to be real and equidirectional.

Global $\P\T$ symmetry is known to allow for parameter regimes of loss/gain elements $\gamma_n$ where a subspace of common eigenstates $\ket{\phi^\nu}$ of $\hat{H}$ and $\hat{P}\hat{T}$ with real $\hat{H}$-eigenvalues exists, originating essentially from the balance between loss and gain in density.
The rest of $\hat{H}$-eigenstates are `spontaneously' symmetry-broken non-$\hat{P}\hat{T}$-eigenstates, which come in $\hat{P}\hat{T}$-related pairs $\ket{\phi^\nu},\ket{\phi^{\nu\prime}}$ with complex conjugate eigenvalues $E_\nu,E_\nu'$.
We now notice that the $\P\T$-unbroken modes are naturally distinguished by vanishing of the corresponding constant NLCs $q_{n,\P}^\pm$, as follows from \cref{eq:q_P_zero} for the case $\DD = \NN$ of global symmetry.
This is illustrated in \cref{fig:nonherm_modeQ_sites}\,(i) for a $\P\T$-symmetric $N=5$-site array with two loss/gain parameters $\gamma,\eta$, where a real central onsite energy $v$ is added to break the $\pm E_\nu^\circ$ symmetry of the $\hat{H}$-eigenvalues $E_\nu = E_\nu^\circ + i \varGamma_\nu/2$ shown in \cref{fig:nonherm_modeQ_sites}\,(a,i).
As we see, all $q_{\NN;\nu}^+$ vanish within the $\gamma$-range (for fixed $\eta$) corresponding to the unbroken phase of real $E_\nu$ in \cref{fig:nonherm_modeQ_sites}\,(a,i).

In the $\P\T$-broken phase of complex $E_\nu$, the $q_{n;\nu}^\pm$ are not spatially constant, since $q_{n \mp 1;\nu}^\pm = q_{n;\nu}^\mp \neq q_{n;\nu}^\pm$ for $\varGamma_\nu \neq 0$ (and $\asym_n = 0$) in \cref{eq:continuity_nonlocal_eigenstate}. 
While generally $q_{n;\nu}^\pm \neq 0$ in the $\P\T$-broken phase, the \textit{remnant} of $\P\T$ symmetry in the occurrence of $\P\T$-related eigenmode pairs is expressed by vanishing sum of NLCs:
By its definition, $q_{n;\nu}^\pm$ will be opposite $q_{n;\nu'}^\pm$ in the paired eigenstate $\ket{\phi^{\nu\prime}} = \hat{P}\hat{T} \ket{\phi^\nu}$ with $E_\nu' = E_\nu^*$,
\begin{equation} \label{eq:q_n_PT_pair_zero_sum}
 q_{n;\nu}^\pm + q_{n;\nu'}^\pm = 0.
\end{equation}
Summing separately the $q_{n;\nu}^\pm$ along the array, the total upper/lower NLCs $Q_{\DD;\nu}^\pm = \sum_{n \in \DD} q_{n;\nu}^\pm$ (here $\DD = \NN$) in state $\ket{\phi^\nu}$ each serve as a single parameter which distinguish $\P\T$-unbroken ($Q_{\DD;\nu}^\pm = 0$) from $\P\T$-broken ($Q_{\DD;\nu}^\pm \neq 0$) modes for a given setup configuration.
This is seen in \cref{fig:nonherm_modeQ_sites}\,(c,i) for $Q_{\DD;\nu}^+$ (note that $Q_{\DD;\nu}^\pm \in i\RR$ due to \cref{eq:q_P_symmetry}).
The $\P\T$ symmetry remnant in the broken phase may further be expressed by the vanishing sum of $Q_{\DD;\nu}^\pm$ over all modes, $\sum_\nu Q_{\DD;\nu}^\pm = 0$.

The nonlocal charge (quasipower) $\varSigma_\psi^\DD = \braket{\psi|\hat{\varSigma}|\psi}$ for a general state $\ket{\psi}$ was shown in \cref{sec:discrete_continuity} to remain constant in time (longitudinal direction) $t$ under under global $\P\T$ symmetry ($\DD = \NN$) following from the vanishing total NLC $Q_\NN = Q_\NN^+ + Q_\NN^- = 0$.
In particular for the eigenmodes $\ket{\phi^\nu}$ addressed here, we have 
\begin{equation}
 \varGamma_\nu \varSigma_\nu^\NN = \partial_t \varSigma_\nu^\NN = Q_{\NN;\nu} = 0,
\end{equation}
which shows that the nonlocal charge $\varSigma_\nu^\NN$ is zero in $\P\T$-broken eigenmodes with $\varGamma_\nu \neq 0$.

The $\P\T$ symmetry of $\hat{H}$ is not a necessary condition for the occurrence of real $E_\nu$ \cite{Klaiman2008_PRA_78_062113_Non-HermitianHamiltonians}, which we now exploit to show the induced invariance of translation NLCs $q_{n,\K}^\pm$ in a $\K\T$-symmetric example array.
Specifically, we implement a non-Hermitian version of the Su-Schrieffer-Heeger model \cite{Zhu2014_PRA_89_062102_MathcalptSymmetry}, which can be considered as a special case of the non-Hermitian Aubry-Andr\'e-Harper model proposed in Ref.\,\cite{Harter2016_PRA_93_062101_Mathcalpt-breakingThreshold}, in an array of odd size $N$ where the hoppings are $\K$-symmetric with period $L = 2$ and the potential is simultaneously $\K\T$- and $\P\T$-symmetric.
The latter property enables the presence of loss/gain-balanced eigenmodes with real $E_\nu$ \cite{Harter2016_PRA_93_062101_Mathcalpt-breakingThreshold}.
Since the array does not consist of an integer number of periods, we set $\DD \equiv \{1,2,\dots,N-L-1\}$, in which $H_{m,n} = H_{m+L,n+L}$.

In \cref{fig:nonherm_modeQ_sites}\,(ii) the $q_{n,\K}^+$ are shown for $N=5$ and a single site pair with loss/gain $\gamma$ supporting a phase of real $E_\nu$; see \cref{fig:nonherm_modeQ_sites}\,(a,ii).
Real onsite energies $v$ are added to break the $\pm E^\circ$-symmetry of the spectrum.
As seen in \cref{fig:nonherm_modeQ_sites}\,(b,ii), for each eigenmode the $\gamma$-region with real $E_\nu$ is distinguished by a spatially constant and imaginary $q_{n,\K;\nu}^+$, where $-q_{n,\K^-}^{- *} = q_{n-L,\K^+}^+$ is assigned to the last three sites $\NN \setminus \DD = \{3,4,5\}$; recall \cref{eq:q_K_-L}.
Outside this $\gamma$-region, the $q_{n,\K;\nu}^\pm$ vary along the array and have a real part, signifying the spontaneous breaking of loss/gain balance in the eigenstates.
In analogy to the $\P\T$-symmetric case above, though, a remnant of the symmetry of the system can be expressed in the broken phase by the imaginary sums $\sum_\nu Q_{\DD;\nu}^\pm \in i\RR$, as anticipated by the $Q_{\DD;\nu}^+$ in \cref{fig:nonherm_modeQ_sites}\,(c,ii).
The characteristic of imaginary NLCs in the unbroken phase thus survives in the sum of total NLCs within $\DD$ over all modes.

In the above non-Hermitian setup examples, we chose arrays with global $\P\T$ (or $\K\T$) symmetry which enable the possibility of a completely real eigenspectrum, in order to reveal the connection of the NLCs to the transition between phases of unitary and nonunitary evolution when varying loss/gain strengths. In particular, we saw that the NLCs vanish for all eigenstates in the $\P\T$-unbroken phase, corresponding to globally symmetric eigenstate densities. As seen from \cref{eq:q_P_zero}, however, the NLCs could also be used as indicators for \textit{locally} symmetric density of any given eigenstate. In a completely locally $\P\T$-symmetric array (i.\,e. decomposable into different, non-overlapping $\P\T$-symmetric domains), the simultaneous vanishing of all the attached domains' NLCs may thus aid in the engineering of selected real eigenvalues induced by balanced gain and loss, even though the setup is globally $\P\T$-asymmetric.

%%%%% fig: ls_driven %%%%%%%%%%%%%%%%%%%%%%%%%%%%%%%%%%%%%%%%%%%%%%%%%%%%%%%%%%%%%%%%%%%%%%%%%
\begin{figure}[t!]
\centering
\includegraphics[width=.95\columnwidth]{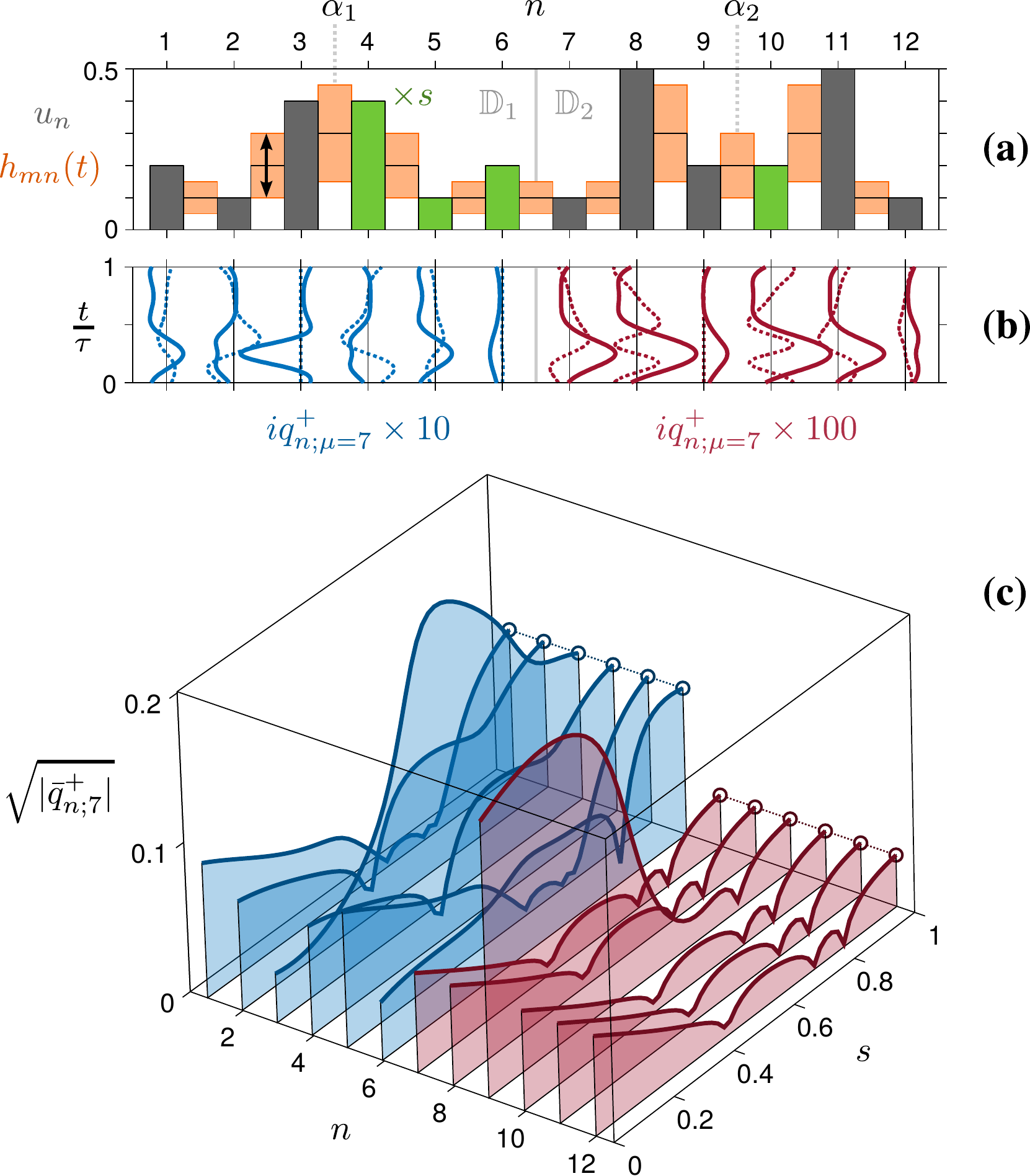}
\caption{\textbf{(a)} Setup with $N = 12$ sites composed of two domains $\DD_1$, $\DD_2$ which are locally $\P$-symmetric for the value $s = 1$ of a scaling parameter $s$ multiplying the onsite potentials $u_{3}, u_{4}, u_{5}, u_{10}$ (colored green), with periodically modulated hoppings $h_{n,n \pm 1}(t) = (1 + \frac{1}{2}\sin\pi t/8) h_{n,n \pm 1}(0)$.
\textbf{(b)} Time evolution over one period $\tau$ of the NLC in Floquet mode $\mu = 7$ associated with the local inversions $\P_{\DD_1}$, $\P_{\DD_2}$.
\textbf{(c)} Period averaged NLC for varying scaling parameter $s$, showing the domainwise constancy in the case $s = 1$ of local symmetry.
}
\label{fig:ls_driven}
\end{figure}
%%%%%%%%%%%%%%%%%%%%%%%%%%%%%%%%%%%%%%%%%%%%%%%%%%%%%%%%%%%%%%%%%%%%%%%%%%%%%%%%%%%%%%%%%%%%%%%

\subsection{Period invariants of bound Floquet states}
\label{sec:inv_per_mod_bound_Floquet}

The invariance of period-averaged NLCs $\bar{q}_{n;\mu}^\pm$ in Floquet states, derived in \cref{sec:inv_per_mod}, applies for arbitrary boundary conditions and symmetry (under $\P$ or $\K$, potentially combined with $\T$).
Their continuous counterpart was very recently shown to expose defects in locally $\K$-symmetric potentials under periodic boundary conditions \cite{Wulf2016_PRE_93_052215_ExposingLocal}.
We here exploit the discrete tight-binding form of $\hat{H}$ to drive the hoppings of a locally $\P$-symmetric array in time, corresponding to periodically modulated \cite{Szameit2008_PRL_101_203902_ObservationDefect-free,Garanovich2012_PR_518_1_LightPropagation} photonic waveguides;  
alternatively, the onsite potential could be driven \cite{Longhi2006_PRB_73_193305_TransmissionLocalization}.

We consider a Hermitian array of $N = 12$ sites with LS controlled by a scaling parameter $s$ multiplying the real potential $u_n$ on sites $n = 4,5,6,10$; see \cref{fig:ls_driven}\,(a).
For $s = 1$ the setup has CLS, with two $6$-site $\P$-symmetric subdomains $\DD_1$ and $\DD_2$, which is broken for $s \neq 1$.
With any type of periodic driving applicable, we here choose an global in-phase harmonic modulation of relative amplitude $f = 1/2$ on the hoppings, 
\begin{equation}
 h_{n,n \pm 1}(t) = (1 + f\sin\omega t) h_{n,n \pm 1}^\circ ; ~~ n, n \pm 1 \in \NN,
\end{equation}
with frequency $\omega = \pi/8$.
The Floquet modes $\ket{\phi^{{\mu}}(t)}$ of the system are computed with the Floquet matrix method \cite{Grifoni1998_PR_304_229_DrivenQuantum} including a total of $6$ sidebands.
The upper NLCs $q_{n;\mu}^+(t)$ in the Floquet state with $\mu = 7$ (chosen simply to yield distinctly varying NLCs) are shown in \cref{fig:ls_driven}\,(b) for the CLS case $s = 1$.
As is seen, at any instant $t$ we have that $iq_{\bar{n} \mp 1;\mu}^\pm(t) = [iq_{n;\mu}^{\pm}(t)]^*$ in each $\P$-symmetric subdomain from \cref{eq:q_P_symmetry}, which clearly holds also for the present dynamical NLCs.
What is not directly anticipated from the time evolution is that their averages $\bar{q}_{n;\mu}^\pm$ over a period $\tau$ are real and spatially constant within $\DD_1$ and $\DD_2$, as was established in \cref{sec:inv_per_mod}.

The profile of the period-averaged NLCs in relation to LS is illustrated in \cref{fig:ls_driven}\,(c) by continuously varying $s$.
For $s < 1$ the system has no symmetry around $\ax_1$ in $\DD_1$, and the $|\bar{q}_{n \in \DD_1;\mu}^+|$ are symmetric on the links around $\ax_1$.
The subdomain $\DD_2$, where only one site is modified by $s$, has a gapped $\P$ symmetry, and the $|\bar{q}_{n \in \DD_2;\mu}^+|$ are spatially constant in the subdomain $\DD_2 \setminus \{ 9,10 \}$ (i.\,e., only $|\bar{q}_{9;\mu}^+| \equiv |\bar{q}_{9,10;\mu}|$ on the central link across $\ax_2$ deviates from the rest).
In both domains we see that $|\bar{q}_{n;\mu}^+|$ becomes spatially constant in the limit of CLS when $s = 1$.
Alternatively to the bound array considered here, an interesting perspective would be to study the behavior of NLCs in driven scattering setups where, e.\,g. Floquet bound states \cite{Longhi2013_SR_3__FloquetBound} may arise in the energy continuum.

\section{Nonlocal currents in stationary scattering}
\label{sec:scattering}

In the previous section the NLCs and their invariance in LS domains were investigated in various types of bound setups with discrete eigenspectra.
We now apply the framework of NLCs to stationary \textit{scattering}, where the energy $E$ (longitudinal propagation constant for photonic waveguides) is a given input parameter, to demonstrate their domainwise invariance and relation to transmission in discrete structures with LSs.

In the following, we will focus on (local) $\P$-symmetric scatterers, whose corresponding NLCs are to be related to perfect transmission ($\mf{T} =1$) of composite setups.
We first consider a general scatterer localized in a domain $\DD$ with transmission and left (right) reflection amplitudes $\mf{t}$ and $\mf{r}$ ($\mf{r}'$ ), respectively, and ingoing plane wave amplitude $\mf{c}_<^+$ ($\mf{c}_>^-$) on the left (right).
As detailed in \cref{app:scattering_NLC}, the NLC under $\P_\DD$ on the first links on the left ($l$) and right ($r$) of the scatterer takes the form
\begin{align} \label{eq:q_scat_in_t_r_general}
  &q_{l-1,l} = q_{r+1,r}^* = 2h \sin k \,\times \\
  &\left\{ \, 
  ( 1 - \mathring{\mf{r}}^* \mathring{\mf{r}}')  \, \mathring{\mf{c}}_+^* \mathring{\mf{c}}_-
  - |\mf{t}|^2 \, \mathring{\mf{c}}_+ \mathring{\mf{c}}_-^*  
  - \mathring{\mf{r}}^* \mf{t} \, |\mathring{\mf{c}}_+|^2
  - \mf{t}^{*} \mathring{\mf{r}}' \, |\mathring{\mf{c}}_-|^2
   \, \right\}, \nonumber 
\end{align}
where $\mathring{\mf{c}}_+ \equiv \zeta^{+\ax} \mf{c}_<^{+}$, $\mathring{\mf{c}}_- \equiv \zeta^{-\ax} \mf{c}_>^{-}$ and $\mathring{\mf{r}} \equiv \zeta^{-2\ax} \mf{r}$ are the ingoing and reflection amplitudes, respectively, for the same scatterer shifted to the origin ($\ax \to 0$), with $\zeta = e^{ik}$ for quasimomentum $k$ away from the scatterer.

%%%%%%%%%%%%%%%%%%%%%%%%%%%%%%%%%%%%%%%%%%%%%%%%%%%%%%%%%%%%%%%%%%%%%%%%%%%%%%%%%%%%%%%%%%%%%%%
\begin{figure}[t!]
\centering
\includegraphics[width=1\columnwidth]{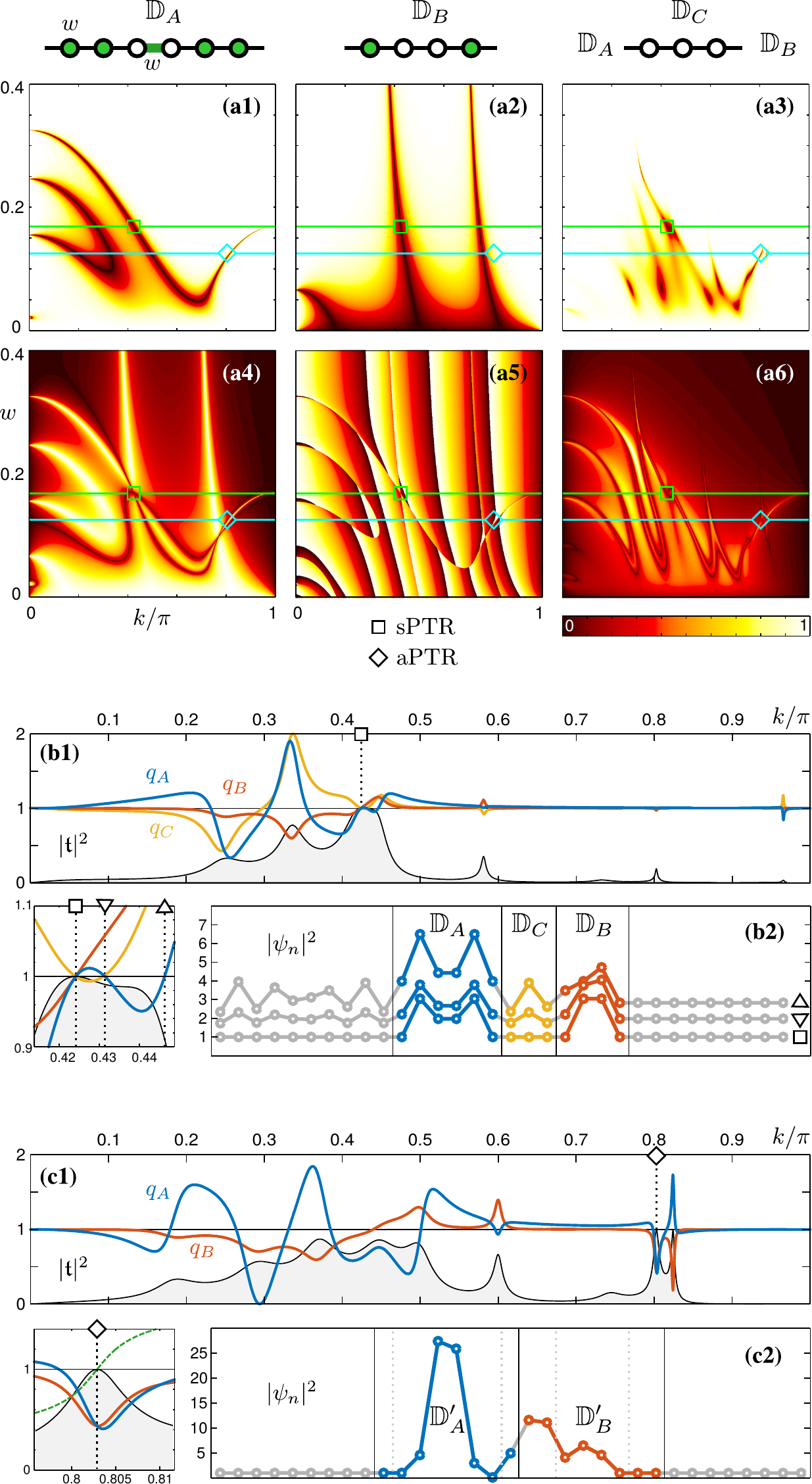}
\caption{Scattering off a locally $\P$-symmetric Hermitian array composed of three LS domains: $\DD_A$ and $\DD_B$ each containing a scatterer with Hamiltonian parameter $w$ (top sketch), and an intervening free propagation domain $\DD_C$.
Reflection magnitudes \textbf{(a1)} $|\mf{r}_A|$ of $\DD_A$, \textbf{(a2)} $|\mf{r}_B|$ of $\DD_B$, and \textbf{(a3)} $|\mf{r}|$ of the composite array $\DD = \{ \DD_A,\DD_C,\DD_B \}$, as well as quantities \textbf{(a4)} $||\mf{r}_A'| - |\mf{r}_B^*||^{1/2}$, \textbf{(a5)}  $|(\arg\mf{r}_A' - \arg\mf{r}_B^*)\,\text{mod}\,2\pi|/2\pi$, and \textbf{(a6)} $||q_A| - |q_B||^{1/2}$, which vanish at unit transmission (PTR, $\mf{r} = 0$), for varying $w$ and $k$.
An sPTR (aPTR) occurs at $w,k$ where $\mf{r}_A' = \mf{r}_B^* = 0$ ($\neq 0$), in which case $|q_A| = |q_B| = 0$ ($\neq 0$), indicated by $\square$ ($\romb$).
\textbf{(b1)} Transmission $|\mf{t}|^2$ and NLCs $q_{A,B,C}$ (offset by $1$ and scaled by overall maximum) in varying $k$ for $w \approx 0.1687$ and \textbf{(b2)} density $|\psi_n|^2$ (offset by $1,2,3$) at $k$-values indicated by $\square,\triangle,\text{\hspace{-.5em}\protect\rotatebox[origin=bl]{60}{$\triangle$}}$ in zoomed inset of (b1) where $q_A = q_B = q_C = 0$ (corresponding to the sPTR in (a)), $q_A = 0 \approx q_C$, $q_A = 0$, respectively.
$|\psi_n|^2$ is $\P_{\DD_X}$-symmetric when $q_X = 0$ ($X=A,B,C$).
\textbf{(c1)} $|\mf{t}|^2$ and $q_{A,B}$ for $w \approx 0.1254$ and \textbf{(c2)} $|\psi_n|^2$ at $k$ indicated by $\romb$ in inset where $q_A = q_B \neq 0$ corresponding to the aPTR in (a); green dashed line shows $\arg[\mf{t}\zeta^{2(\ax_2 - \ax_1)}]/\pi$.
$\DD_A$ ($\DD_B$) is here augmented symmetrically by 1 (2) site(s) into $\DD_A'$ ($\DD_B'$).
}
\label{fig:ptr}
\end{figure}
%%%%%%%%%%%%%%%%%%%%%%%%%%%%%%%%%%%%%%%%%%%%%%%%%%%%%%%%%%%%%%%%%%%%%%%%%%%%%%%%%%%%%%%%%%%%%%%

\subsection{Local $\P$ symmetry and perfect transmission}
\label{sec:ptr}

Let us now assume the scatterer to be $\P$-symmetric within a connected domain $\DD$ and Hermitian with equidirectional hoppings, meaning that $q_\DD^+ = q_{l-1,l} \in i \RR$ is translation invariant along $\DD$.
We then also have that $\mathring{\mf{r}} = \mathring{\mf{r}}'$ from the $\P$ symmetry of the scatterer (around the origin) and $1 - |\mf{r}|^2 = |\mf{t}|^2$, $\mf{t}^{*} \mathring{\mf{r}}' = - \mathring{\mf{r}}^* \mf{t}$ from the unitarity of the $\mathsf{S}$-matrix (see \cref{app:scattering_NLC}), so that the NLC expression reduces to
\begin{align} \label{eq:q_scat_in_t_r_herm_Psymm}
  q_\DD^+ &= q_\DD^{-*} = 2h \sin k \,\times \nonumber \\
  &\left\{ \, |\mf{t}|^2 \, \left( \mathring{\mf{c}}_+^* \mathring{\mf{c}}_-  -  \mathring{\mf{c}}_+ \mathring{\mf{c}}_-^* \right)  - 
  \mathring{\mf{r}}^* \mf{t} \, \left( |\mathring{\mf{c}}_+|^2 - |\mathring{\mf{c}}_-|^2 \right) \, \right\},
\end{align}
where we note that $\mathring{\mf{r}}^* \mf{t} \in i\RR$.
This expression shows how the $q_\DD^\pm$ depend on the characteristics $\mf{t}$, $\mf{r}$ and position $\ax$ of the scatterer (domain $\DD$) as well as the amplitudes incident on it.

At the center $k = 0$ and edges $k = \pm \pi$ of the leads' Brillouin zone, the scattering NLCs vanish generically.
Away from those $k$-points, $q_\DD^\pm = 0$ and thereby symmetric density $\rho_n$ in $\DD$ (see \cref{eq:q_P_zero}) can be achieved by tuning the ingoing amplitudes to satisfy $\left( |\mathring{\mf{c}}_+|^2 - |\mathring{\mf{c}}_-|^2 \right)/\text{Im}[\mathring{\mf{c}}_+^* \mathring{\mf{c}}_-] = \pm 2 |\mf{t}|/|\mf{r}|$ for $\mathring{\mf{r}}^* \mf{t} \in i\RR^\pm$.
For example, this is satisfied for $\mathring{\mf{c}}_+ = 1$ and $|\mathring{\mf{c}}_-| = \sqrt{2} - 1$ at all $E$ for which $\arg{\mathring{\mf{c}}_-} = \pm \arcsin \, |\mf{r}/\mf{t}|$. 
A special situation occurs for $\mathsf{S}$-matrix eigenstates, in which $\mathring{\mf{c}}_+ = \pm \mathring{\mf{c}}_-$ and thus $q_\DD^\pm = 0$ for any $\mf{t},\mf{r}$.
For a globally $\P$-symmetric (about $\ax = 0$) Hermitian scatterer, those states are simultaneously $\hat{P}$ eigenstates with eigenvalues $\lambda_\P = \pm 1$ and $\hat{P}\hat{T}$ eigenstates with eigenvalues $\lambda_{\P\T} = e^{i\vartheta_\lambda}$ where $\vartheta_\lambda = -2\arg(\mf{c}) - \arg(\mf{t} \pm \mf{r})$ (note that $|\mf{t} \pm \mf{r}| = 1$ in this case).

In the most common setting for a single scatterer, however, there is only one incoming wave from the left ($\mathring{\mf{c}}_- = 0$) or from the right ($\mathring{\mf{c}}_+ = 0$).
In this case, from \cref{eq:q_scat_in_t_r_herm_Psymm} $q_\DD^\pm = 0$ only if the scatterer is transparent at the given $k$, $\mf{r} = 0$, and vice versa (or if $\mf{t} = 0$, which strictly only occurs for singular potentials in unbiased 1D systems, in contrast to Fano transmission zeros when multiple interference paths are available \cite{Miroshnichenko2005_PRE_71_036626_NonlinearFano}).
Thus, perfect transmission $\mf{T} = 1$ is accompanied by a $\P_\DD$-symmetric density $\rho_n$ within $\DD$ ($\P_\DD \T$-symmetric $\psi_{n;E}$ up to a phase).

The assumption $\mathring{\mf{c}}_{-(+)} = 0$ for left (right) incidence is compatible with the connection to another, generally different, locally symmetric scatterer on the right (left) of $\DD$ which is also transparent at the same $k$.
By adjusting the Hamiltonian parameters, a CLS setups may be assembled from multiple, simultaneously transparent scatterers in LS domains $\DD_d$ ($d = 1,2,\dots$), with the state density following the LS of the setup.
We refer to such a state as a `symmetric' perfect transmission resonance (sPTR);
an example is illustrated in \cref{fig:ptr}.
We here have two scatterers localized in domains $\DD_A$ and $\DD_B$, connected via a free (with $h,u$ like in the leads) gap $\DD_C$, with their scattering properties shown in \cref{fig:ptr}\,(a).
As we see in \cref{fig:ptr}\,(b), the vanishing of the individual $q_A^+,q_B^+,q_C^+$ reveals the $k$ at which the density $\rho_n$ is locally symmetric in $\DD_A,\DD_B,\DD_C$, respectively, and an sPTR occurs when $q_A^+ = q_B^+ = q_C^+ = 0$.

In contrast to an sPTR, transparency ($\mf{T} = 1$) may also occur in a state whose density does not follow any LS of the setup, which we refer to as an `asymmetric' PTR (aPTR).
To see the relation of the NLCs to the occurrence of an aPTR, we consider a Hermitian setup of two different scatterers $\DD_1$ and $\DD_2$ with corresponding scattering matrices $\mathsf{S}_1$ and $\mathsf{S}_2$.
The total $\mathsf{S}$-matrix has elements $\mf{r} = \mf{r}_1 + \mf{p}\mf{t}_1'\mf{r}_2\mf{t}_1$, $\mf{r}' = \mf{r}_2' + \mf{p}\mf{t}_2\mf{r}_1'\mf{t}_2'$, and $\mf{t} = \mf{t}' = \mf{p}\mf{t}_1\mf{t}_2$, where $\mf{p} = (1 - \mf{r}_2\mf{r}_1')$ accounts for all orders of multiple reflections between the scatterers.
The condition $|\mf{t}| = 1$ for a PTR leads to $|\mf{r}_1'-\mf{r}_2|^2 = -4|\mf{r}_1'||\mf{r}_2|\sin\vartheta_1'\sin\vartheta_2$, with $\vartheta_{1,2}^{(\prime)}$ being the phases of $\mf{r}_{1,2}^{(\prime)}$, which in turn is equivalent to
\begin{equation} \label{eq:PTR_condition}
 \mf{r}_1' = \mf{r}_2^*,
\end{equation}
as can be seen geometrically in the complex plane.
Thus, in the special case $|\mf{t}_1| = |\mf{t}_2| = 1$, the phases $\vartheta_1',\vartheta_2$ are arbitrary, while $(\vartheta_1' + \vartheta_2)\,\text{mod}\,2\pi = 0$ must hold for $|\mf{t}_1| = |\mf{t}_2| \neq 1$ \cite{Zhukovsky2010_PRA_81_053808_PerfectTransmission}.
If both scatterers are $\P$-symmetric about their centers, it additionally holds that $\mathring{\mf{r}}_d^*\mf{t}^{ }_d \in i\RR$ ($d=1,2$), and it can then be shown that 
\begin{equation} \label{eq:t_PTR}
 \mf{t} = \pm \zeta^{2(\ax_1 - \ax_2)}
\end{equation}
at the PTR.
That is, the phase of the transmission amplitude depends (up to $\pi$) only on the distance between the centers of \textit{any} two arbitrary but locally $\P$-symmetric scatterers.

Assuming the scatterers to be $\P$-symmetric, each will also have spatially constant NLC $q_{\DD_d}^\pm \in i\RR$ ($d = 1,2$) at a given $E$.
As shown in \cref{app:derivation_q}, at a PTR it then holds that
\begin{equation} \label{eq:PTR_condition_q}
 q_{\DD_1}^+ = \pm q_{\DD_2}^+,
\end{equation}
the sign being determined by the details of a given setup.
This is demonstrated in the example setup of \cref{fig:ptr}\,(c), featuring an aPTR where the nonzero NLCs $q_{\DD_1}^+ \equiv q_A^+$ and $q_{\DD_2}^+ \equiv q_B^+$ cross.
The PTR relation in \cref{eq:t_PTR} can also be derived using \cref{eq:PTR_condition_q}; see \cref{app:derivation_q}.

To summarize, if the setup of two LS scatterers is perfectly transmitting, then it always holds that $|q_{\DD_1}^+| = |q_{\DD_2}^+|$, with two possible cases with respect to the symmetry of $\rho_n$:
(i) If $|q_{\DD_1}^+| = |q_{\DD_2}^+| = 0$, then we have an sPTR with $\rho_n$ being $\P$-symmetric in the LS domains. 
(ii) If $|q_{\DD_1}^+| = |q_{\DD_2}^+| \neq 0$, then we have an aPTR and $\rho_n$ is asymmetric with respect to the LS domains.
The NLCs thus connect the transport properties of globally asymmetric systems to the spatially resolved profile of the scattering state amplitude with respect to its possible LSs.
In composite CLS setup of multiple scatterers, there may be parts of locally symmetric and asymmetric $\rho_n$ at a PTR indicated by zero and nonzero equal NLC invariants, respectively. 

%%%%%%%%%%%%%%%%%%%%%%%%%%%%%%%%%%%%%%%%%%%%%%%%%%%%%%%%%%%%%%%%%%%%%%%%%%%%%%%%%%%%%%%%%%%%%%%
\begin{figure}[t!]
\centering
\includegraphics[width=1\columnwidth]{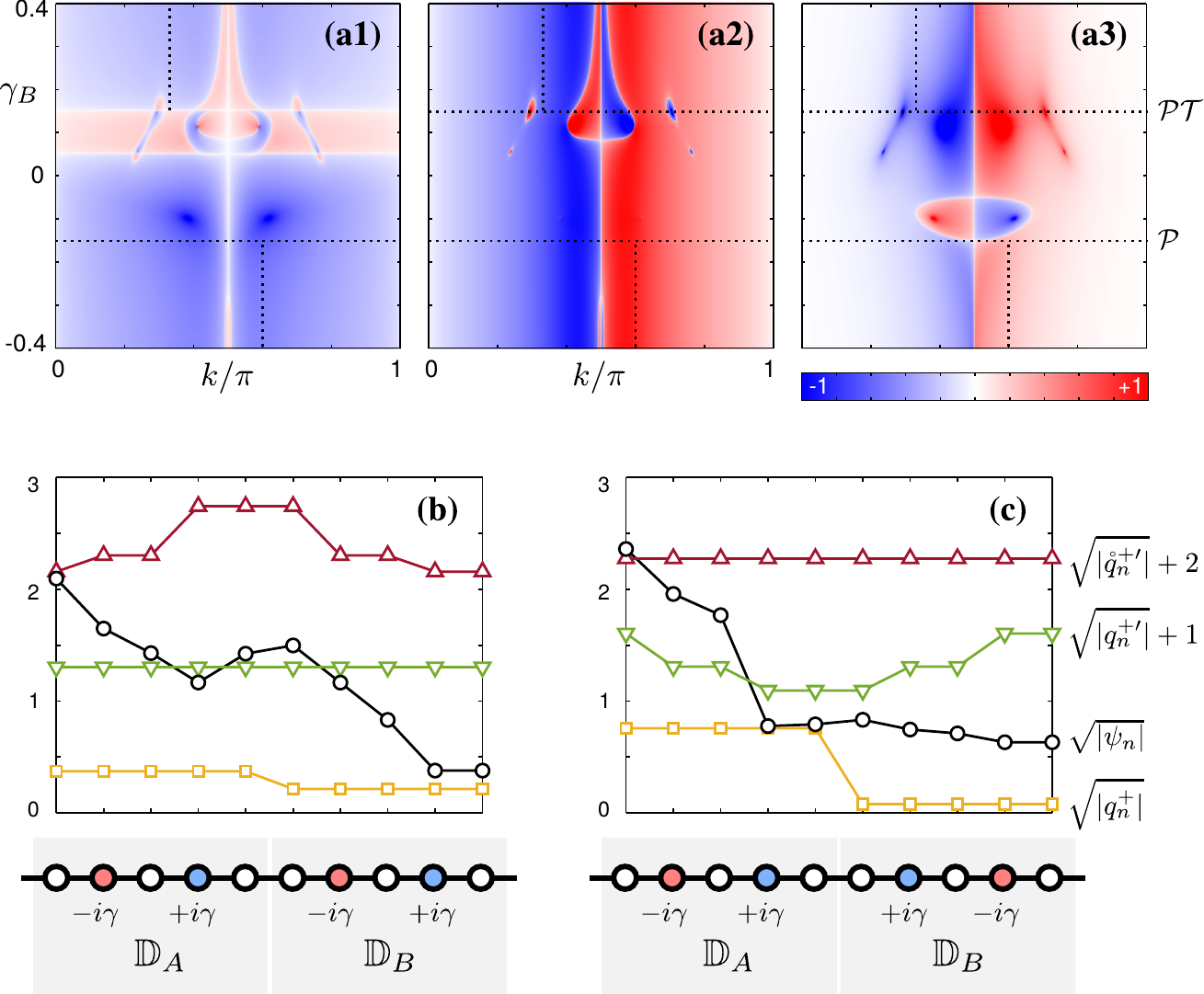}
\caption{Scattering off a locally $\P\T$-symmetric non-Hermitian array $\DD = \{\DD_A,\DD_B\}$ with onsite elements $\{v_n\}_X = \{0,-i\gamma_X,0,+i\gamma_X,0 \}$ ($X = A,B$) and common hoppings $\{h_{n,n+1} \} = 0.1\times\{ 1.0,1.5,1.5,1.0\}$. 
\textbf{(a1)} Quantity  $|\mf{T} - 1| - \sqrt{\mf{R}\mf{R}'}$ of $\DD$ and NLCs \textbf{(a2)} $q^+_A/i$ of $\DD_A$ and \textbf{(a3)} $q^+_B/i$ of $\DD_B$ for unit incoming amplitude from the left, in varying $k$ and $\gamma_B$ for fixed $\gamma_A = 0.15$.
The colormap is normalized to $1$ in (a1) and truncated at $\pm 1$ in (a2,a3); it scales with the power of $1/4$ to increase contrast.
\textbf{(b)} Density $|\psi_n|^2$ ($\circ$) and spatially resolved NLCs $q^+_{n}$ ({\tiny $\square$}) for local $\P_{\DD_{A,B}}$ transforms, $q^{+\prime}_{n}$ ({\scriptsize $\triangledown$}) and $\mathring{q}^{+\prime}_{n}$ ({\scriptsize $\vartriangle$}) for global $\P$ transform, for $\gamma_B = \gamma_A \equiv \gamma$ (global $\P\T$ symmetry) at $k = \pi/3$ indicated by upper dotted line in (a1).
\textbf{(c)} Same as (b) but for $\gamma_B = -\gamma_A \equiv \gamma$ (global $\P$ symmetry) at $k = 0.6\pi$ indicated by lower dotted line in (a1).
}
\label{fig:local_pt}
\end{figure}
%%%%%%%%%%%%%%%%%%%%%%%%%%%%%%%%%%%%%%%%%%%%%%%%%%%%%%%%%%%%%%%%%%%%%%%%%%%%%%%%%%%%%%%%%%%%%%%

\subsection{Invariants in locally $\P\T$-symmetric scattering}
\label{sec:non-Hermitian_scattering}

We close this section by addressing the NLCs of scattering states in non-Hermitian setups.
Since the energy is here a free input parameter which is chosen real, the translation invariance of the NLCs in locally $\P\T$-symmetric domains is manifest for any choice of loss/gain parameters.
In the case of a $\P\T$-symmetric scatterer, the $\mathsf{S}$-matrix elements are subject to the generalized unitarity relation \cite{Ge2012_PRA_85_023802_ConservationRelations} 
\begin{equation} \label{eq:conservation_PT}
 |\mf{t}|^2 - \frac{\mf{t}^*}{\mf{t}~}\mathring{\mf{r}} \, \mathring{\mf{r}}' = 1 ~~ \Rightarrow ~~ |\mf{T} - 1| = \sqrt{\mf{R}\mf{R}'}
\end{equation}
for $\mf{t} \neq 0$, with $\mf{t}^*\mf{r},\mf{t}^*\mf{r}' \in i \RR$.
The general expression (\ref{eq:q_scat_in_t_r_general}) for the NLC then becomes 
\begin{align} \label{eq:q_scat_in_t_r_PTsymm}
  q_\DD^+ &= q_\DD^{-*} = 2h \sin k \,\times \nonumber \\
  &\left\{ \, |\mf{t}|^2 \, \left( \mathring{\mf{c}}_+^* \mathring{\mf{c}}_-  -  \mathring{\mf{c}}_+ \mathring{\mf{c}}_-^* \right)  - 
  \mathring{\mf{r}}^* \mf{t} \, |\mathring{\mf{c}}_+|^2 - \mathring{\mf{r}}' \mf{t}^* \, |\mathring{\mf{c}}_-|^2 \, \right\},
\end{align}
which further reduces to \cref{eq:q_scat_in_t_r_herm_Psymm} if the $\P\T$-symmetric scatterer is also Hermitian.
A $\P\T$-unbroken phase for a globally $\P\T$-symmetric, two-terminal scattering setup may be identified as the parametric region of $\hat{H}$ with unimodular $\mathsf{S}$-matrix eigenvalues $\lambda_\mathsf{S}^{1,2}$ at given energy \cite{Chong2011_PRL_106_093902_Mathcalpmathcalt-symBreaking}. 
Indeed, for $\mathsf{S}$-matrix eigenstates, where $\mf{c}_{<(>)}^{-(+)} = \lambda_\mathsf{S}^{1,2} \mf{c}_{<(>)}^{+(-)}$, the NLC becomes 
\begin{equation} \label{eq:q_scat_Seig}
  q_{\DD}^+ = q_\DD^{-*} =  2h \sin k \mathring{\mf{c}}_<^{+*} \, \mathring{\mf{c}}_>^- \left( 1 - |\lambda_\mathsf{S}^{1,2}|^2 \right),
\end{equation}
yielding a $\hat{P}\hat{T}$ eigenstate when $|\lambda_\mathsf{S}^{1,2}| = 1$ since then $q_{\DD}^\pm = 0$.
The latter condition, however, can be met in \cref{eq:q_scat_in_t_r_PTsymm} also for non-$\mathsf{S}$-eigenstates in dependence of the incoming amplitudes $\mathring{\mf{c}}_<^{+}, \mathring{\mf{c}}_>^-$ for a given $\mathsf{S}(E)$.
The crossing from zero to nonzero NLC thus provides an extended identification of a $\P\T$ `phase transition' in scattering beyond $\mathsf{S}$-matrix eigenstates \cite{Kalozoumis2014_PRA_90_043809_SystematicPathway}, identical to that of bound systems (as discussed above in \cref{sec:non-Hermitian}).

More importantly, the vanishing of NLCs can be employed to identify configurations with \textit{locally} $\P\T$-invariant states, that is, where \cref{eq:q_P_zero} is fulfilled in a $\P\T$-symmetric subdomain of the setup.
As an example, in \cref{fig:local_pt} we consider a non-Hermitian setup of two attached domains $\DD_A$, $\DD_B$ with loss/gain parameter pairs $\pm \gamma_A$, $\pm \gamma_B$, respectively.
The setup is designed such that it is globally $\P\T$-symmetric ($\P$-symmetric) for $\gamma_B = \gamma_A$ ($\gamma_B = -\gamma_A$).
As we see in \cref{fig:local_pt}\,(a1), there are values of $\gamma_B \neq \gamma_A$ where \cref{eq:conservation_PT} is fulfilled at certain energies (see white contours), such that the system behaves as being $\P\T$-symmetric in terms of its $\mathsf{S}$-matrix.
Interestingly, it is possible to fulfill \cref{eq:conservation_PT} at all $k$ with a locally $\P\T$-symmetric setup for appropriately chosen loss/gain parameters, as seen for the present setup for $\gamma_B = \gamma_A/3 = 0.05$ (lower horizontal white line in \cref{fig:local_pt}\,(a1)).
In \cref{fig:local_pt}\,(a2) and (a3) we see that the NLCs $q_A^+$ and $q_B^+$, corresponding to local $\P\T$-symmetry in $\DD_A$ and $\DD_B$, vanish at different contours in the $(k,\gamma_B)$-plane, revealing the occurrence of locally symmetric density embedded into the globally $\P\T$-asymmetric system.

Leaving a more comprehensive investigation of NLCs in relation to scattering properties of locally $\P\T$-symmetric setups for future work, we focus now on their spatial constancy in 1D domains.
This is illustrated in \cref{fig:local_pt}\,(b) and (c), where the considered setup is, apart from locally $\P\T$-symmetric, also globally $\P\T$- and $\P$-symmetric, respectively.
At the chosen generic $k$-values, the density profile $|\psi_n|^2$ is completely asymmetric under any of those symmetry transformations at the chosen (generic) $k$-values.
For the local $\P$ transforms in $\DD_A$ and $\DD_B$, the domainwise constancy of the associated NLC $q_n^+$ reveals the local $\P\T$ symmetry of $\hat{H}$ encoded in the scattering state.
Note that this is achievable because $E$ is set real as an input parameter, in contrast to non-Hermitian \textit{bound} 1D systems where local $\P\T$ symmetry typically does not support real eigenenergies.
In \cref{fig:local_pt}\,(b), the constant NLC $q_n^{+\prime}$ (associated with global $\P$ transform) reflects the global $\P\T$ symmetry, while the dual NLC $\mathring{q}_n^{+\prime}$ is (varying but) symmetric; see \cref{eq:q_P_symmetry}.
In \cref{fig:local_pt}\,(c), we have the opposite situation: $q_n^{+\prime}$ is symmetric, while the constant $\mathring{q}_n^{+\prime}$ reflects the global $\P$ symmetry, even in the presence of loss and gain.

\section{Conclusions and perspectives}
\label{sec:conclusions}

A discrete nonlocal current-density continuity formalism was developed to relate localized spatial symmetries to Schr\"odinger wave properties in Hermitian and non-Hermitian lattice systems.
Sources and sinks of the nonlocal currents (NLCs) are identified in the breaking of the local symmetries (LSs), with the associated charge evolution governed by the NLCs at symmetry domain boundaries.
For stationary states, the symmetry-adapted NLCs are translationally invariant in one-dimensional finite domains with $\S(\T)$-symmetric sub-Hamiltonians, where $\S = \P$ (inversion) or $\K$ (translation) is potentially combined with $\T$ (time reversal).
These invariant NLCs  reflect the local symmetries of generic single-particle lattice Hamiltonians encoded into arbitrarily irregular field profiles, and enable the mapping between wave amplitudes of symmetry-related sites in finite domains.
We illustrated the NLC invariance in cases of complete, overlapping, and gapped domainwise symmetry, as well as for non-Hermitian (locally) $\P\T$- and $\K\T$-symmetric (scattering) systems with balanced gain and loss.
In scattering setups, the NLCs were further shown to classify perfect transmission resonances with respect to their spatially resolved density profile, being symmetric (asymmetric) in adjacent locally symmetric lattice subdomains with equal and zero (nonzero) NLC magnitude.
Further, in periodically driven setups the NLC invariance was shown to be retained for quasi-stationary (Floquet) states on period-average, thus revealing LSs also far from equilibrium. 
To summarize, the developed theoretical framework generalizes the concept of global symmetries with globally induced (state-independent) properties to local symmetries with locally induced (state-dependent) properties of lattice wave excitations.

We have here utilized photonic waveguide arrays as an applicational platform since they are well described by a discrete Schr\"odinger equation with nearest neighbor couplings.
The formalism and general results apply equally to other lattice systems modeled effectively by a discrete Schr\"odinger equation, like superlattice nanostructures or quantum dot arrays \cite{Ferry1997____TransportNanostructures}.
For stationary states the applicability is further broadened to generic wave mechanical systems described by a discretized Helmholtz equation, such as e.\,g. granular acoustic media with domainwise homogeneities \cite{Boechler2011_JAP_109_074906_TunableVibrational}.
In the limit of large site density with constant inter-site hopping, the same formalism can be equally employed for spatially continuous systems (with the lattice used as a numerical grid), treating simulations of, e.\,g., photonic multilayer devices \cite{Zhukovsky2010_PRA_81_053808_PerfectTransmission,Peng2002_APL_80_3063_Symmetry-inducedPerfect} or acoustic waveguides \cite{Hladky-hennion2013_JAP_113_154901_AcousticWave,Theocharis2014_NJP_16_093017_LimitsSlow}.

The generic character of the proposed framework, based on a continuity equation, \cref{eq:continuity_nonlocal}, allows for a unified description of LS in Hermitian and non-Hermitian setups, independently of the boundary conditions imposed.
Further, the present operator formulation of the NLCs, expressed in \cref{eq:q_n_+-_operator}, enables a straightforward extension beyond nearest neighbor couplings or to higher dimensions, simply by including the additional couplings into the hopping operators $\hat{H}^\pm$ accordingly.
In general, the net NLC assigned to a site with (multidimensional) index $\bs{n}$ will then consist of the sum 
of nonlocal link currents $iq_{\bs{n}\bs{m}}^\S = \psi_\bs{n}^* h_{\S(\bs{n},\bs{m})} \psi_{\S(\bs{m})} - \psi_{\S(\bs{n})}h_{\bs{n},\bs{m}}^* \psi_{\bs{m}}^*$ over connected sites $\bs{m}$ for a given symmetry operation $\S$, defined in analogy with the 1D NLCs in \cref{eq:q_n+-}.
A thorough account on symmetry considerations and associated NLCs in planar discrete systems, where rotations and plane reflections add to the possible $\S$-transforms, is presented in Ref.\,\cite{Roentgen2017_AOP_380_135_Non-localCurrents}.
There, also the connectivity of sites under general local transformations plays a decisive role in defining NLCs \cite{connectivity}.
Multidimensional divergence naturally prevents individual $q_{\bs{n}\bs{m}}^\S$ from being spatially constant along a selected dimension, just as is the case for the usual current.
NLC invariance does apply if the sites $\bs{m}$ above represent next-nearest or (selected) remote neighbors along a \textit{single} dimension, as is effectively the case, e.\,g., for zigzag chains with embedded scatterers \cite{Stockhofe2015_PRA_92_023605_Sub-Supercritical} or discrete helical structures \cite{Stockhofe2015_PRA_91_023606_BlochDynamics}, respectively.
Extended locally symmetric domains in such setups will then be distinguished by constant NLCs in arbitrary stationary states.
The domainwise invariant stationary NLCs may further be generalized to discrete \textit{interacting} systems where the interaction itself is represented by a locally $\S\T$-symmetric function on the lattice sites.
A promising case would be the discrete version of self-induced $\P\T$-symmetric potentials \cite{Sarma2014_PRE_89_052918_ContinuousDiscrete,Sinha2015_PRE_91_042908_SymmetriesExact} in lattice subdomains, with the mean-field nonlinear term $\rho_n = |\psi_n|^2$ replaced by the nonlocal density $\sigma_n = \psi_n^*\psi_{\bar{n}}$.
Finally, an intriguing prospect would be to investigate local \textit{dynamical} symmetries, i.\,e. time reversal or translation symmetries within finite temporal intervals, potentially combined with local spatial symmetries, and their implications for nonequilibrium states.
With the above perspectives, we believe that the developed connection between discrete nonlocal continuity and local symmetries may contribute to the understanding of wave structure and response in discrete systems with localized regularities.

\section*{Acknowledgements}

We are thankful to M.\,R\"ontgen for fruitful suggestions and critical reading of the manuscript, as well as to S.\,Weimann, A.\,Szameit, and A.\,V.\,Zampetaki for helpful discussions.
C.\,V.\,M. and P.\,S. gratefully acknowledge funding by the Deutsche Forschungsgemeinschaft under grant DFG SCHM 885/29-1.

\appendix

\section{Combination of spatial symmetry \\ with time reversal}
\label{app:time_reversal}

We extend the defined local spatial transformations to their combination with the operation of time-reversal $\T: t \rightarrow -t$, represented here as usual by an antiunitary operator $\hat{T} = \hat{T}^{-1}$ performing complex conjugation \cite{Haake2001_QSC__15_TimeReversal},
\begin{equation} \label{eq:time_reversal_operator}
 \hat{T}: ~ i \rightarrow -i; ~~~ \hat{T} \, i \, \hat{T} = -i,
\end{equation}
in the spatial representation.
Time-reversal thus acts on a state in the basis $\{\ket{n}\}$ as 
\begin{equation} \label{eq:psi_star}
 \hat{T}\ket{\psi}(t) = \sum_n \psi_n^*(-t) \ket{n} \equiv \ket{\psi^\sT}
\end{equation}
and transforms the Hamiltonian as $\hat{T}\hat{H}\hat{T} = \hat{H}^*$ (defined by $\braket{m|\hat{H}^*|n} = H_{mn}^*$).
The latter becomes relevant for complex onsite elements 
\begin{equation} \label{eq:complex_onsite}
 v_n \equiv u_n + i \frac{\gamma_n}{2} ~~~~ (u_n,\gamma_n \in \RR),
\end{equation}
where positive (negative) $\gamma_n$ models the gain (loss) rate of the state density at site $n$ (as clarified in the next section), while complex hoppings $h_{n,n'} = |h_{n,n'}|e^{i\varphi_{n,n'}}$ with Peierls phase $\varphi_{n,n'}$ generally arise from the presence of gauge fields (effectively also in 1D systems \cite{Golshani2014_PRL_113_123903_ImpactLoss}).
If the Hamiltonian obeys (cf. \cref{eq:Hmn_local_Ssym})
\begin{equation} \label{eq:Hmn_local_STsym}
H_{\bar{m}\bar{n}} = H_{mn}^* ~~~\forall~ m,n \in \DD,
\end{equation}
then we will call the system locally $\S\T$-symmetric, that is, $\P\T$-symmetric or $\K\T$-symmetric within the finite spatial domain $\UU = \DD \cup \bar{\DD}$.
Note here that we treat (local) non-Hermiticity, expressed through the $\gamma_n$, solely as an \textit{effective} simple model to express gain or loss of density from or to the surrounding medium, rather than a fundamental description of the physical system.

\section{Local discrete continuity equation}
\label{app:continuity_local}

In equivalence to the Schr\"odinger equation (\ref{eq:Schr}) for a general state $\ket{\psi}$, the temporal evolution of the local density $\rho_n = \braket{\psi|\hat{\rho}_n|\psi} = \psi_n^*\psi_n$ fulfills the discrete continuity equation 
\begin{equation} \label{eq:continuity_local}
 \partial_t \rho_n = j_n - i(v_n-v_n^*)\rho_n = j_n + \gamma_n \rho_n, 
\end{equation}
with $\hat{\rho}_n \equiv \ket{n}\bra{n}$ being the local density operator in the Schr\"odinger picture, where $j_n = j_n^+ + j_n^- $ is the sum of the magnitudes
\begin{align} \label{eq:j_n+-}
j_n^\pm & \equiv j_{n,n \pm 1} \nonumber \\ 
 & = \frac{1}{i} \left( \psi_n^* h_{n,n \pm 1} \psi_{n \pm 1} - \psi_{n \pm 1}^* h_{n,n \pm 1}^* \psi_n \right)
\end{align}
of the currents $\bs{j}_n^\pm$ flowing inwards from sites $n \pm 1$ to site $n$.
Under the assumption of Hermitian hoppings, $h_{n,n \pm 1}^* = h_{n \pm 1,n}$, the total current magnitude can be written as the expectation value $j_n = \braket{\psi|\hat{j}_n|\psi}$ of a current operator
\begin{equation} \label{eq:j_n_tot_operator}
 \hat{j}_n = \hat{j}_n^+ + \hat{j}_n^- = \frac{1}{i}[\hat{\rho}_n,\hat{H}]
\end{equation}
assigned to site $n$.
Note that the discrete form of the current divergence entering the continuity equation is generally given by the sum of the projections $j_{nm} = \bs{j}_{nm} \cdot \bs{e}_{nm}$ of the link currents $\bs{j}_{nm}$ from point $\bs{r}_m$ to point $\bs{r}_n$ onto their directions $\bs{e}_{nm}$, scaled by the distance $|\bs{r}_n - \bs{r}_m|$ \cite{Boykin2010_EJP_31_1077_CurrentDensity}.
The local (referring to site $n$) continuity equation (\ref{eq:continuity_local}) shows that the imaginary part $2\gamma_n$ of the onsite element $v_n$ acts like a source ($\gamma_n > 0$) or sink ($\gamma_n < 0$) of density $\rho_n$ at site $n$, in addition to the inflow of current $j_n$.
The partial link current operators producing the $\bs{j}_n^\pm$ can be written as \cite{Baranger1991_PRB_44_10637_ClassicalQuantum} $\hat{\bs{j}}_n^\pm = \hat{\rho}_n \hat{\bs{v}}^\pm + \hat{\bs{v}}^\mp \hat{\rho}_n$ in terms of the partial velocity operators $\hat{\bs{v}}^\pm = \frac{1}{i}[\hat{\bs{r}},\hat{H}^\pm]$ with $\hat{\bs{r}} = \sum_n \bs{r}_n \hat{\rho}_n$ being the position operator, such that the operators producing the projections $j_n^\pm$ are given by
\begin{equation} \label{eq:j_n_+-_operator}
 \hat{j}_n^\pm = \frac{1}{i} (\hat{\rho}_n\hat{H}^\pm - \hat{H}^\mp\hat{\rho}_n).
\end{equation}
Here, the $\hat{H}^\pm$ are the `hopping operators' with matrix elements $\braket{n|\hat{H}^\pm|m} = h_{n,m} \delta_{n \pm 1,m} $, that is, the upper/lower diagonal of the tridiagonal matrix $H$ in the tight-binding case.

\section{Bitemporal nonlocal current}
\label{app:bitemporal_nlc}

Projecting $\hat{\sigma}\ket{\psi}$ onto the time-reversed state $\T\ket{\psi} = \ket{\psi^\sT}$ instead of $\ket{\psi}$ in \cref{eq:nonlocal_density} yields a \textit{bitemporal} nonlocal `density'
\begin{equation} \label{eq:nonlocal_density_bitemporal}
 \sigma^\sT_n \equiv \braket{\psi^\sT|\hat{\sigma}_n|\psi} = \psi_n(-t) \psi_{\bar{n}}(t).
\end{equation}
Using the SE, the evolution of $\sigma^\sT_n$ can be shown to obey the continuity equation 
\begin{equation} \label{eq:continuity_nonlocal_bitemporal}
 \partial_t \sigma^\sT_n = q^\sT_n - i (v_{\bar{n}} - v_n) \sigma^\sT_n = q^\sT_n + \xi_n \sigma^\sT_n,
\end{equation}
now with asymmetry function $\xi_n \equiv (v_{\bar{n}} - v_n)/i$ and with bitemporal net NLC $q^\sT_n = q^{\sT+}_n + q^{\sT-}_n$ flowing into site $n$, where
\begin{align} \label{eq:qTrev_n+-}
&iq^{\sT\pm}_{n} \equiv 
iq_{n,\S}^{\sT\pm} \equiv 
iq^\sT_{n,n \pm 1;\S} = \\
&\psi_n(-t) h_{\S(n),\S(n \pm 1)} \psi_{\S(n \pm 1)}(t) - \psi_{n \pm 1}(-t) h_{n,n \pm 1} \psi_{\S(n)}(t), \nonumber 
\end{align}
defined in similarity to the NLCs in \cref{eq:q_n+-}.
Under the assumption of equidirectional hoppings, $H_{mn} = H_{nm}$, the $q_{n,\S}^{\sT\pm}$ are given by 
\begin{equation} \label{eq:qTrev_as_pTrev_q_psi}
q_{n,\S}^{\sT\pm} = \braket{\psi^\sT|\hat{q}_{n,\S}^\pm|\psi} ~~~~ (\S = \P,\K)
\end{equation}
which resembles an expectation value as in \cref{eq:q_n_+-_expval}, though with projection onto $\ket{\psi^\sT}$.
The potential asymmetries $\xi_n$ again act like sources for the dual NLC, with $\xi_n = \beta_n$ for real $v_n$.
In contrast to \cref{eq:continuity_nonlocal}, however, an $\S\T$-symmetric potential now yields $\xi_n = -\gamma_n$, with the sources (sinks) for $\rho_n$ representing sinks (sources) for $\sigma^\sT_n$.
An $\S$-symmetric potential ($\xi_n = 0$), on the other hand, leads to vanishing $q^\sT_{n,\S}$ for energy eigenstates within a symmetry domain $\DD$, even in the presence of unbalanced loss and/or gain where $q_{n,\S}$ may vary spatially.

For an energy eigenstate, a local invariance analogous to that of $q_{n,\S}^\pm$ can be derived also for the bitemporal NLCs $q_{n,\S}^{\sT\pm}$ in a locally $\S$-symmetric domain.
The associated continuity equation (\ref{eq:continuity_nonlocal_bitemporal}) reduces to
\begin{equation} \label{eq:continuity_nonlocal_bitemporal_eigenstate}
 q_n^{\sT} + \xi_n \sigma_n^{\sT} = 0 
\end{equation}
for an energy eigenstate $\ket{\phi}$, so that the Kirchoff law $q^{\sT+}_{n,\S} + q^{\sT-}_{n,\S} = 0$ holds in a domain $\DD$ with $\S$-symmetric potential, that is, with $\xi_{n\in\DD} = 0$.
If also the hoppings are $\S$-symmetric in $\DD$, then $q_{n \mp 1,\S}^{\sT\pm} = -\eta_{n \mp 1,n}q_{n,\S}^{\sT\mp}$ from \cref{eq:qTrev_n+-}, yielding
\begin{equation} \label{eq:qTrev_n_+-_transfer}
 q_{n,\S}^{\sT\pm} = \eta_{n,n \mp 1} \, q_{n \mp 1,\S}^{\sT\pm} ~~~~ (\S = \P,\K)
\end{equation}
for $n,n \pm 1 \in \DD$, in analogy to \cref{eq:q_n-1_+_equal_-q_n_-}.
Thus, for equidirectional hoppings ($\eta_{n,n \mp 1} = 1$) the $q_{n,\S}^{\sT\pm}$ are translation invariant within the $\S$-symmetric domain $\DD$, while accumulating a Peierls-like phase along it for complex Hermitian hoppings.
Note that the above applies also to complex eigenenergies $E$, since the `width' $\varGamma$ is absent in \cref{eq:continuity_nonlocal_bitemporal_eigenstate}, and thereby to general \textit{non-Hermitian} arrays with locally $\S$-symmetric onsite loss/gain elements.

Finally, with the substitution $\{\sigma_n,q_n,\eta_{m,n}^*\} \to \{\sigma^\sT_n,q^\sT_n,\eta_{m,n} \}$, the considerations on the nonlocal charge evolution in \cref{eq:charge_evolution} can be equally applied to the evolution of a bitemporal nonlocal charge $\varSigma_\psi^{\DD,\sT} = \sum_{n \in \DD} \sigma^\sT_n$.
This is then constant in time for a globally $\P$-symmetric (non-Hermitian) array $\DD = \NN$ with unit hopping ratios where, e.\,g., only lossy elements $\gamma_n < 0$ are present. 

\section{Dual nonlocal current}
\label{app:dual_nlc}

We introduce the `dual' NLC $\mathring{q}_{n}^\pm$ as an equal-time alternative to the bitemporal NLC $q_{n,\S}^{\sT\pm}$, defined by
\begin{align} \label{eq:qtilde_n+-}
&i\mathring{q}_{n}^\pm \equiv i\mathring{q}_{n,\S}^\pm \equiv i\mathring{q}_{n,n \pm 1;\S} = \nonumber \\
&\psi_n h_{\S(n),\S(n \pm 1)} \psi_{\S(n \pm 1)} - \psi_{n \pm 1} h_{n,n \pm 1} \psi_{\S(n)},
\end{align}
where $\psi_m \equiv \psi_m(t)$, in similarity to \cref{eq:q_n+-} but with $\psi_n,\psi_{n \pm 1}h_{n,n \pm 1}$ replacing their complex conjugates in $q_{n,\S}^\pm$.
For equidirectional hoppings, they can be expressed via the NLC operators as
\begin{equation} \label{eq:qtilde_as_psistar_q_psi}
\mathring{q}_{n,\S}^\pm = \braket{\psi^*|\hat{q}_{n,\S}|\psi} ~~~~ (\S = \P,\K)
\end{equation}
by projecting onto the complex conjugated state $\ket{\psi^*} \equiv \sum_n \psi_n^*\ket{n}$ (that is, without the time reversal $t \to -t$ applied for $\ket{\psi^\sT}$).
In contrast to the $q_{n,\S}^\pm$ and $q_{n,\S}^{\sT\pm}$, the $\mathring{q}_{n,\S}^\pm$ are time-dependent even for a stationary state (with $\varGamma = 0$) through the harmonic factor $e^{-2iEt}$, and do not obey a nonlocal continuity equation like \cref{eq:continuity_nonlocal,eq:continuity_nonlocal_bitemporal}.
Their conditional \textit{spatial} constancy for an energy eigenstate $\ket{\phi}$ in an $\S$-symmetric domain $\DD$ can, nevertheless, be derived directly noticing that
\begin{align} 
i(\mathring{q}_{n,\S}^+ + \mathring{q}_{n,\S}^-) - (v_n - v_{\bar{n}})\phi_n \phi_{\bar{n}} &= \nonumber \\ 
(a_n g_{\bar{n}} - g_n a_{\bar{n}}) \, e^{-2iEt} &=  0      \label{eq:Schr_diff}
\end{align}
by using the close-coupling equations 
\begin{align} \label{eq:Schr_cc}
0 &= h_{n,n-1} a_{n-1} + (v_n - E) a_n + h_{n,n+1} a_{n+1} \nonumber \\
  & \equiv g_n, ~~~ n \in \NN.
\end{align}
expressed at transform-related points $n$ and $\bar{n} = \S(n)$.
In \cref{eq:Schr_cc}, pertaining from the time-independent Schr\"odinger equation $\hat{H} \ket{\phi} = E\ket{\phi}$, the set of sites $\NN$ covers the system of interest: 
For an isolated system closed boundary conditions $a_0 = a_{N+1} = 0$ are imposed by setting $h_{0,1} = h_{N,N+1} = 0$, while a scattering setup is realized by appending homogeneous semi-infinite chains with $v_n = v$ and $h_{n,n+1} = h$ on the left ($n<1$) and right ($n>N$) of the given array.

The LS condition (\ref{eq:Hmn_local_Ssym}) together with \cref{eq:Schr_diff} and \cref{eq:qtilde_n+-} then lead to $\mathring{q}_{n}^\pm = \eta_{n,n \mp 1} \, \mathring{q}_{n \mp 1}^\pm$, just like in \cref{eq:qTrev_n_+-_transfer}.
Indeed, since $\mathring{q}_{n}^\pm = q_{n}^{\sT\pm} \, e^{-2iEt}$ for an energy eigenstate, it will have the same spatial characteristics as the stationary $q_{n}^{\sT\pm}$.

\section{Derivation of general mapping relations and current connection}
\label{app:mapping_derivation}

The NLCs $q_{n,\S}^\pm$ and their duals $\mathring{q}_{n,\S}^\pm$ for a general state $\ket{\psi}$ can be expressed together via a matrix product acting on a column vector of amplitudes in $\bar{\DD}$, as
\begin{align} \label{eq:q_vec}
 i\begin{bmatrix} q_{n,\S}^\pm \\ \mathring{q}_{n,\S}^\pm \end{bmatrix} = 
   &\begin{bmatrix} \psi_n^* & \psi_{n \pm 1}^*h_{n,n \pm 1}^* \\ \psi_n & \psi_{n \pm 1}h_{n,n \pm 1} \end{bmatrix} \\
  &\times \begin{bmatrix} h_{\S(n),\S(n \pm 1)} & 0 \\ 0 & -1 \end{bmatrix} 
 \begin{bmatrix} \psi_{\S(n \pm 1)} \\ \psi_{\S(n)} \end{bmatrix} \nonumber 
\end{align}
for arbitrary Hamiltonian, with $\S = \P,\K$.
For nonzero local currents $j_n^\pm$, the first matrix above (with determinant $ij_n^\pm$) is invertible and we can solve for the symmetry-transformed amplitudes to obtain
\begin{align} \label{eq:current_mapping_matrix_general}
 \begin{bmatrix} \psi_{\S(n \pm 1)} \\ \psi_{\S(n)} \end{bmatrix} &= 
 \frac{1}{j_n^\pm} 
 \begin{bmatrix} h_{\S(n),\S(n \pm 1)}^{-1} & 0 \\ 0 & 1 \end{bmatrix} \\ &\times
 \begin{bmatrix} \psi_{n \pm 1}h_{n,n \pm 1} & -\psi_{n \pm 1}^*h_{n,n \pm 1}^* \\ \psi_n & -\psi_n^* \end{bmatrix}
 \begin{bmatrix} q_{n,\S}^\pm \\ \mathring{q}_{n,\S}^\pm \end{bmatrix}. \nonumber
\end{align}
The lower equation above yields the mapping relation in \cref{eq:current_mapping} from $\psi_n,\psi_n^*$ to $\psi_{\S(n)}$, where the upper ($+$) \textit{or} lower ($-$) NLCs can be equally used.
The upper equation yields the mapping relation between the sites adjacent to $n,\S(n)$ using the \textit{same} (correspondingly upper or lower) NLCs, provided that the involved hoppings are real and $\S$-symmetric, $h_{\S(n),\S(n \pm 1)} = h_{n,n \pm 1} = h_{n,n \pm 1}^*$.
Multiplying \cref{eq:current_mapping_matrix_general} from the left by the row vector 
\begin{equation}
 i[-h_{\S(n),\S(n \pm 1)}\psi_{\S(n)}^*, h_{\S(n),\S(n \pm 1)}^*\psi_{\S(n \pm 1)}^*] \nonumber
\end{equation}
we arrive the current identity in \cref{eq:current_connection}.
Note that for locally $\S$-symmetric and real equidirectional hoppings $h_{n,n \pm 1} = h_{n,n \pm 1}^* = h_{\S(n),\S(n \pm 1)}$ ($n \in \DD$), also the \textit{neighboring} amplitudes at $n \pm 1$ and $\S(n \pm 1)$ are related via the \textit{same} currents (now with corresponding sign in $\pm$); see \cref{eq:current_mapping_matrix_general}.
In other words we have 
$j_n^\pm \psi_{\S(n)} = 
 [ q_{n,\S}^\pm,-\mathring{q}_{n,\S}^\pm] [\psi_n,\psi_n^* ]^\top = 
 [ q_{n \mp 1,\S}^\pm,-\mathring{q}_{n \mp 1,\S}^\pm] [\psi_n,\psi_n^* ]^\top ,$
so that the $\S$-transformed amplitudes at $n,\S(n)$ are related by \textit{both} $\pm$ NLCs `flowing' on \textit{either} of the attached link pairs $(n,n \pm 1)$, $(\S(n),\S(n \pm 1))$; see nonlocal `flow' in \cref{fig:nonlocal_flow}.

Using the bitemporal NLC $q_{n,\S}^{\sT\pm}$ instead of $\mathring{q}_{n,\S}^\pm$ would also lead to a mapping relation analogous to \cref{eq:current_mapping}, though mapping $\psi_n(-t),\psi_n^*(t)$ to $\psi_{\S(n)}(t)$ and involving accordingly a bitemporal local current
\begin{align} \label{eq:j_n+-_bitemporal}
ij_n^{\sT\pm} \equiv
\psi_n^*(t) h_{n,n \pm 1} \psi_{n \pm 1}(-t) - \psi_{n \pm 1}^*(t) h_{n,n \pm 1}^* \psi_n(-t) .
\end{align}
In the case of a stationary state $\ket{\phi}$ (with real $E$), the stationary mapping relation of \cref{eq:current_mapping_stat} is identically recovered, since the temporal factors $e^{-iEt}$ are factored out.

The global version of the parity or Bloch theorem, $\phi_{\P(n)} = \pm \phi_n$ or $\phi_{\K(n)} = e^{ikL}\phi_n$, from the mapping relation (\ref{eq:current_mapping_stat}) is recovered directly by inserting Eqs.\,(\ref{eq:qtilde_P_zero}),(\ref{eq:q_P_j_Peig}) or \cref{eq:q_K_Bloch} with $\DD = \NN$ for a parity or Bloch eigenstate, respectively (or Eqs.\,(\ref{eq:q_P_zero}),(\ref{eq:qtilde_P_j_PTeig}) for a $\P\T$ eigenstate).
Note that, to avoid $j_n^\pm = 0$ for $\P$ eigenstates, a non-Hermitian $\P$-symmetric system can be considered, subsequently taking the limit of vanishing loss/gain to reproduce the Hermitian case.

We finally derive the general form of the \textit{summation} mapping relation leading to \cref{eq:sum_map_PT} as a special case.
Assuming $\P\T$-symmetric hoppings $h_{\S(n),\S(n \pm 1)} = h_{n,n \pm 1}^*$ for $n \in \DD$, we notice that the NLCs $q_n^\pm$ for a state $\ket{\psi}$ can be written (times $i$) as 
\begin{align} \label{eq:q_n_diff_operator}
 iq^\pm_n &= h_{n,n \pm 1}^* \left( \psi_{n}^* \varDelta^\pm \psi_{\bar{n}} - \psi_{\bar{n}}\varDelta^\pm \psi_{n}^* \right) \\
 &= h_{n,n \pm 1}^* \psi_{n}^* \psi_{n+1}^* \varDelta^\pm \left( \frac{\psi_{\bar{n}}}{\psi_{n}^*} \right), \label{eq:q_n_disc_quotient}
\end{align}
with the forward/backward discrete difference operator $\varDelta^{\pm}$ defined here by
\begin{equation}
 \varDelta^{\pm} \psi_{f(n)} = \psi_{f(n \pm 1)} - \psi_{n}
\end{equation}
for any mapping $f(n)$ (set to $f(n) = n$ or $f(n) = \bar{n} = \S(n)$ above).
Summing the discrete quotient difference \cite{Kelley2000____DifferenceEquations} in \cref{eq:q_n_disc_quotient} from site $n$ to  $n_0 \mp 1$ for a given site $n_0$ and solving for the mapped amplitude $\psi_{\bar{n}}$ we obtain
\begin{equation} \label{eq:sum_map_gen}
 \psi_{\bar{n}} = \psi_n^* \left( \frac{\psi_{\bar{n}_0}}{\psi_{n_0}^*} - i \sum_{m=n}^{n_0 \mp 1} \frac{q^\pm_m}{\psi_{m}^* h_{m,m \pm 1}^*  \psi_{m \pm 1}^*} \right),
\end{equation}
applicable for nonzero $\psi_{n_0},\psi_{m}, \psi_{m \pm 1}$.

The summation mapping presented here holds also for zero $j_n$, implying real $h_{n,n+1}$ for real $a_n$ in a stationary state $\ket{\phi}$, and can thus be seen as a complement to the current mapping in \cref{eq:current_mapping}. 
For real $a_n$ within a locally $\P\T$-symmetric domain $\DD$ with odd number of sites $D$, choosing the fixed point to be the symmetry center $n_0 = \ax$ (so that $a_{\bar{n}_0} = a_{n_0}$) simplifies the above expression to \cref{eq:sum_map_PT}.

\section{Nonlocal currents from scattering amplitudes}
\label{app:scattering_NLC}

To simulate scattering of an incident monochromatic wave on a localized potential in the discrete system, we attach semi-infinite chains with real constant onsite elements $v_n = v$ and hoppings $h_n^\pm = h$ on the left ($n \leqslant 0$) and right ($n\geqslant N+1$) of the $N$-site scatterer.
In the homogeneous chain leads, stationary solutions of the SE (\ref{eq:Schr}) are linear combinations of plain wave states $\ket{\pm} e^{-iEt}$ where $\braket{n|\pm} = e^{\pm ikn} \equiv \zeta^{\pm n}$ with quasimomentum $k$ for energies obeying the dispersion relation 
\begin{equation} \label{eq:dispersion}
 E = v + 2h \cos k .
\end{equation}
For `biased' setups with different lead Hamiltonians ($v_{n<1} \neq v_{n > N}$ and/or $h_{n<1}^\pm \neq h_{n > N}^\pm$) the plain wave states are flux normalized by the respective factor $[h \sin k]^{1/2}$ with $k$ given by the lead dispersion;
we here consider unbiased setups with equal $v,h$ in the two leads for simplicity.
The spatial amplitudes $a_n$ of a scattering state $\ket{\psi_E}$ are expanded in the plain waves of the leads as ($\zeta = e^{ik}$)
\begin{align} \label{eq:scat_bc}
  a_n = \begin{cases} \mf{c}_<^+ \, \zeta^{+n} + \mf{c}_<^- \, \zeta^{-n}, &~n < 1 \\ 
		      \mf{c}_>^+ \, \zeta^{+n} + \mf{c}_>^- \, \zeta^{-n}, &~n > N \end{cases}
\end{align}
on the left and right of the scatterer, respectively.
The computation of the amplitudes $a_n$ along the scattering region as well as the output amplitudes $\mf{c}_<^-,\mf{c}_<^+$ for a given input is presented in compact form in \cref{app:scattering}.

An incoming plain wave with unit amplitude from the left, setting $\mf{c}_<^+ = 1, \mf{c}_>^- = 0$ (from the right, setting $\mf{c}_<^+ = 0, \mf{c}_>^- = 1$), is reflected with amplitude $\mf{c}_<^- = \mf{r}$ ($\mf{c}_>^+ = \mf{r}'$) and transmitted with amplitude $\mf{c}_>^+ = \mf{t}$ ($\mf{c}_<^- = \mf{t}'$).
These scattering amplitudes define the scattering matrix $\mathsf{S}$ which maps general input amplitudes to their output in the $\ket{\pm}$ basis,
\begin{equation} \label{eq:Smatrix}
 \begin{bmatrix}\, \mf{c}_<^- \, \vspace{.7ex} \\ \mf{c}_>^+ \end{bmatrix} = \mathsf{S}
 \begin{bmatrix}\, \mf{c}_<^+ \, \vspace{.7ex} \\ \mf{c}_>^- \end{bmatrix}, ~~
 \mathsf{S} = \begin{bmatrix} \, \mf{r} & \mf{t}' \, \vspace{.7ex} \\ \, \mf{t}  & \mf{r}' \, \end{bmatrix}.
\end{equation}
The reflection and transmission coefficients are given by $\mf{R} = |\mf{r}|^2$ and $\mf{T} = |\mf{t}|^2$, respectively, with $\mf{R} + \mf{T} = 1$ dictated by current conservation in the absence of (imbalanced) loss and gain.

For a (local) $\P$ transform, the NLC on the first links on the left and right of a scatterer localized in a domain $\DD$ can generally be written in terms of the in- and outgoing amplitudes in the $\ket{\pm}$ basis as
\begin{align} \label{eq:q_scat_in-out-amps}
  q_{l-1,l} = q_{r+1,r}^* &=  2h \sin k 
  \left( \, \mathring{\mf{c}}_<^{+*} \, \mathring{\mf{c}}_>^-   -  \mathring{\mf{c}}_<^{-*} \, \mathring{\mf{c}}_>^+ \, \right) \nonumber \\[1.5ex]  
 &= ih (\zeta^* - \zeta) 
 \begin{bmatrix}\, {\mf{c}}_<^+ \, \vspace{.7ex} \\ {\mf{c}}_<^- \end{bmatrix}^\dagger
 \mathsf{A}^\dagger \mathsf{\Lambda} \, \mathsf{A}
 \begin{bmatrix}\, {\mf{c}}_>^+ \, \vspace{.7ex} \\ {\mf{c}}_>^- \end{bmatrix},
\end{align}
where $\mathsf{\Lambda} = {\footnotesize \begin{bmatrix}~ 0 & 1 \vspace{.7ex} \\[-1ex] -1 & 0 ~\end{bmatrix} }$ and $\mathring{\mf{c}}_{<(>)}^\pm = \zeta^{\pm \ax} \mf{c}_{<(>)}^\pm$ correspond to a shift of the scatterer by $-\ax$ to the origin via the matrix $\mathsf{A}(\ax) = \text{d}[\zeta^{+\ax},\zeta^{-\ax}]$, with the center $\ax$ of $\DD$ taken to coincide with the center of inversion (we suppress the subscript $\P$ in $q_{n;\P}^\pm$ from here on).
Note that, since the scattering matrix $\mathsf{S}$ connects $\ket{\pm}$-amplitudes, the above expression for the NLC remains essentially the same for a continuous system: 
The discreteness of the system enters merely through the factor $h (\zeta - \zeta^*) = 2ih \sin k$, which is replaced by $ik$ in the continuum limit (with real $h$) where the NLC becomes a two-point quantity $q(x,\bar{x})$.

Substituting the outgoing amplitudes $\mf{c}_<^-,\mf{c}_>^+$ from \cref{eq:Smatrix}, the NLC above takes the form of \cref{eq:q_scat_in_t_r_general} in the main text, where it is used that $\mf{t} = \mf{t}'$ in general \cite{Ge2012_PRA_85_023802_ConservationRelations,Lieb1966____MathematicalPhysics} for nonsingular Hamiltonians.

\section{Computation of scattering state for arbitrary input and scatterer type}
\label{app:scattering}

Augmenting the scatterer domain $\NN$ by two lead sites on each side ($n=-1,0$ and $n=N+1,N+2$ in the present indexing), we can solve the close-coupling equations (\ref{eq:Schr_cc}) for the outgoing plane wave amplitudes $\mf{c}_<^-,\mf{c}_>^+$ as well as the site amplitudes 
\begin{equation}
 \bs{a} \equiv [a_1,a_2,\dots,a_N]^\top
\end{equation}
along the scatterer by sparse matrix multiplication and inversion of size $N+2$.
Specifically, using the asymptotic ansatz (\ref{eq:scat_bc}) for the two sites on either side of $\NN$, we rearrange the Schr\"odinger system of $N+2$ next-neighbor equations to solve for $\mf{c}_<^-,\mf{c}_>^+$ and $\bs{a}$ in the form of the column vector
\begin{align} \label{eq:scat_general_output}
 \begin{bmatrix}\, \mf{c}_<^- \, \\ \bs{a} \\ \mf{c}_>^+ \end{bmatrix} =
 -\left[ \, W Z \, \right]^{-1}
 \begin{bmatrix}\, 
  (h^*\zeta^* + \tilde{v}) \zeta^{+l}\, \mf{c}_<^+ \, \\ 
 \bs{\zeta}^<_> \\ 
  (h^*\zeta^* + \tilde{v}) \zeta^{-r} \, \mf{c}_>^- \,\end{bmatrix}.
\end{align}
Here, $l$ ($r$) is the index of the lead site attached to the leftmost (rightmost) scatterer site, $\tilde{v} = v - E$, and
\begin{equation} \label{eq:HZ}
 \bs{\zeta}^<_>  = \begin{bmatrix}\, h^* \zeta^{+l} \mf{c}_<^+ \, \\ \bs{0}_{N-2} \\ \, h^* \zeta^{-r} \mf{c}_>^- \,\end{bmatrix},~~
 Z = \begin{bmatrix}\, \zeta^{-l+1} & \bs{0}_N^\top & 0 \\ \zeta^{-l} & \bs{0}_N^\top & 0 \\ \bs{0}_N & I_N & \bs{0}_N \\ 0 & \bs{0}_N^\top & \zeta^{r} \\ 0 & \bs{0}_N^\top & \zeta^{r+1} \,\end{bmatrix},
 \end{equation}
with $\bs{0}_N$ being a $1 \times N$ row vector of zeros and $I_N$ the $N \times N$ unit matrix.
The $(N+2) \times (N+4)$ matrix $W$ is the $(N+2) \times (N+2)$ block of $H - EI$ corresponding to the domain $\{l,\NN,r\}$, augmented by the column $[h ~ 0 \cdots 0]^\top$ on the left and $[0 \cdots 0 ~ h]^\top$ on the right.

With the scatterer domain defined as $\NN = \{1,2,...,N \}$ we here have $l = 0$ and $r = N+1$.
In the case $N = 1$ of a single site scatterer, $\bs{\zeta}^<_>$ reduces to a single element 
\begin{equation}
 {\zeta}^<_> = h^* (\zeta^{+l} \mf{c}_<^+ + \zeta^{-r} \mf{c}_>^-).
\end{equation}
Note also that for biased setups, the $\tilde{v},h$ and $k$ entering the top (bottom) row elements in \cref{eq:scat_general_output} are accordingly replaced by those of the left (right) lead;
in the present work we consider unbiased setups.

The above method is essentially equivalent to propagating the input signal via the resolvent (lattice Green function) of the effective lead-coupled Hamiltonian of the scattering region.
Here, however, we explicitly solve outgoing $\ket{\pm}$-amplitudes and combine them compactly into a column vector with the spatially resolved field in the scatterer domain $\NN$.
This formulation of the scattering problem has an advantage over a standard site-by-site transfer matrix method in that the scatterer can be of arbitrary type:
Keeping a two-terminal geometry, the scatterer may contain, e.\,g., side-coupled sites or a 2D cluster of interconnected sites.
No explicit single-node transfer matrices propagating the wave along the scatterer need to be specified, rather the complete Hamiltonian under appropriate indexing is utilized directly.

\section{Derivation of \cref{eq:PTR_condition_q,eq:t_PTR}}
\label{app:derivation_q}

Assuming the two scatterers in $\DD_1$ and $\DD_2$ to be $\P$-symmetric, each will have spatially constant NLC $q_{\DD_d}^\pm \in i\RR$ ($d = 1,2$) at a given $E$.
Using the general expression (\ref{eq:q_scat_in-out-amps}), the NLCs in $\DD_1$ can be written as
\begin{align} \label{eq:q_scat_D1}
  &q_{\DD_1}^+ = - q_{\DD_1}^- = ih (\zeta^* - \zeta) \times \nonumber \\
 &\begin{bmatrix}\, \mf{c}_+ \, \vspace{.7ex} \\ \mf{r}\mf{c}_+ \end{bmatrix}^\dagger
 \mathsf{A}^\dagger \mathsf{\Lambda} \, \mathsf{A} \, \mathsf{C}
 \begin{bmatrix}\, \mf{c}_+ \, \vspace{.7ex} \\ \mf{c}_- \end{bmatrix},
 ~~~ \mathsf{C} = \mf{p} \begin{bmatrix} \mf{t}_1 & \mf{r}_1'\mf{t}_2' \\ \mf{r}_2 \mf{t}_1 & \mf{t}_2' \end{bmatrix},
\end{align}
in terms of the global in- and outgoing $\ket{\pm}$-amplitudes $\mf{c}_+ = \mf{c}_<^+, \mf{c}_- = \mf{c}_>^-$, where the `connection' matrix $\mathsf{C}$ was used to eliminate the amplitudes between the scatterers (note that the $\DD_1$ and $\DD_2$ may be directly attached in the above formulation).
Considering now the case of a PTR ($\mf{r} = 0$, $\mf{r}_1' = \mf{r}_2^*$) under left incidence only (${\mf{c}}_>^- = 0$), the upper NLC becomes 
\begin{equation} \label{eq:q_scat_D1_PTR}
  q_{\DD_1}^+ = \pm 2 i h \sin k \, \frac{|\mf{r}_1|}{|\mf{t}_1|} \, |\mf{c}_+|^2 .
\end{equation}
On the other hand, its difference from $q_{\DD_2}^+$ is generally given by 
\begin{align} \label{eq:q_scat_D1_D2_diff}
 q_{\DD_1}^+ &- q_{\DD_2}^+ = i h (\zeta^* - \zeta) \times \\ 
 &\mf{p} \mf{t}_1 
 \left\{ \, \left(1 - \mf{t}^* \zeta^{2(\ax_1 - \ax_2)} \right) \zeta^{-2\ax_1}\,\mf{r}_2  
	  - \zeta^{2\ax_1}\,\mf{r}^* \, \right\} |\mf{c}_+|^2 \nonumber 
\end{align}
for left incidence.
In the case of a PTR, applying conditions (\ref{eq:PTR_condition}) and (\ref{eq:t_PTR}) necessarily leads to \cref{eq:PTR_condition_q} for \cref{eq:q_scat_D1_D2_diff} to be consistent with \cref{eq:q_scat_D1}, recalling that $\mathring{\mf{r}}_d^*\mf{t}^{ }_d \in i\RR^\pm$.

The PTR condition in \cref{eq:t_PTR} for two scatterers can, finally, be derived using the domain NLCs.
Indeed, the relation between NLCs and transparency of a two-scatterer setup can be verified for a setup with a gap of two `free' sites $c-1,c$ ($v_{c-1} = v_c = v$, $h_{c-1,c} = h$) between $\DD_1$ and $\DD_2$, for which an explicit calculation yields
\begin{equation}
 q_{\DD_1}^+ = \mf{t} \, \zeta^{2(\ax_2 - \ax_1)} q_{\DD_2}^+
\end{equation}
for $|\mf{t}| = 1$.
Since the $q_{\DD_1}^+ \equiv q_{l-1;\P_{\DD_1}}^+$, $q_{\DD_2}^+ \equiv q_{r;\P_{\DD_2}}^+$ are constant and imaginary, we have that $\mf{t} \, \zeta^{2(\ax_2 - \ax_1)} = \pm 1$, which is \cref{eq:t_PTR}.

%%%%%%%%%%%%%%%%%%%%%%%%%%%%%%%%%%%%%%%%%%%%%%%%%%%%%

%


\begin{thebibliography}{71}%
\makeatletter
\providecommand \@ifxundefined [1]{%
 \@ifx{#1\undefined}
}%
\providecommand \@ifnum [1]{%
 \ifnum #1\expandafter \@firstoftwo
 \else \expandafter \@secondoftwo
 \fi
}%
\providecommand \@ifx [1]{%
 \ifx #1\expandafter \@firstoftwo
 \else \expandafter \@secondoftwo
 \fi
}%
\providecommand \natexlab [1]{#1}%
\providecommand \enquote  [1]{``#1''}%
\providecommand \bibnamefont  [1]{#1}%
\providecommand \bibfnamefont [1]{#1}%
\providecommand \citenamefont [1]{#1}%
\providecommand \href@noop [0]{\@secondoftwo}%
\providecommand \href [0]{\begingroup \@sanitize@url \@href}%
\providecommand \@href[1]{\@@startlink{#1}\@@href}%
\providecommand \@@href[1]{\endgroup#1\@@endlink}%
\providecommand \@sanitize@url [0]{\catcode `\\12\catcode `\$12\catcode
  `\&12\catcode `\#12\catcode `\^12\catcode `\_12\catcode `\%12\relax}%
\providecommand \@@startlink[1]{}%
\providecommand \@@endlink[0]{}%
\providecommand \url  [0]{\begingroup\@sanitize@url \@url }%
\providecommand \@url [1]{\endgroup\@href {#1}{\urlprefix }}%
\providecommand \urlprefix  [0]{URL }%
\providecommand \Eprint [0]{\href }%
\providecommand \doibase [0]{http://dx.doi.org/}%
\providecommand \selectlanguage [0]{\@gobble}%
\providecommand \bibinfo  [0]{\@secondoftwo}%
\providecommand \bibfield  [0]{\@secondoftwo}%
\providecommand \translation [1]{[#1]}%
\providecommand \BibitemOpen [0]{}%
\providecommand \bibitemStop [0]{}%
\providecommand \bibitemNoStop [0]{.\EOS\space}%
\providecommand \EOS [0]{\spacefactor3000\relax}%
\providecommand \BibitemShut  [1]{\csname bibitem#1\endcsname}%
\let\auto@bib@innerbib\@empty
%</preamble>
\bibitem [{\citenamefont {Zettili}(2009)}]{Zettili2009____QuantumMechanics}%
  \BibitemOpen
  \bibfield  {author} {\bibinfo {author} {\bibfnamefont {N.}\
  \bibnamefont {Zettili}},\ }\href@noop {} {\emph {\bibinfo {title} {Quantum
  Mechanics: Concepts and Applications}}}\ 
  (\bibinfo  {publisher} {Wiley},\ \bibinfo {address} {Chichester, {U.K}},\
  \bibinfo {year} {2009})\BibitemShut {NoStop}%
\bibitem [{\citenamefont {Jiles}(1994)}]{Jiles1994____IntroductionElectronic}%
  \BibitemOpen
  \bibfield  {author} {\bibinfo {author} {\bibfnamefont {D.~C.}\
  \bibnamefont {Jiles}},\ }\href@noop {} {\emph {\bibinfo {title} {Introduction
  to the Electronic Properties of Materials}}},\ 
  ed.\ (\bibinfo  {publisher} {Springer},\ \bibinfo {address} {London},\ \bibinfo {year} {1994})\BibitemShut {NoStop}%
\bibitem [{\citenamefont {Hasan}\ and\ \citenamefont
  {Kane}(2010)}]{Hasan2010_RMP_82_3045_TextitcolloquiumTopological}%
  \BibitemOpen
  \bibfield  {author} {\bibinfo {author} {\bibfnamefont {M.~Z.}\ \bibnamefont
  {Hasan}}\ and\ \bibinfo {author} {\bibfnamefont {C.~L.}\ \bibnamefont
  {Kane}},\ }\bibfield  {title} {\enquote {\bibinfo {title}
  {{\textit{Colloquium}} : Topological insulators},}\ }\href {\doibase
  10.1103/RevModPhys.82.3045} {\bibfield  {journal} {\bibinfo  {journal}
  {Rev. Mod. Phys.}\ }\textbf {\bibinfo {volume} {82}},\ \bibinfo
  {pages} {3045} (\bibinfo {year} {2010})}\BibitemShut {NoStop}%
\bibitem [{\citenamefont {Burkov}\ \emph {et~al.}(2011)\citenamefont {Burkov},
  \citenamefont {Hook},\ and\ \citenamefont
  {Balents}}]{Burkov2011_PRB_84_235126_TopologicalNodal}%
  \BibitemOpen
  \bibfield  {author} {\bibinfo {author} {\bibfnamefont {A.~A.}\ \bibnamefont
  {Burkov}}, \bibinfo {author} {\bibfnamefont {M.~D.}\ \bibnamefont {Hook}}, \
  and\ \bibinfo {author} {\bibfnamefont {L.}\ \bibnamefont {Balents}},\
  }\bibfield  {title} {\enquote {\bibinfo {title} {Topological nodal
  semimetals},}\ }\href {\doibase 10.1103/PhysRevB.84.235126} {\bibfield
  {journal} {\bibinfo  {journal} {Phys. Rev. B}\ }\textbf {\bibinfo
  {volume} {84}},\ \bibinfo {pages} {235126} (\bibinfo {year}
  {2011})}\BibitemShut {NoStop}%
\bibitem [{\citenamefont {Vicencio}\ \emph {et~al.}(2015)\citenamefont
  {Vicencio}, \citenamefont {Cantillano}, \citenamefont {{Morales-Inostroza}},
  \citenamefont {Real}, \citenamefont {{Mej\'{i}a-Cort\'{e}s}}, \citenamefont
  {Weimann}, \citenamefont {Szameit},\ and\ \citenamefont
  {Molina}}]{Vicencio2015_PRL_114_245503_ObservationLocalized}%
  \BibitemOpen
  \bibfield  {author} {\bibinfo {author} {\bibfnamefont {R.~A.}\
  \bibnamefont {Vicencio}}, \bibinfo {author} {\bibfnamefont {C.}\
  \bibnamefont {Cantillano}}, \bibinfo {author} {\bibfnamefont {L.}\
  \bibnamefont {{Morales-Inostroza}}}, \bibinfo {author} {\bibfnamefont
  {B.}\ \bibnamefont {Real}}, \bibinfo {author} {\bibfnamefont
  {C.}\ \bibnamefont {{Mej\'{i}a-Cort\'{e}s}}}, \bibinfo {author}
  {\bibfnamefont {S.}\ \bibnamefont {Weimann}}, \bibinfo {author}
  {\bibfnamefont {A.}\ \bibnamefont {Szameit}}, \ and\ \bibinfo {author}
  {\bibfnamefont {M.~I.}\ \bibnamefont {Molina}},\ }\bibfield  {title}
  {\enquote {\bibinfo {title} {Observation of localized states in Lieb photonic
  lattices},}\ }\href {\doibase 10.1103/PhysRevLett.114.245503} {\bibfield
  {journal} {\bibinfo  {journal} {Phys. Rev. Lett.}\ }\textbf {\bibinfo
  {volume} {114}},\ \bibinfo {pages} {245503} (\bibinfo {year}
  {2015})}\BibitemShut {NoStop}%
\bibitem [{\citenamefont {Grossmann}\ \emph {et~al.}(1991)\citenamefont
  {Grossmann}, \citenamefont {Dittrich}, \citenamefont {Jung},\ and\
  \citenamefont {H\"{a}nggi}}]{Grossmann1991_PRL_67_516_CoherentDestruction}%
  \BibitemOpen
  \bibfield  {author} {\bibinfo {author} {\bibfnamefont {F.}~\bibnamefont
  {Grossmann}}, \bibinfo {author} {\bibfnamefont {T.}~\bibnamefont {Dittrich}},
  \bibinfo {author} {\bibfnamefont {P.}~\bibnamefont {Jung}}, \ and\ \bibinfo
  {author} {\bibfnamefont {P.}~\bibnamefont {H\"{a}nggi}},\ }\bibfield  {title}
  {\enquote {\bibinfo {title} {Coherent destruction of tunneling},}\ }\href
  {\doibase 10.1103/PhysRevLett.67.516} {\bibfield  {journal} {\bibinfo
  {journal} {Phys. Rev. Lett.}\ }\textbf {\bibinfo {volume} {67}},\
  \bibinfo {pages} {516} (\bibinfo {year} {1991})}\BibitemShut {NoStop}%
\bibitem [{\citenamefont {Lehmann}\ \emph {et~al.}(2003)\citenamefont
  {Lehmann}, \citenamefont {Kohler}, \citenamefont {H\"{a}nggi},\ and\
  \citenamefont
  {Nitzan}}]{Lehmann2003_JCP_118_3283_RectificationLaser-induced}%
  \BibitemOpen
  \bibfield  {author} {\bibinfo {author} {\bibfnamefont {J.}\
  \bibnamefont {Lehmann}}, \bibinfo {author} {\bibfnamefont {S.}\
  \bibnamefont {Kohler}}, \bibinfo {author} {\bibfnamefont {P.}\
  \bibnamefont {H\"{a}nggi}}, \ and\ \bibinfo {author} {\bibfnamefont
  {A.}\ \bibnamefont {Nitzan}},\ }\bibfield  {title} {\enquote {\bibinfo
  {title} {Rectification of laser-induced electronic transport through
  molecules},}\ }\href {\doibase 10.1063/1.1536639} {\bibfield  {journal}
  {\bibinfo  {journal} {J. Chem. Phys.}\ }\textbf {\bibinfo
  {volume} {118}},\ \bibinfo {pages} {3283} (\bibinfo {year}
  {2003})}\BibitemShut {NoStop}%
\bibitem [{\citenamefont {Denisov}\ \emph {et~al.}(2007)\citenamefont
  {Denisov}, \citenamefont {{Morales-Molina}}, \citenamefont {Flach},\ and\
  \citenamefont {H\"{a}nggi}}]{Denisov2007_PRA_75_063424_PeriodicallyDriven}%
  \BibitemOpen
  \bibfield  {author} {\bibinfo {author} {\bibfnamefont {S.}~\bibnamefont
  {Denisov}}, \bibinfo {author} {\bibfnamefont {L.}~\bibnamefont
  {{Morales-Molina}}}, \bibinfo {author} {\bibfnamefont {S.}~\bibnamefont
  {Flach}}, \ and\ \bibinfo {author} {\bibfnamefont {P.}~\bibnamefont
  {H\"{a}nggi}},\ }\bibfield  {title} {\enquote {\bibinfo {title} {Periodically
  driven quantum ratchets: Symmetries and resonances},}\ }\href {\doibase
  10.1103/PhysRevA.75.063424} {\bibfield  {journal} {\bibinfo  {journal}
  {Phys. Rev. A}\ }\textbf {\bibinfo {volume} {75}},\ \bibinfo {pages}
  {063424} (\bibinfo {year} {2007})}\BibitemShut {NoStop}%
\bibitem [{\citenamefont
  {Maci\'{a}}(2006)}]{Macia2006_RPP_69_397_RoleAperiodic}%
  \BibitemOpen
  \bibfield  {author} {\bibinfo {author} {\bibfnamefont {Enrique}\ \bibnamefont
  {Maci\'{a}}},\ }\bibfield  {title} {\enquote {\bibinfo {title} {The role of
  aperiodic order in science and technology},}\ }\href {\doibase
  10.1088/0034-4885/69/2/R03} {\bibfield  {journal} {\bibinfo  {journal}
  {Rep. Prog. Phys.}\ }\textbf {\bibinfo {volume} {69}},\
  \bibinfo {pages} {397} (\bibinfo {year} {2006})}\BibitemShut {NoStop}%
\bibitem [{\citenamefont {Albuquerque}\ and\ \citenamefont
  {Cottam}(2003)}]{Albuquerque2003_PR_376_225_TheoryElementary}%
  \BibitemOpen
  \bibfield  {author} {\bibinfo {author} {\bibfnamefont {E.~L.}\ \bibnamefont
  {Albuquerque}}\ and\ \bibinfo {author} {\bibfnamefont {M.~G.}\ \bibnamefont
  {Cottam}},\ }\bibfield  {title} {\enquote {\bibinfo {title} {Theory of
  elementary excitations in quasiperiodic structures},}\ }\href {\doibase
  10.1016/S0370-1573(02)00559-8} {\bibfield  {journal} {\bibinfo  {journal}
  {Phys. Rep.}\ }\textbf {\bibinfo {volume} {376}},\ \bibinfo {pages}
  {225} (\bibinfo {year} {2003})}\BibitemShut {NoStop}%
\bibitem [{\citenamefont {Lahini}\ \emph {et~al.}(2009)\citenamefont {Lahini},
  \citenamefont {Pugatch}, \citenamefont {Pozzi}, \citenamefont {Sorel},
  \citenamefont {Morandotti}, \citenamefont {Davidson},\ and\ \citenamefont
  {Silberberg}}]{Lahini2009_PRL_103_013901_ObservationLocalization}%
  \BibitemOpen
  \bibfield  {author} {\bibinfo {author} {\bibfnamefont {Y.}~\bibnamefont
  {Lahini}}, \bibinfo {author} {\bibfnamefont {R.}~\bibnamefont {Pugatch}},
  \bibinfo {author} {\bibfnamefont {F.}~\bibnamefont {Pozzi}}, \bibinfo
  {author} {\bibfnamefont {M.}~\bibnamefont {Sorel}}, \bibinfo {author}
  {\bibfnamefont {R.}~\bibnamefont {Morandotti}}, \bibinfo {author}
  {\bibfnamefont {N.}~\bibnamefont {Davidson}}, \ and\ \bibinfo {author}
  {\bibfnamefont {Y.}~\bibnamefont {Silberberg}},\ }\bibfield  {title}
  {\enquote {\bibinfo {title} {Observation of a localization transition in
  quasiperiodic photonic lattices},}\ }\href {\doibase
  10.1103/PhysRevLett.103.013901} {\bibfield  {journal} {\bibinfo  {journal}
  {Phys. Rev. Lett.}\ }\textbf {\bibinfo {volume} {103}},\ \bibinfo
  {pages} {013901} (\bibinfo {year} {2009})}\BibitemShut {NoStop}%
\bibitem [{\citenamefont {Morfonios}\ \emph {et~al.}(2014)\citenamefont
  {Morfonios}, \citenamefont {Schmelcher}, \citenamefont {Kalozoumis},\ and\
  \citenamefont {Diakonos}}]{Morfonios2014_ND_78_71_LocalSymmetry}%
  \BibitemOpen
  \bibfield  {author} {\bibinfo {author} {\bibfnamefont {C.}~\bibnamefont
  {Morfonios}}, \bibinfo {author} {\bibfnamefont {P.}~\bibnamefont
  {Schmelcher}}, \bibinfo {author} {\bibfnamefont {P.~A.}\ \bibnamefont
  {Kalozoumis}}, \ and\ \bibinfo {author} {\bibfnamefont {F.~K.}\ \bibnamefont
  {Diakonos}},\ }\bibfield  {title} {\enquote {\bibinfo {title} {Local symmetry
  dynamics in one-dimensional aperiodic lattices: a numerical study},}\ }\href
  {\doibase 10.1007/s11071-014-1422-1} {\bibfield  {journal} {\bibinfo
  {journal} {Nonlin. Dyn.}\ }\textbf {\bibinfo {volume} {78}},\ \bibinfo
  {pages} {71} (\bibinfo {year} {2014})}\BibitemShut {NoStop}%
\bibitem [{\citenamefont {Wochner}\ \emph {et~al.}(2009)\citenamefont
  {Wochner}, \citenamefont {Gutt}, \citenamefont {Autenrieth}, \citenamefont
  {Demmer}, \citenamefont {Bugaev}, \citenamefont {Ortiz}, \citenamefont
  {Duri}, \citenamefont {Zontone}, \citenamefont {Gr\"{u}bel},\ and\
  \citenamefont {Dosch}}]{Wochner2009_PNAS_106_11511_X-rayCross}%
  \BibitemOpen
  \bibfield  {author} {\bibinfo {author} {\bibfnamefont {P.}\ \bibnamefont
  {Wochner}}, \bibinfo {author} {\bibfnamefont {C.}\ \bibnamefont
  {Gutt}}, \bibinfo {author} {\bibfnamefont {T.}\ \bibnamefont {Autenrieth}},
  \bibinfo {author} {\bibfnamefont {T.}\ \bibnamefont {Demmer}}, \bibinfo
  {author} {\bibfnamefont {V.}\ \bibnamefont {Bugaev}}, \bibinfo
  {author} {\bibfnamefont {A.~D.}\ \bibnamefont {Ortiz}}, \bibinfo
  {author} {\bibfnamefont {A.}\ \bibnamefont {Duri}}, \bibinfo {author}
  {\bibfnamefont {F.}\ \bibnamefont {Zontone}}, \bibinfo {author}
  {\bibfnamefont {G.}\ \bibnamefont {Gr\"{u}bel}}, \ and\ \bibinfo
  {author} {\bibfnamefont {H.}\ \bibnamefont {Dosch}},\ }\bibfield  {title}
  {\enquote {\bibinfo {title} {X-ray cross correlation analysis uncovers hidden
  local symmetries in disordered matter},}\ }\href {\doibase
  10.1073/pnas.0905337106} {\bibfield  {journal} {\bibinfo  {journal}
  {Proc. Natl. Acad. Sci.}\ }\textbf {\bibinfo
  {volume} {106}},\ \bibinfo {pages} {11511} (\bibinfo {year}
  {2009})}\BibitemShut {NoStop}%
\bibitem [{\citenamefont {Lahini}\ \emph {et~al.}(2008)\citenamefont {Lahini},
  \citenamefont {Avidan}, \citenamefont {Pozzi}, \citenamefont {Sorel},
  \citenamefont {Morandotti}, \citenamefont {Christodoulides},\ and\
  \citenamefont {Silberberg}}]{Lahini2008_PRL_100_013906_AndersonLocalization}%
  \BibitemOpen
  \bibfield  {author} {\bibinfo {author} {\bibfnamefont {Y.}\ \bibnamefont
  {Lahini}}, \bibinfo {author} {\bibfnamefont {A.}\ \bibnamefont {Avidan}},
  \bibinfo {author} {\bibfnamefont {F.}\ \bibnamefont {Pozzi}}, \bibinfo
  {author} {\bibfnamefont {M.}\ \bibnamefont {Sorel}}, \bibinfo {author}
  {\bibfnamefont {R.}\ \bibnamefont {Morandotti}}, \bibinfo {author}
  {\bibfnamefont {D.~N.}\ \bibnamefont {Christodoulides}}, \ and\
  \bibinfo {author} {\bibfnamefont {Y.}\ \bibnamefont {Silberberg}},\
  }\bibfield  {title} {\enquote {\bibinfo {title} {Anderson localization and
  nonlinearity in {One-Dimensional} disordered photonic lattices},}\ }\href
  {\doibase 10.1103/PhysRevLett.100.013906} {\bibfield  {journal} {\bibinfo
  {journal} {Phys. Rev. Lett.}\ }\textbf {\bibinfo {volume} {100}},\
  \bibinfo {pages} {013906} (\bibinfo {year} {2008})}\BibitemShut {NoStop}%
\bibitem [{\citenamefont {Mustapha}\ \emph {et~al.}(2013)\citenamefont
  {Mustapha}, \citenamefont {{D{\textquoteright}Arco}}, \citenamefont {Pierre},
  \citenamefont {No\"{e}l}, \citenamefont {Ferrabone},\ and\ \citenamefont
  {Dovesi}}]{Mustapha2013_JPCM_25_105401_UseSymmetry}%
  \BibitemOpen
  \bibfield  {author} {\bibinfo {author} {\bibfnamefont {S.}\ \bibnamefont
  {Mustapha}}, \bibinfo {author} {\bibfnamefont {P.}\ \bibnamefont
  {{D{\textquoteright}Arco}}}, \bibinfo {author} {\bibfnamefont {M.~De~La}\
  \bibnamefont {Pierre}}, \bibinfo {author} {\bibfnamefont {Y.}\ \bibnamefont
  {No\"{e}l}}, \bibinfo {author} {\bibfnamefont {M.}\ \bibnamefont
  {Ferrabone}}, \ and\ \bibinfo {author} {\bibfnamefont {R.}\ \bibnamefont
  {Dovesi}},\ }\bibfield  {title} {\enquote {\bibinfo {title} {On the use of
  symmetry in configurational analysis for the simulation of disordered
  solids},}\ }\href {\doibase 10.1088/0953-8984/25/10/105401} {\bibfield
  {journal} {\bibinfo  {journal} {J. Phys. Condens. Matter.}\
  }\textbf {\bibinfo {volume} {25}},\ \bibinfo {pages} {105401} (\bibinfo
  {year} {2013})}\BibitemShut {NoStop}%
\bibitem [{\citenamefont {Pascal}(2001)}]{Pascal2001_JPCA_105_9040_ConciseSet}%
  \BibitemOpen
  \bibfield  {author} {\bibinfo {author} {\bibfnamefont {R.~A.}\
  \bibnamefont {Pascal}},\ }\bibfield  {title} {\enquote {\bibinfo {title} {A
  concise set of {{\textquotedblleft}Large{\textquotedblright}}, symmetric
  molecules for evaluation of modern computational methods},}\ }\href {\doibase
  10.1021/jp011482j} {\bibfield  {journal} {\bibinfo  {journal} {J.
  Phys. Chem. A}\ }\textbf {\bibinfo {volume} {105}},\ \bibinfo {pages}
  {9040} (\bibinfo {year} {2001})}\BibitemShut {NoStop}%
\bibitem [{\citenamefont {Domaga\l{}a}\ and\ \citenamefont
  {Jelsch}(2008)}]{Domagaa2008_JAC_41_1140_OptimalLocal}%
  \BibitemOpen
  \bibfield  {author} {\bibinfo {author} {\bibfnamefont {S.}\
  \bibnamefont {Domaga\l{}a}}\ and\ \bibinfo {author} {\bibfnamefont
  {C.}\ \bibnamefont {Jelsch}},\ }\bibfield  {title} {\enquote {\bibinfo
  {title} {Optimal local axes and symmetry assignment for charge-density
  refinement},}\ }\href {\doibase 10.1107/S0021889808033384} {\bibfield
  {journal} {\bibinfo  {journal} {J. Appl. Cryst.}\ }\textbf
  {\bibinfo {volume} {41}},\ \bibinfo {pages} {1140} (\bibinfo {year}
  {2008})}\BibitemShut {NoStop}%
\bibitem [{\citenamefont {Ferry}\ and\ \citenamefont
  {Goodnick}(1997)}]{Ferry1997____TransportNanostructures}%
  \BibitemOpen
  \bibfield  {author} {\bibinfo {author} {\bibfnamefont {D.~K.}\ \bibnamefont
  {Ferry}}\ and\ \bibinfo {author} {\bibfnamefont {S.~M.}\
  \bibnamefont {Goodnick}},\ }\href
  {http://ebooks.cambridge.org/ebook.jsf?bid=CBO9780511626128} {\emph {\bibinfo
  {title} {Transport in Nanostructures}}}\ (\bibinfo  {publisher} {Cambridge
  University Press},\ \bibinfo {address} {Cambridge},\ \bibinfo {year}
  {1997})\BibitemShut {NoStop}%
\bibitem [{\citenamefont
  {Zhukovsky}(2010)}]{Zhukovsky2010_PRA_81_053808_PerfectTransmission}%
  \BibitemOpen
  \bibfield  {author} {\bibinfo {author} {\bibfnamefont {S.~V.}\
  \bibnamefont {Zhukovsky}},\ }\bibfield  {title} {\enquote {\bibinfo {title}
  {Perfect transmission and highly asymmetric light localization in photonic
  multilayers},}\ }\href {\doibase 10.1103/PhysRevA.81.053808} {\bibfield
  {journal} {\bibinfo  {journal} {Phys. Rev. A}\ }\textbf {\bibinfo
  {volume} {81}},\ \bibinfo {pages} {053808} (\bibinfo {year}
  {2010})}\BibitemShut {NoStop}%
\bibitem [{\citenamefont {Peng}\ \emph {et~al.}(2002)\citenamefont {Peng},
  \citenamefont {Huang}, \citenamefont {Qiu}, \citenamefont {Wang},
  \citenamefont {Hu}, \citenamefont {Jiang},\ and\ \citenamefont
  {Mazzer}}]{Peng2002_APL_80_3063_Symmetry-inducedPerfect}%
  \BibitemOpen
  \bibfield  {author} {\bibinfo {author} {\bibfnamefont {R.~W.}\ \bibnamefont
  {Peng}}, \bibinfo {author} {\bibfnamefont {X.~Q.}\ \bibnamefont {Huang}},
  \bibinfo {author} {\bibfnamefont {F.}~\bibnamefont {Qiu}}, \bibinfo {author}
  {\bibfnamefont {Mu}~\bibnamefont {Wang}}, \bibinfo {author} {\bibfnamefont
  {A.}~\bibnamefont {Hu}}, \bibinfo {author} {\bibfnamefont {S.~S.}\
  \bibnamefont {Jiang}}, \ and\ \bibinfo {author} {\bibfnamefont
  {M.}~\bibnamefont {Mazzer}},\ }\bibfield  {title} {\enquote {\bibinfo {title}
  {Symmetry-induced perfect transmission of light waves in quasiperiodic
  dielectric multilayers},}\ }\href {\doibase 10.1063/1.1468895} {\bibfield
  {journal} {\bibinfo  {journal} {Appl. Phys. Lett.}\ }\textbf {\bibinfo
  {volume} {80}},\ \bibinfo {pages} {3063} (\bibinfo {year}
  {2002})}\BibitemShut {NoStop}%
\bibitem [{\citenamefont {{Hladky-Hennion}}\ \emph {et~al.}(2013)\citenamefont
  {{Hladky-Hennion}}, \citenamefont {Vasseur}, \citenamefont {Degraeve},
  \citenamefont {Granger},\ and\ \citenamefont
  {Billy}}]{Hladky-hennion2013_JAP_113_154901_AcousticWave}%
  \BibitemOpen
  \bibfield  {author} {\bibinfo {author} {\bibfnamefont {A.~C.}\ \bibnamefont
  {{Hladky-Hennion}}}, \bibinfo {author} {\bibfnamefont {J.~O.}\ \bibnamefont
  {Vasseur}}, \bibinfo {author} {\bibfnamefont {S.}~\bibnamefont {Degraeve}},
  \bibinfo {author} {\bibfnamefont {C.}~\bibnamefont {Granger}}, \ and\
  \bibinfo {author} {\bibfnamefont {M.~de}\ \bibnamefont {Billy}},\ }\bibfield
  {title} {\enquote {\bibinfo {title} {Acoustic wave localization in
  one-dimensional Fibonacci phononic structures with mirror symmetry},}\ }\href
  {\doibase 10.1063/1.4801890} {\bibfield  {journal} {\bibinfo  {journal}
  {J. Appl. Phys.}\ }\textbf {\bibinfo {volume} {113}},\ \bibinfo
  {pages} {154901} (\bibinfo {year} {2013})}\BibitemShut {NoStop}%
\bibitem [{\citenamefont {Theocharis}\ \emph {et~al.}(2014)\citenamefont
  {Theocharis}, \citenamefont {Richoux}, \citenamefont {Garc\'{i}a},
  \citenamefont {Merkel},\ and\ \citenamefont
  {Tournat}}]{Theocharis2014_NJP_16_093017_LimitsSlow}%
  \BibitemOpen
  \bibfield  {author} {\bibinfo {author} {\bibfnamefont {G.}~\bibnamefont
  {Theocharis}}, \bibinfo {author} {\bibfnamefont {O.}~\bibnamefont {Richoux}},
  \bibinfo {author} {\bibfnamefont {V.~Romero}\ \bibnamefont {Garc\'{i}a}},
  \bibinfo {author} {\bibfnamefont {A.}~\bibnamefont {Merkel}}, \ and\ \bibinfo
  {author} {\bibfnamefont {V.}~\bibnamefont {Tournat}},\ }\bibfield  {title}
  {\enquote {\bibinfo {title} {Limits of slow sound propagation and
  transparency in lossy, locally resonant periodic structures},}\ }\href
  {\doibase 10.1088/1367-2630/16/9/093017} {\bibfield  {journal} {\bibinfo
  {journal} {New J. Phys.}\ }\textbf {\bibinfo {volume} {16}},\
  \bibinfo {pages} {093017} (\bibinfo {year} {2014})}\BibitemShut {NoStop}%
\bibitem [{\citenamefont {Hsueh}\ \emph {et~al.}(2011)\citenamefont {Hsueh},
  \citenamefont {Wun}, \citenamefont {Lin},\ and\ \citenamefont
  {Cheng}}]{Hsueh2011_JOSAB_28_2584_FeaturesPerfect}%
  \BibitemOpen
  \bibfield  {author} {\bibinfo {author} {\bibfnamefont {W.~J.}\ \bibnamefont
  {Hsueh}}, \bibinfo {author} {\bibfnamefont {S.~J.}\ \bibnamefont {Wun}},
  \bibinfo {author} {\bibfnamefont {Z.~J.}\ \bibnamefont {Lin}}, \ and\
  \bibinfo {author} {\bibfnamefont {Y.~H.}\ \bibnamefont {Cheng}},\ }\bibfield
  {title} {\enquote {\bibinfo {title} {Features of the perfect transmission in
  {Thue-Morse} dielectric multilayers},}\ }\href {\doibase
  10.1364/JOSAB.28.002584} {\bibfield  {journal} {\bibinfo  {journal} {Journal
  of the Optical Society of America B}\ }\textbf {\bibinfo {volume} {28}},\
  \bibinfo {pages} {2584} (\bibinfo {year} {2011})}\BibitemShut {NoStop}%
\bibitem [{\citenamefont {Kalozoumis}\ \emph
  {et~al.}(2013{\natexlab{a}})\citenamefont {Kalozoumis}, \citenamefont
  {Morfonios}, \citenamefont {Diakonos},\ and\ \citenamefont
  {Schmelcher}}]{Kalozoumis2013_PRA_87_032113_LocalSymmetries}%
  \BibitemOpen
  \bibfield  {author} {\bibinfo {author} {\bibfnamefont {P.~A.}\ \bibnamefont
  {Kalozoumis}}, \bibinfo {author} {\bibfnamefont {C.}~\bibnamefont
  {Morfonios}}, \bibinfo {author} {\bibfnamefont {F.~K.}\ \bibnamefont
  {Diakonos}}, \ and\ \bibinfo {author} {\bibfnamefont {P.}~\bibnamefont
  {Schmelcher}},\ }\bibfield  {title} {\enquote {\bibinfo {title} {Local
  symmetries in one-dimensional quantum scattering},}\ }\href {\doibase
  10.1103/PhysRevA.87.032113} {\bibfield  {journal} {\bibinfo  {journal}
  {Phys. Rev. A}\ }\textbf {\bibinfo {volume} {87}},\ \bibinfo {pages}
  {032113} (\bibinfo {year} {2013}{\natexlab{a}})}\BibitemShut {NoStop}%
\bibitem [{\citenamefont {Kalozoumis}\ \emph
  {et~al.}(2014{\natexlab{a}})\citenamefont {Kalozoumis}, \citenamefont
  {Morfonios}, \citenamefont {Diakonos},\ and\ \citenamefont
  {Schmelcher}}]{Kalozoumis2014_PRL_113_050403_InvariantsBroken}%
  \BibitemOpen
  \bibfield  {author} {\bibinfo {author} {\bibfnamefont {P.~A.}\ \bibnamefont
  {Kalozoumis}}, \bibinfo {author} {\bibfnamefont {C.}~\bibnamefont
  {Morfonios}}, \bibinfo {author} {\bibfnamefont {F.~K.}\ \bibnamefont
  {Diakonos}}, \ and\ \bibinfo {author} {\bibfnamefont {P.}~\bibnamefont
  {Schmelcher}},\ }\bibfield  {title} {\enquote {\bibinfo {title} {Invariants
  of broken discrete symmetries},}\ }\href {\doibase
  10.1103/PhysRevLett.113.050403} {\bibfield  {journal} {\bibinfo  {journal}
  {Phys. Rev. Lett.}\ }\textbf {\bibinfo {volume} {113}},\ \bibinfo
  {pages} {050403} (\bibinfo {year} {2014}{\natexlab{a}})}\BibitemShut
  {NoStop}%
\bibitem [{\citenamefont {Zampetakis}\ \emph {et~al.}(2016)\citenamefont
  {Zampetakis}, \citenamefont {Diakonou}, \citenamefont {Morfonios},
  \citenamefont {Kalozoumis}, \citenamefont {Diakonos},\ and\ \citenamefont
  {Schmelcher}}]{Zampetakis2016_JPAMT_49_195304_InvariantCurrent}%
  \BibitemOpen
  \bibfield  {author} {\bibinfo {author} {\bibfnamefont {V.~E.}\ \bibnamefont
  {Zampetakis}}, \bibinfo {author} {\bibfnamefont {M.~K.}\ \bibnamefont
  {Diakonou}}, \bibinfo {author} {\bibfnamefont {C.~V.}\ \bibnamefont
  {Morfonios}}, \bibinfo {author} {\bibfnamefont {P.~A.}\ \bibnamefont
  {Kalozoumis}}, \bibinfo {author} {\bibfnamefont {F.~K.}\ \bibnamefont
  {Diakonos}}, \ and\ \bibinfo {author} {\bibfnamefont {P.}~\bibnamefont
  {Schmelcher}},\ }\bibfield  {title} {\enquote {\bibinfo {title} {Invariant
  current approach to wave propagation in locally symmetric structures},}\
  }\href {\doibase 10.1088/1751-8113/49/19/195304} {\bibfield  {journal}
  {\bibinfo  {journal} {J. Phys. A: Math. Theor.}\
  }\textbf {\bibinfo {volume} {49}},\ \bibinfo {pages} {195304} (\bibinfo
  {year} {2016})}\BibitemShut {NoStop}%
\bibitem [{\citenamefont {Kalozoumis}\ \emph
  {et~al.}(2013{\natexlab{b}})\citenamefont {Kalozoumis}, \citenamefont
  {Morfonios}, \citenamefont {Palaiodimopoulos}, \citenamefont {Diakonos},\
  and\ \citenamefont
  {Schmelcher}}]{Kalozoumis2013_PRA_88_033857_LocalSymmetries}%
  \BibitemOpen
  \bibfield  {author} {\bibinfo {author} {\bibfnamefont {P.~A.}\ \bibnamefont
  {Kalozoumis}}, \bibinfo {author} {\bibfnamefont {C.}~\bibnamefont
  {Morfonios}}, \bibinfo {author} {\bibfnamefont {N.}~\bibnamefont
  {Palaiodimopoulos}}, \bibinfo {author} {\bibfnamefont {F.~K.}\ \bibnamefont
  {Diakonos}}, \ and\ \bibinfo {author} {\bibfnamefont {P.}~\bibnamefont
  {Schmelcher}},\ }\bibfield  {title} {\enquote {\bibinfo {title} {Local
  symmetries and perfect transmission in aperiodic photonic multilayers},}\
  }\href {\doibase 10.1103/PhysRevA.88.033857} {\bibfield  {journal} {\bibinfo
  {journal} {Phys. Rev. A}\ }\textbf {\bibinfo {volume} {88}},\ \bibinfo
  {pages} {033857} (\bibinfo {year} {2013}{\natexlab{b}})}\BibitemShut
  {NoStop}%
\bibitem [{\citenamefont {Kalozoumis}\ \emph
  {et~al.}(2014{\natexlab{b}})\citenamefont {Kalozoumis}, \citenamefont
  {Pappas}, \citenamefont {Diakonos},\ and\ \citenamefont
  {Schmelcher}}]{Kalozoumis2014_PRA_90_043809_SystematicPathway}%
  \BibitemOpen
  \bibfield  {author} {\bibinfo {author} {\bibfnamefont {P.~A.}\ \bibnamefont
  {Kalozoumis}}, \bibinfo {author} {\bibfnamefont {G.}~\bibnamefont {Pappas}},
  \bibinfo {author} {\bibfnamefont {F.~K.}\ \bibnamefont {Diakonos}}, \ and\
  \bibinfo {author} {\bibfnamefont {P.}~\bibnamefont {Schmelcher}},\ }\bibfield
   {title} {\enquote {\bibinfo {title} {Systematic pathway to
  {$\mathcal{PT}$-symmetry} breaking in scattering systems},}\ }\href {\doibase
  10.1103/PhysRevA.90.043809} {\bibfield  {journal} {\bibinfo  {journal}
  {Phys. Rev. A}\ }\textbf {\bibinfo {volume} {90}},\ \bibinfo {pages}
  {043809} (\bibinfo {year} {2014}{\natexlab{b}})}\BibitemShut {NoStop}%
\bibitem [{\citenamefont {Kalozoumis}\ \emph {et~al.}(2016)\citenamefont
  {Kalozoumis}, \citenamefont {Morfonios}, \citenamefont {Diakonos},\ and\
  \citenamefont
  {Schmelcher}}]{Kalozoumis2016_PRA_93_063831_Mathcalpt-symmetryBreaking}%
  \BibitemOpen
  \bibfield  {author} {\bibinfo {author} {\bibfnamefont {P.~A.}\ \bibnamefont
  {Kalozoumis}}, \bibinfo {author} {\bibfnamefont {C.~V.}\ \bibnamefont
  {Morfonios}}, \bibinfo {author} {\bibfnamefont {F.~K.}\ \bibnamefont
  {Diakonos}}, \ and\ \bibinfo {author} {\bibfnamefont {P.}~\bibnamefont
  {Schmelcher}},\ }\bibfield  {title} {\enquote {\bibinfo {title}
  {{$\mathcal{PT}$-symmetry} breaking in waveguides with competing loss-gain
  pairs},}\ }\href {\doibase 10.1103/PhysRevA.93.063831} {\bibfield  {journal}
  {\bibinfo  {journal} {Phys. Rev. A}\ }\textbf {\bibinfo {volume} {93}},\
  \bibinfo {pages} {063831} (\bibinfo {year} {2016})}\BibitemShut {NoStop}%
\bibitem [{\citenamefont {Kalozoumis}\ \emph {et~al.}(2015)\citenamefont
  {Kalozoumis}, \citenamefont {Richoux}, \citenamefont {Diakonos},
  \citenamefont {Theocharis},\ and\ \citenamefont
  {Schmelcher}}]{Kalozoumis2015_PRB_92_014303_InvariantCurrents}%
  \BibitemOpen
  \bibfield  {author} {\bibinfo {author} {\bibfnamefont {P.~A.}\ \bibnamefont
  {Kalozoumis}}, \bibinfo {author} {\bibfnamefont {O.}~\bibnamefont {Richoux}},
  \bibinfo {author} {\bibfnamefont {F.~K.}\ \bibnamefont {Diakonos}}, \bibinfo
  {author} {\bibfnamefont {G.}~\bibnamefont {Theocharis}}, \ and\ \bibinfo
  {author} {\bibfnamefont {P.}~\bibnamefont {Schmelcher}},\ }\bibfield  {title}
  {\enquote {\bibinfo {title} {Invariant currents in lossy acoustic waveguides
  with complete local symmetry},}\ }\href {\doibase 10.1103/PhysRevB.92.014303}
  {\bibfield  {journal} {\bibinfo  {journal} {Phys. Rev. B}\ }\textbf
  {\bibinfo {volume} {92}},\ \bibinfo {pages} {014303} (\bibinfo {year}
  {2015})}\BibitemShut {NoStop}%
\bibitem [{\citenamefont {Bagchi}\ \emph {et~al.}(2001)\citenamefont {Bagchi},
  \citenamefont {Quesne},\ and\ \citenamefont
  {Znojil}}]{Bagchi2001_MPLA_16_2047_GeneralizedContinuity}%
  \BibitemOpen
  \bibfield  {author} {\bibinfo {author} {\bibfnamefont {B.}\ \bibnamefont
  {Bagchi}}, \bibinfo {author} {\bibfnamefont {C.}\ \bibnamefont
  {Quesne}}, \ and\ \bibinfo {author} {\bibfnamefont {M.}~\bibnamefont
  {Znojil}},\ }\bibfield  {title} {\enquote {\bibinfo {title} {Generalized
  continuity equation and modified normalization in {{$\mathcal{PT}$}-symmetric} quantum
  mechanics},}\ }\href
  {http://www.worldscientific.com/doi/abs/10.1142/S0217732301005333} {\bibfield
   {journal} {\bibinfo  {journal} {Mod. Phys. Lett. A}\ }\textbf
  {\bibinfo {volume} {16}},\ \bibinfo {pages} {2047} (\bibinfo
  {year} {2001})}\BibitemShut {NoStop}%
\bibitem [{\citenamefont
  {Japaridze}(2002)}]{Japaridze2002_JPAMG_35_1709_SpaceState}%
  \BibitemOpen
  \bibfield  {author} {\bibinfo {author} {\bibfnamefont {G.~S.}\ \bibnamefont
  {Japaridze}},\ }\bibfield  {title} {\enquote {\bibinfo {title} {Space of
  state vectors in {{$\mathcal{PT}$}-symmetric} quantum mechanics},}\ }\href {\doibase
  10.1088/0305-4470/35/7/315} {\bibfield  {journal} {\bibinfo  {journal}
  {J. Phys. A: Math. Gen.}\ }\textbf {\bibinfo {volume}
  {35}},\ \bibinfo {pages} {1709} (\bibinfo {year} {2002})}\BibitemShut
  {NoStop}%
\bibitem [{\citenamefont {Jr}\ and\ \citenamefont
  {Narducci}(2015)}]{Jr2015_JPAMT_48_155304_BilocalPicture}%
  \BibitemOpen
  A related bilocal current has also been used in an alternative description of double slit interference, see
  \bibfield  {author} {\bibinfo {author} {\bibfnamefont {L.~P.}\
  \bibnamefont {Withers Jr.}}\ and\ \bibinfo {author} {\bibfnamefont {F.~A.}\
  \bibnamefont {Narducci}},\ }\bibfield  {title} {\enquote {\bibinfo {title} {A
  bilocal picture of quantum mechanics},}\ }\href {\doibase
  10.1088/1751-8113/48/15/155304} {\bibfield  {journal} {\bibinfo  {journal}
  {J. Phys. A: Math. Theor.}\ }\textbf {\bibinfo
  {volume} {48}},\ \bibinfo {pages} {155304} (\bibinfo {year}
  {2015})}\BibitemShut {NoStop}%
\bibitem [{\citenamefont
  {Schomerus}(2013)}]{Schomerus2013_PTRSL_371_20120194_ScatteringTheory}%
  \BibitemOpen
  \bibfield  {author} {\bibinfo {author} {\bibfnamefont {H.}\ \bibnamefont
  {Schomerus}},\ }\bibfield  {title} {\enquote {\bibinfo {title} {From
  scattering theory to complex wave dynamics in {non-Hermitian} {{$\mathcal{PT}$}-symmetric}
  resonators},}\ }\href {\doibase 10.1098/rsta.2012.0194} {\bibfield  {journal}
  {\bibinfo  {journal} {Phil. Trans. R. Soc. A}\ }\textbf
  {\bibinfo {volume} {371}},\ \bibinfo {pages} {20120194} (\bibinfo {year}
  {2013})}\BibitemShut {NoStop}%
\bibitem [{\citenamefont {Baranger}\ \emph {et~al.}(1991)\citenamefont
  {Baranger}, \citenamefont {{DiVincenzo}}, \citenamefont {Jalabert},\ and\
  \citenamefont {Stone}}]{Baranger1991_PRB_44_10637_ClassicalQuantum}%
  \BibitemOpen
  \bibfield  {author} {\bibinfo {author} {\bibfnamefont {H.~U.}\
  \bibnamefont {Baranger}}, \bibinfo {author} {\bibfnamefont {D.~P.}\
  \bibnamefont {{DiVincenzo}}}, \bibinfo {author} {\bibfnamefont {R.~A.}\
  \bibnamefont {Jalabert}}, \ and\ \bibinfo {author} {\bibfnamefont
  {A.~D.}\ \bibnamefont {Stone}},\ }\bibfield  {title} {\enquote {\bibinfo
  {title} {Classical and quantum ballistic-transport anomalies in
  microjunctions},}\ }\href {\doibase 10.1103/PhysRevB.44.10637} {\bibfield
  {journal} {\bibinfo  {journal} {Phys. Rev. B}\ }\textbf {\bibinfo
  {volume} {44}},\ \bibinfo {pages} {10637} (\bibinfo {year}
  {1991})}\BibitemShut {NoStop}%
\bibitem [{\citenamefont {Boykin}\ \emph {et~al.}(2010)\citenamefont {Boykin},
  \citenamefont {Luisier},\ and\ \citenamefont
  {Klimeck}}]{Boykin2010_EJP_31_1077_CurrentDensity}%
  \BibitemOpen
  \bibfield  {author} {\bibinfo {author} {\bibfnamefont {T.~B.}\
  \bibnamefont {Boykin}}, \bibinfo {author} {\bibfnamefont {M.}\
  \bibnamefont {Luisier}}, \ and\ \bibinfo {author} {\bibfnamefont {G.}\
  \bibnamefont {Klimeck}},\ }\bibfield  {title} {\enquote {\bibinfo {title}
  {Current density and continuity in discretized models},}\ }\href {\doibase
  10.1088/0143-0807/31/5/010} {\bibfield  {journal} {\bibinfo  {journal}
  {Eur. J. Phys.}\ }\textbf {\bibinfo {volume} {31}},\ \bibinfo
  {pages} {1077} (\bibinfo {year} {2010})}\BibitemShut
  {NoStop}%
\bibitem [{\citenamefont {Sinha}\ and\ \citenamefont
  {Ghosh}(2015)}]{Sinha2015_PRE_91_042908_SymmetriesExact}%
  \BibitemOpen
  \bibfield  {author} {\bibinfo {author} {\bibfnamefont {D.}\ \bibnamefont
  {Sinha}}\ and\ \bibinfo {author} {\bibfnamefont {P.~K.}\ \bibnamefont
  {Ghosh}},\ }\bibfield  {title} {\enquote {\bibinfo {title} {Symmetries and
  exact solutions of a class of nonlocal nonlinear Schr\"odinger equations with
  self-induced parity-time-symmetric potential},}\ }\href {\doibase
  10.1103/PhysRevE.91.042908} {\bibfield  {journal} {\bibinfo  {journal}
  {Phys. Rev. E}\ }\textbf {\bibinfo {volume} {91}},\ \bibinfo {pages}
  {042908} (\bibinfo {year} {2015})}\BibitemShut {NoStop}%
\bibitem [{\citenamefont {Bender}\ and\ \citenamefont
  {Boettcher}(1998)}]{Bender1998_PRL_80_5243_RealSpectra}% 
  \BibitemOpen
  \bibfield  {author} {\bibinfo {author} {\bibfnamefont {C.~M.}\ \bibnamefont
  {Bender}}\ and\ \bibinfo {author} {\bibfnamefont {S.}\ \bibnamefont
  {Boettcher}},\ }\bibfield  {title} {\enquote {\bibinfo {title} {Real spectra
  in {Non-Hermitian} Hamiltonians having {{$\mathcal{PT}$}} symmetry},}\ }\href {\doibase
  10.1103/PhysRevLett.80.5243} {\bibfield  {journal} {\bibinfo  {journal}
  {Phys. Rev. Lett.}\ }\textbf {\bibinfo {volume} {80}},\ \bibinfo
  {pages} {5243} (\bibinfo {year} {1998})}\BibitemShut {NoStop}%
\bibitem [{\citenamefont {Bender}(2007)}]{Bender2007_RPP_70_947_MakingSense}%
  \BibitemOpen
  \bibfield  {author} {\bibinfo {author} {\bibfnamefont {Carl~M.}\ \bibnamefont
  {Bender}},\ }\bibfield  {title} {\enquote {\bibinfo {title} {Making sense of
  {non-Hermitian} hamiltonians},}\ }\href {\doibase 10.1088/0034-4885/70/6/R03}
  {\bibfield  {journal} {\bibinfo  {journal} {Reports on Progress in Physics}\
  }\textbf {\bibinfo {volume} {70}},\ \bibinfo {pages} {947} (\bibinfo {year}
  {2007})}\BibitemShut {NoStop}%
\bibitem [{\citenamefont
  {Weigert}(2004)}]{Weigert2004_CJP_54_1139_PhysicalInterpretation}%
  \BibitemOpen
  \bibfield  {author} {\bibinfo {author} {\bibfnamefont {S.}\ \bibnamefont
  {Weigert}},\ }\bibfield  {title} {\enquote {\bibinfo {title} {The physical
  interpretation of {{$\mathcal{PT}$}}}-invariant Potentials,}\ }\href {\doibase
  10.1023/B:CJOP.0000044016.95629.a7} {\bibfield  {journal} {\bibinfo
  {journal} {Czech. J. Phys.}\ }\textbf {\bibinfo {volume}
  {54}},\ \bibinfo {pages} {1139} (\bibinfo {year} {2004})}\BibitemShut
  {NoStop}%
\bibitem [{\citenamefont {Cannata}\ \emph {et~al.}(2007)\citenamefont
  {Cannata}, \citenamefont {Dedonder},\ and\ \citenamefont
  {Ventura}}]{Cannata2007_AP_322_397_Scattering-symmetric}%
  \BibitemOpen
  \bibfield  {author} {\bibinfo {author} {\bibfnamefont {F.}\
  \bibnamefont {Cannata}}, \bibinfo {author} {\bibfnamefont {{J.-P.}}\
  \bibnamefont {Dedonder}}, \ and\ \bibinfo {author} {\bibfnamefont {A.}\
  \bibnamefont {Ventura}},\ }\bibfield  {title} {\enquote {\bibinfo {title}
  {Scattering in -symmetric quantum mechanics},}\ }\href {\doibase
  10.1016/j.aop.2006.05.011} {\bibfield  {journal} {\bibinfo  {journal} {Ann. 
  Phys.}\ }\textbf {\bibinfo {volume} {322}},\ \bibinfo {pages} {397}
  (\bibinfo {year} {2007})}\BibitemShut {NoStop}%
\bibitem [{\citenamefont {Chong}\ \emph {et~al.}(2011)\citenamefont {Chong},
  \citenamefont {Ge},\ and\ \citenamefont
  {Stone}}]{Chong2011_PRL_106_093902_Mathcalpmathcalt-symBreaking}%
  \BibitemOpen
  \bibfield  {author} {\bibinfo {author} {\bibfnamefont {Y.~D.}\ \bibnamefont
  {Chong}}, \bibinfo {author} {\bibfnamefont {L.}~\bibnamefont {Ge}}, \ and\
  \bibinfo {author} {\bibfnamefont {A.~D.}\ \bibnamefont {Stone}},\
  }\bibfield  {title} {\enquote {\bibinfo {title}
  {{$\mathcal{P}\mathcal{T}$-Symmetry} breaking and {Laser-Absorber} modes in
  optical scattering systems},}\ }\href {\doibase
  10.1103/PhysRevLett.106.093902} {\bibfield  {journal} {\bibinfo  {journal}
  {Phys. Rev. Lett.}\ }\textbf {\bibinfo {volume} {106}},\ \bibinfo
  {pages} {093902} (\bibinfo {year} {2011})}\BibitemShut {NoStop}%
\bibitem [{\citenamefont {Ambichl}\ \emph {et~al.}(2013)\citenamefont
  {Ambichl}, \citenamefont {Makris}, \citenamefont {Ge}, \citenamefont {Chong},
  \citenamefont {Stone},\ and\ \citenamefont
  {Rotter}}]{Ambichl2013_PRX_3_041030_BreakingPt}%
  \BibitemOpen
  \bibfield  {author} {\bibinfo {author} {\bibfnamefont {P.}\ \bibnamefont
  {Ambichl}}, \bibinfo {author} {\bibfnamefont {K.~G.}\ \bibnamefont
  {Makris}}, \bibinfo {author} {\bibfnamefont {L.}~\bibnamefont {Ge}}, \bibinfo
  {author} {\bibfnamefont {Y.}\ \bibnamefont {Chong}}, \bibinfo {author}
  {\bibfnamefont {A.~D.}\ \bibnamefont {Stone}}, \ and\ \bibinfo {author}
  {\bibfnamefont {S.}\ \bibnamefont {Rotter}},\ }\bibfield  {title}
  {\enquote {\bibinfo {title} {Breaking of {{$\mathcal{PT}$}} symmetry in bounded and
  unbounded scattering systems},}\ }\href {\doibase 10.1103/PhysRevX.3.041030}
  {\bibfield  {journal} {\bibinfo  {journal} {Phys. Rev. X}\ }\textbf
  {\bibinfo {volume} {3}},\ \bibinfo {pages} {041030} (\bibinfo {year}
  {2013})}\BibitemShut {NoStop}%
\bibitem [{\citenamefont {Garmon}\ \emph {et~al.}(2015)\citenamefont {Garmon},
  \citenamefont {Gianfreda},\ and\ \citenamefont
  {Hatano}}]{Garmon2015_PRA_92_022125_BoundStates}%
  \BibitemOpen
  \bibfield  {author} {\bibinfo {author} {\bibfnamefont {S.}\
  \bibnamefont {Garmon}}, \bibinfo {author} {\bibfnamefont {M.}\
  \bibnamefont {Gianfreda}}, \ and\ \bibinfo {author} {\bibfnamefont
  {N.}\ \bibnamefont {Hatano}},\ }\bibfield  {title} {\enquote {\bibinfo
  {title} {Bound states, scattering states, and resonant states in
  {$\mathcal{PT}$-symmetric} open quantum systems},}\ }\href {\doibase
  10.1103/PhysRevA.92.022125} {\bibfield  {journal} {\bibinfo  {journal}
  {Phys. Rev. A}\ }\textbf {\bibinfo {volume} {92}},\ \bibinfo {pages}
  {022125} (\bibinfo {year} {2015})}\BibitemShut {NoStop}%
\bibitem [{\citenamefont
  {Longhi}(2016)}]{Longhi2016_PRA_93_022102_Non-HermitianTight-binding}%
  \BibitemOpen
  \bibfield  {author} {\bibinfo {author} {\bibfnamefont {S.}\ \bibnamefont
  {Longhi}},\ }\bibfield  {title} {\enquote {\bibinfo {title} {{Non-Hermitian}
  tight-binding network engineering},}\ }\href {\doibase
  10.1103/PhysRevA.93.022102} {\bibfield  {journal} {\bibinfo  {journal}
  {Phys. Rev. A}\ }\textbf {\bibinfo {volume} {93}},\ \bibinfo {pages}
  {022102} (\bibinfo {year} {2016})}\BibitemShut {NoStop}%
\bibitem [{\citenamefont {Makris}\ \emph {et~al.}(2016)\citenamefont {Makris},
  \citenamefont {Musslimani}, \citenamefont {Christodoulides},\ and\
  \citenamefont {Rotter}}]{Makris2016_IJSTQ_22_42_ConstantIntensity}%
  \BibitemOpen
  \bibfield  {author} {\bibinfo {author} {\bibfnamefont {K.~G.}\ \bibnamefont
  {Makris}}, \bibinfo {author} {\bibfnamefont {Z.~H.}\ \bibnamefont
  {Musslimani}}, \bibinfo {author} {\bibfnamefont {D.~N.}\ \bibnamefont
  {Christodoulides}}, \ and\ \bibinfo {author} {\bibfnamefont {S.}~\bibnamefont
  {Rotter}},\ }\bibfield  {title} {\enquote {\bibinfo {title} {Constant
  intensity supermodes in {Non-Hermitian} lattices},}\ }\href {\doibase
  10.1109/JSTQE.2016.2593866} {\bibfield  {journal} {\bibinfo  {journal}
  {{IEEE} J. Sel. Top. Quantum Electron.}\ }\textbf
  {\bibinfo {volume} {22}},\ \bibinfo {pages} {42} (\bibinfo {year}
  {2016})}\BibitemShut {NoStop}%
\bibitem [{\citenamefont {Eisenberg}\ \emph {et~al.}(2000)\citenamefont
  {Eisenberg}, \citenamefont {Silberberg}, \citenamefont {Morandotti},\ and\
  \citenamefont {Aitchison}}]{Eisenberg2000_PRL_85_1863_DiffractionManagement}%
  \BibitemOpen
  \bibfield  {author} {\bibinfo {author} {\bibfnamefont {H.~S.}\ \bibnamefont
  {Eisenberg}}, \bibinfo {author} {\bibfnamefont {Y.}~\bibnamefont
  {Silberberg}}, \bibinfo {author} {\bibfnamefont {R.}~\bibnamefont
  {Morandotti}}, \ and\ \bibinfo {author} {\bibfnamefont {J.~S.}\ \bibnamefont
  {Aitchison}},\ }\bibfield  {title} {\enquote {\bibinfo {title} {Diffraction
  management},}\ }\href {\doibase 10.1103/PhysRevLett.85.1863} {\bibfield
  {journal} {\bibinfo  {journal} {Phys. Rev. Lett.}\ }\textbf {\bibinfo
  {volume} {85}},\ \bibinfo {pages} {1863} (\bibinfo {year}
  {2000})}\BibitemShut {NoStop}%
\bibitem [{\citenamefont {Teimourpour}\ \emph {et~al.}(2014)\citenamefont
  {Teimourpour}, \citenamefont {{El-Ganainy}}, \citenamefont {Eisfeld},
  \citenamefont {Szameit},\ and\ \citenamefont
  {Christodoulides}}]{Teimourpour2014_PRA_90_053817_LightTransport}% 
  \BibitemOpen
  \bibfield  {author} {\bibinfo {author} {\bibfnamefont {M.~H.}\ \bibnamefont
  {Teimourpour}}, \bibinfo {author} {\bibfnamefont {R.}~\bibnamefont
  {{El-Ganainy}}}, \bibinfo {author} {\bibfnamefont {A.}~\bibnamefont
  {Eisfeld}}, \bibinfo {author} {\bibfnamefont {A.}~\bibnamefont {Szameit}}, \
  and\ \bibinfo {author} {\bibfnamefont {D.~N.}\ \bibnamefont
  {Christodoulides}},\ }\bibfield  {title} {\enquote {\bibinfo {title} {Light
  transport in {$\mathcal{PT}$-invariant} photonic structures with hidden
  symmetries},}\ }\href {\doibase 10.1103/PhysRevA.90.053817} {\bibfield
  {journal} {\bibinfo  {journal} {Phys. Rev. A}\ }\textbf {\bibinfo
  {volume} {90}},\ \bibinfo {pages} {053817} (\bibinfo {year}
  {2014})}\BibitemShut {NoStop}%
\bibitem [{\citenamefont {Beygi}\ \emph {et~al.}(2015)\citenamefont {Beygi},
  \citenamefont {Klevansky},\ and\ \citenamefont
  {Bender}}]{Beygi2015_PRA_91_062101_CoupledOscillator}%
  \BibitemOpen
  \bibfield  {author} {\bibinfo {author} {\bibfnamefont {A.}\ \bibnamefont
  {Beygi}}, \bibinfo {author} {\bibfnamefont {S.~P.}\ \bibnamefont
  {Klevansky}}, \ and\ \bibinfo {author} {\bibfnamefont {C.~M.}\ \bibnamefont
  {Bender}},\ }\bibfield  {title} {\enquote {\bibinfo {title} {Coupled
  oscillator systems having partial {$\mathcal{PT}$} symmetry},}\ }\href
  {\doibase 10.1103/PhysRevA.91.062101} {\bibfield  {journal} {\bibinfo
  {journal} {Phys. Rev. A}\ }\textbf {\bibinfo {volume} {91}},\ \bibinfo
  {pages} {062101} (\bibinfo {year} {2015})}\BibitemShut {NoStop}%
\bibitem [{\citenamefont {Bender}\ \emph {et~al.}(2015)\citenamefont {Bender},
  \citenamefont {Li}, \citenamefont {Ellis},\ and\ \citenamefont
  {Kottos}}]{Bender2015_PRA_92_041803_Wave-packetSelf-imaging}%
  \BibitemOpen
  \bibfield  {author} {\bibinfo {author} {\bibfnamefont {N.}~\bibnamefont
  {Bender}}, \bibinfo {author} {\bibfnamefont {H.}~\bibnamefont {Li}}, \bibinfo
  {author} {\bibfnamefont {F.~M.}\ \bibnamefont {Ellis}}, \ and\ \bibinfo
  {author} {\bibfnamefont {T.}~\bibnamefont {Kottos}},\ }\bibfield  {title}
  {\enquote {\bibinfo {title} {Wave-packet self-imaging and giant
  recombinations via stable {Bloch-Zener} oscillations in photonic lattices
  with local {$\mathcal{P}\mathcal{T}$} symmetry},}\ }\href {\doibase
  10.1103/PhysRevA.92.041803} {\bibfield  {journal} {\bibinfo  {journal}
  {Phys. Rev. A}\ }\textbf {\bibinfo {volume} {92}},\ \bibinfo {pages}
  {041803} (\bibinfo {year} {2015})}\BibitemShut {NoStop}%
\bibitem [{\citenamefont {Wimmer}\ \emph {et~al.}(2015)\citenamefont {Wimmer},
  \citenamefont {Regensburger}, \citenamefont {Miri}, \citenamefont {Bersch},
  \citenamefont {Christodoulides},\ and\ \citenamefont
  {Peschel}}]{Wimmer2015_NC_6_7782_ObservationOptical}%
  \BibitemOpen
  \bibfield  {author} {\bibinfo {author} {\bibfnamefont {M.}\ \bibnamefont
  {Wimmer}}, \bibinfo {author} {\bibfnamefont {A.}\ \bibnamefont
  {Regensburger}}, \bibinfo {author} {\bibfnamefont {{M.-A.}}\
  \bibnamefont {Miri}}, \bibinfo {author} {\bibfnamefont {C.}\
  \bibnamefont {Bersch}}, \bibinfo {author} {\bibfnamefont {D.~N.}\
  \bibnamefont {Christodoulides}}, \ and\ \bibinfo {author} {\bibfnamefont
  {U.}\ \bibnamefont {Peschel}},\ }\bibfield  {title} {\enquote {\bibinfo
  {title} {Observation of optical solitons in {{$\mathcal{PT}$}-symmetric} lattices},}\
  }\href {\doibase 10.1038/ncomms8782} {\bibfield  {journal} {\bibinfo
  {journal} {Nat. Comm.}\ }\textbf {\bibinfo {volume} {6}},\
  \bibinfo {pages} {7782} (\bibinfo {year} {2015})}\BibitemShut {NoStop}%
\bibitem [{\citenamefont {Szameit}\ \emph {et~al.}(2007)\citenamefont
  {Szameit}, \citenamefont {Dreisow}, \citenamefont {Pertsch}, \citenamefont
  {Nolte},\ and\ \citenamefont
  {T\"{u}nnermann}}]{Szameit2007_OE_15_1579_ControlDirectional}%
  \BibitemOpen
  \bibfield  {author} {\bibinfo {author} {\bibfnamefont {A.}\
  \bibnamefont {Szameit}}, \bibinfo {author} {\bibfnamefont {F.}\
  \bibnamefont {Dreisow}}, \bibinfo {author} {\bibfnamefont {T.}\
  \bibnamefont {Pertsch}}, \bibinfo {author} {\bibfnamefont {S.}\
  \bibnamefont {Nolte}}, \ and\ \bibinfo {author} {\bibfnamefont {A.}\
  \bibnamefont {T\"{u}nnermann}},\ }\bibfield  {title} {\enquote {\bibinfo
  {title} {Control of directional evanescent coupling in fs laser written
  waveguides},}\ }\href {\doibase 10.1364/OE.15.001579} {\bibfield  {journal}
  {\bibinfo  {journal} {Opt. Express}\ }\textbf {\bibinfo {volume} {15}},\
  \bibinfo {pages} {1579} (\bibinfo {year} {2007})}\BibitemShut {NoStop}%
\bibitem {note}%
  \BibitemOpen
  The discrete parameters $\ax$ and $L$ here refer to the site indexing and not to the continuous coordinate along the array:
  The center $\ax$ is integer (half integer) for odd (even) number of sites $D$ in $\DD$, and the integer period $L$ is the shift in site number to the right. 
  Intersite distances in the physical system are mapped to corresponding hopping element magnitudes in the model, determined by evanescent field overlap between waveguides in photonics \cite{Szameit2007_OE_15_1579_ControlDirectional}
  \BibitemShut {NoStop}%
\bibitem [{\citenamefont {Sarma}\ \emph {et~al.}(2014)\citenamefont {Sarma},
  \citenamefont {Miri}, \citenamefont {Musslimani},\ and\ \citenamefont
  {Christodoulides}}]{Sarma2014_PRE_89_052918_ContinuousDiscrete}%
  \BibitemOpen
  \bibfield  {author} {\bibinfo {author} {\bibfnamefont {A.~K.}\
  \bibnamefont {Sarma}}, \bibinfo {author} {\bibfnamefont {{M.-A.}}\
  \bibnamefont {Miri}}, \bibinfo {author} {\bibfnamefont {Z.~H.}\
  \bibnamefont {Musslimani}}, \ and\ \bibinfo {author} {\bibfnamefont
  {D.~N.}\ \bibnamefont {Christodoulides}},\ }\bibfield  {title}
  {\enquote {\bibinfo {title} {Continuous and discrete Schr\"odinger systems
  with parity-time-symmetric nonlinearities},}\ }\href {\doibase
  10.1103/PhysRevE.89.052918} {\bibfield  {journal} {\bibinfo  {journal}
  {Phys. Rev. E}\ }\textbf {\bibinfo {volume} {89}},\ \bibinfo {pages}
  {052918} (\bibinfo {year} {2014})}\BibitemShut {NoStop}%
\bibitem [{\citenamefont
  {Moiseyev}(2011)}]{Moiseyev2011____Non-hermitianQuantum}%
  \BibitemOpen
  \bibfield  {author} {\bibinfo {author} {\bibfnamefont {N.}\ \bibnamefont
  {Moiseyev}},\ }\href@noop {} {\emph {\bibinfo {title} {{Non-Hermitian}
  Quantum Mechanics}}}\ (\bibinfo  {publisher} {Cambridge University Press},\
  \bibinfo {year} {2011})\BibitemShut {NoStop}%
\bibitem [{\citenamefont {Golshani}\ \emph {et~al.}(2014)\citenamefont
  {Golshani}, \citenamefont {Weimann}, \citenamefont {Jafari}, \citenamefont
  {Nezhad}, \citenamefont {Langari}, \citenamefont {Bahrampour}, \citenamefont
  {Eichelkraut}, \citenamefont {Mahdavi},\ and\ \citenamefont
  {Szameit}}]{Golshani2014_PRL_113_123903_ImpactLoss}%
  \BibitemOpen
  \bibfield  {author} {\bibinfo {author} {\bibfnamefont {M.}~\bibnamefont
  {Golshani}}, \bibinfo {author} {\bibfnamefont {S.}~\bibnamefont {Weimann}},
  \bibinfo {author} {\bibfnamefont {Kh.}\ \bibnamefont {Jafari}}, \bibinfo
  {author} {\bibfnamefont {M.~K.}\ \bibnamefont {Nezhad}}, \bibinfo
  {author} {\bibfnamefont {A.}~\bibnamefont {Langari}}, \bibinfo {author}
  {\bibfnamefont {A.~R.}\ \bibnamefont {Bahrampour}}, \bibinfo {author}
  {\bibfnamefont {T.}~\bibnamefont {Eichelkraut}}, \bibinfo {author}
  {\bibfnamefont {S.~M.}\ \bibnamefont {Mahdavi}}, \ and\ \bibinfo {author}
  {\bibfnamefont {A.}~\bibnamefont {Szameit}},\ }\bibfield  {title} {\enquote
  {\bibinfo {title} {Impact of loss on the wave dynamics in photonic waveguide
  lattices},}\ }\href {\doibase 10.1103/PhysRevLett.113.123903} {\bibfield
  {journal} {\bibinfo  {journal} {Phys. Rev. Lett.}\ }\textbf {\bibinfo
  {volume} {113}},\ \bibinfo {pages} {123903} (\bibinfo {year}
  {2014})}\BibitemShut {NoStop}%
\bibitem [{\citenamefont {Klaiman}\ and\ \citenamefont
  {Cederbaum}(2008)}]{Klaiman2008_PRA_78_062113_Non-HermitianHamiltonians}%
  \BibitemOpen
  \bibfield  {author} {\bibinfo {author} {\bibfnamefont {S.}\ \bibnamefont
  {Klaiman}}\ and\ \bibinfo {author} {\bibfnamefont {L.~S.}\ \bibnamefont
  {Cederbaum}},\ }\bibfield  {title} {\enquote {\bibinfo {title}
  {{Non-Hermitian} Hamiltonians with space-time symmetry},}\ }\href {\doibase
  10.1103/PhysRevA.78.062113} {\bibfield  {journal} {\bibinfo  {journal}
  {Phys. Rev. A}\ }\textbf {\bibinfo {volume} {78}},\ \bibinfo {pages}
  {062113} (\bibinfo {year} {2008})}\BibitemShut {NoStop}%
\bibitem [{\citenamefont {Zhu}\ \emph {et~al.}(2014)\citenamefont {Zhu},
  \citenamefont {L\"{u}},\ and\ \citenamefont
  {Chen}}]{Zhu2014_PRA_89_062102_MathcalptSymmetry}%
  \BibitemOpen
  \bibfield  {author} {\bibinfo {author} {\bibfnamefont {B.}\ \bibnamefont
  {Zhu}}, \bibinfo {author} {\bibfnamefont {R.}\ \bibnamefont {L\"{u}}}, \
  and\ \bibinfo {author} {\bibfnamefont {S.}\ \bibnamefont {Chen}},\
  }\bibfield  {title} {\enquote {\bibinfo {title} {{$\mathcal{PT}$} symmetry in
  the {non-Hermitian} {Su-Schrieffer-Heeger} model with complex boundary
  potentials},}\ }\href {\doibase 10.1103/PhysRevA.89.062102} {\bibfield
  {journal} {\bibinfo  {journal} {Phys. Rev. A}\ }\textbf {\bibinfo
  {volume} {89}},\ \bibinfo {pages} {062102} (\bibinfo {year}
  {2014})}\BibitemShut {NoStop}%
\bibitem [{\citenamefont {Harter}\ \emph {et~al.}(2016)\citenamefont {Harter},
  \citenamefont {Lee},\ and\ \citenamefont
  {Joglekar}}]{Harter2016_PRA_93_062101_Mathcalpt-breakingThreshold}%
  \BibitemOpen
  \bibfield  {author} {\bibinfo {author} {\bibfnamefont {A.~K.}\
  \bibnamefont {Harter}}, \bibinfo {author} {\bibfnamefont {T.~E.}\
  \bibnamefont {Lee}}, \ and\ \bibinfo {author} {\bibfnamefont {Y.~N.}\
  \bibnamefont {Joglekar}},\ }\bibfield  {title} {\enquote {\bibinfo {title}
  {{$\mathcal{PT}$-breaking} threshold in spatially asymmetric {Aubry-Andr\'e}
  and Harper models: Hidden symmetry and topological states},}\ }\href
  {\doibase 10.1103/PhysRevA.93.062101} {\bibfield  {journal} {\bibinfo
  {journal} {Phys. Rev. A}\ }\textbf {\bibinfo {volume} {93}},\ \bibinfo
  {pages} {062101} (\bibinfo {year} {2016})}\BibitemShut {NoStop}%
\bibitem [{\citenamefont {Wulf}\ \emph {et~al.}(2016)\citenamefont {Wulf},
  \citenamefont {Morfonios}, \citenamefont {Diakonos},\ and\ \citenamefont
  {Schmelcher}}]{Wulf2016_PRE_93_052215_ExposingLocal}%
  \BibitemOpen
  \bibfield  {author} {\bibinfo {author} {\bibfnamefont {T.}~\bibnamefont
  {Wulf}}, \bibinfo {author} {\bibfnamefont {C.~V.}\ \bibnamefont {Morfonios}},
  \bibinfo {author} {\bibfnamefont {F.~K.}\ \bibnamefont {Diakonos}}, \ and\
  \bibinfo {author} {\bibfnamefont {P.}~\bibnamefont {Schmelcher}},\ }\bibfield
   {title} {\enquote {\bibinfo {title} {Exposing local symmetries in distorted
  driven lattices via time-averaged invariants},}\ }\href {\doibase
  10.1103/PhysRevE.93.052215} {\bibfield  {journal} {\bibinfo  {journal}
  {Phys. Rev. E}\ }\textbf {\bibinfo {volume} {93}},\ \bibinfo {pages}
  {052215} (\bibinfo {year} {2016})}\BibitemShut {NoStop}%
\bibitem [{\citenamefont {Szameit}\ \emph {et~al.}(2008)\citenamefont
  {Szameit}, \citenamefont {Garanovich}, \citenamefont {Heinrich},
  \citenamefont {Sukhorukov}, \citenamefont {Dreisow}, \citenamefont {Pertsch},
  \citenamefont {Nolte}, \citenamefont {T\"{u}nnermann},\ and\ \citenamefont
  {Kivshar}}]{Szameit2008_PRL_101_203902_ObservationDefect-free}%
  \BibitemOpen
  \bibfield  {author} {\bibinfo {author} {\bibfnamefont {A.}\
  \bibnamefont {Szameit}}, \bibinfo {author} {\bibfnamefont {I.~L.}\
  \bibnamefont {Garanovich}}, \bibinfo {author} {\bibfnamefont {M.}\
  \bibnamefont {Heinrich}}, \bibinfo {author} {\bibfnamefont {A.~A.}\
  \bibnamefont {Sukhorukov}}, \bibinfo {author} {\bibfnamefont {F.}\
  \bibnamefont {Dreisow}}, \bibinfo {author} {\bibfnamefont {T.}\
  \bibnamefont {Pertsch}}, \bibinfo {author} {\bibfnamefont {S.}\
  \bibnamefont {Nolte}}, \bibinfo {author} {\bibfnamefont {A.}\
  \bibnamefont {T\"{u}nnermann}}, \ and\ \bibinfo {author} {\bibfnamefont
  {Y.~S.}\ \bibnamefont {Kivshar}},\ }\bibfield  {title} {\enquote {\bibinfo
  {title} {Observation of {Defect-Free} surface modes in optical waveguide
  arrays},}\ }\href {\doibase 10.1103/PhysRevLett.101.203902} {\bibfield
  {journal} {\bibinfo  {journal} {Phys. Rev. Lett.}\ }\textbf {\bibinfo
  {volume} {101}},\ \bibinfo {pages} {203902} (\bibinfo {year}
  {2008})}\BibitemShut {NoStop}%
\bibitem [{\citenamefont {Garanovich}\ \emph {et~al.}(2012)\citenamefont
  {Garanovich}, \citenamefont {Longhi}, \citenamefont {Sukhorukov},\ and\
  \citenamefont {Kivshar}}]{Garanovich2012_PR_518_1_LightPropagation}%
  \BibitemOpen
  \bibfield  {author} {\bibinfo {author} {\bibfnamefont {I.~L.}\ \bibnamefont
  {Garanovich}}, \bibinfo {author} {\bibfnamefont {S.}\ \bibnamefont
  {Longhi}}, \bibinfo {author} {\bibfnamefont {A.~A.}\ \bibnamefont
  {Sukhorukov}}, \ and\ \bibinfo {author} {\bibfnamefont {Y.~S.}\
  \bibnamefont {Kivshar}},\ }\bibfield  {title} {\enquote {\bibinfo {title}
  {Light propagation and localization in modulated photonic lattices and
  waveguides},}\ }\href {\doibase 10.1016/j.physrep.2012.03.005} {\bibfield
  {journal} {\bibinfo  {journal} {Phys. Rep.}\ } \
  \textbf {\bibinfo {volume} {518}},\ \bibinfo {pages} {1} (\bibinfo {year}
  {2012})}\BibitemShut {NoStop}%
\bibitem [{\citenamefont
  {Longhi}(2006)}]{Longhi2006_PRB_73_193305_TransmissionLocalization}%
  \BibitemOpen
  \bibfield  {author} {\bibinfo {author} {\bibfnamefont {S.}\ \bibnamefont
  {Longhi}},\ }\bibfield  {title} {\enquote {\bibinfo {title} {Transmission and
  localization control by ac fields in tight-binding lattices with an
  impurity},}\ }\href {\doibase 10.1103/PhysRevB.73.193305} {\bibfield
  {journal} {\bibinfo  {journal} {Phys. Rev. B}\ }\textbf {\bibinfo
  {volume} {73}},\ \bibinfo {pages} {193305} (\bibinfo {year}
  {2006})}\BibitemShut {NoStop}%
\bibitem [{\citenamefont {Grifoni}\ and\ \citenamefont
  {H\"{a}nggi}(1998)}]{Grifoni1998_PR_304_229_DrivenQuantum}%
  \BibitemOpen
  \bibfield  {author} {\bibinfo {author} {\bibfnamefont {M.}\ \bibnamefont
  {Grifoni}}\ and\ \bibinfo {author} {\bibfnamefont {P.}\ \bibnamefont
  {H\"{a}nggi}},\ }\bibfield  {title} {\enquote {\bibinfo {title} {Driven
  quantum tunneling},}\ }\href {\doibase 10.1016/S0370-1573(98)00022-2}
  {\bibfield  {journal} {\bibinfo  {journal} {Phys. Rep.}\ }\textbf
  {\bibinfo {volume} {304}},\ \bibinfo {pages} {229} (\bibinfo {year}
  {1998})}\BibitemShut {NoStop}%
\bibitem [{\citenamefont {Longhi}\ and\ \citenamefont
  {Valle}(2013)}]{Longhi2013_SR_3__FloquetBound}%
  \BibitemOpen
  \bibfield  {author} {\bibinfo {author} {\bibfnamefont {S.}\ \bibnamefont
  {Longhi}}\ and\ \bibinfo {author} {\bibfnamefont {G.~D.}\
  \bibnamefont {Valle}},\ }\bibfield  {title} {\enquote {\bibinfo {title}
  {Floquet bound states in the continuum},}\ }\href {\doibase
  10.1038/srep02219} {\bibfield  {journal} {\bibinfo  {journal} {Sci. 
  Rep.}\ }\textbf {\bibinfo {volume} {3}}, \bibinfo {pages} {2219} (\bibinfo {year} {2013})}\BibitemShut {NoStop}%
\bibitem [{\citenamefont {Ge}\ \emph {et~al.}(2012)\citenamefont {Ge},
  \citenamefont {Chong},\ and\ \citenamefont
  {Stone}}]{Ge2012_PRA_85_023802_ConservationRelations}%
  \BibitemOpen
  \bibfield  {author} {\bibinfo {author} {\bibfnamefont {L.}~\bibnamefont
  {Ge}}, \bibinfo {author} {\bibfnamefont {Y.~D.}\ \bibnamefont {Chong}}, \
  and\ \bibinfo {author} {\bibfnamefont {A.~D.}\ \bibnamefont {Stone}},\
  }\bibfield  {title} {\enquote {\bibinfo {title} {Conservation relations and
  anisotropic transmission resonances in one-dimensional
  {$\mathcal{PT}$-symmetric} photonic heterostructures},}\ }\href {\doibase
  10.1103/PhysRevA.85.023802} {\bibfield  {journal} {\bibinfo  {journal}
  {Phys. Rev. A}\ }\textbf {\bibinfo {volume} {85}},\ \bibinfo {pages}
  {023802} (\bibinfo {year} {2012})}\BibitemShut {NoStop}%
\bibitem [{\citenamefont {Miroshnichenko}\ \emph {et~al.}(2005)\citenamefont
  {Miroshnichenko}, \citenamefont {Mingaleev}, \citenamefont {Flach},\ and\
  \citenamefont {Kivshar}}]{Miroshnichenko2005_PRE_71_036626_NonlinearFano}%
  \BibitemOpen
  \bibfield  {author} {\bibinfo {author} {\bibfnamefont {A.~E.}\
  \bibnamefont {Miroshnichenko}}, \bibinfo {author} {\bibfnamefont {S.~F.}\
  \bibnamefont {Mingaleev}}, \bibinfo {author} {\bibfnamefont {S.}\
  \bibnamefont {Flach}}, \ and\ \bibinfo {author} {\bibfnamefont {Y.~S.}\
  \bibnamefont {Kivshar}},\ }\bibfield  {title} {\enquote {\bibinfo {title}
  {Nonlinear Fano resonance and bistable wave transmission},}\ }\href {\doibase
  10.1103/PhysRevE.71.036626} {\bibfield  {journal} {\bibinfo  {journal}
  {Phys. Rev. E}\ }\textbf {\bibinfo {volume} {71}},\ \bibinfo {pages}
  {036626} (\bibinfo {year} {2005})}\BibitemShut {NoStop}%
\bibitem [{\citenamefont {Boechler}\ \emph {et~al.}(2011)\citenamefont
  {Boechler}, \citenamefont {Yang}, \citenamefont {Theocharis}, \citenamefont
  {Kevrekidis},\ and\ \citenamefont
  {Daraio}}]{Boechler2011_JAP_109_074906_TunableVibrational}%
  \BibitemOpen
  \bibfield  {author} {\bibinfo {author} {\bibfnamefont {N.}~\bibnamefont
  {Boechler}}, \bibinfo {author} {\bibfnamefont {J.}~\bibnamefont {Yang}},
  \bibinfo {author} {\bibfnamefont {G.}~\bibnamefont {Theocharis}}, \bibinfo
  {author} {\bibfnamefont {P.~G.}\ \bibnamefont {Kevrekidis}}, \ and\ \bibinfo
  {author} {\bibfnamefont {C.}~\bibnamefont {Daraio}},\ }\bibfield  {title}
  {\enquote {\bibinfo {title} {Tunable vibrational band gaps in one-dimensional
  diatomic granular crystals with three-particle unit cells},}\ }\href
  {\doibase 10.1063/1.3556455} {\bibfield  {journal} {\bibinfo  {journal}
  {J. Appl. Phys.}\ }\textbf {\bibinfo {volume} {109}},\ \bibinfo
  {pages} {074906} (\bibinfo {year} {2011})}\BibitemShut {NoStop}%
\bibitem [{\citenamefont {{M. R\"ontgen \textit{et al.}}}()}]{Roentgen2017_AOP_380_135_Non-localCurrents}%
  \BibitemOpen
  \bibfield  {author} {\bibinfo {author} {\bibnamefont {M. R\"ontgen}}}, 
  \bibfield  {author} {\bibinfo {author} {\bibnamefont {C.~V. Morfonios}}},
  \bibfield  {author} {\bibinfo {author} {\bibnamefont {F.~K. Diakonos}}}, and
  \bibfield  {author} {\bibinfo {author} {\bibnamefont {P. Schmelcher}}}, 
  {\enquote {\bibinfo {title} { Non-local currents and the structure of eigenstates in planar discrete systems with local symmetries},}\ }
  \href{\doibase 10.1063/1.3556455} {\bibfield  {journal} {\bibinfo  {journal}
  {Ann. Phys.}\ }\textbf {\bibinfo {volume} {380}},\ \bibinfo
  {pages} {135} (\bibinfo {year} {2017})}\BibitemShut {NoStop}
\bibitem {connectivity}%
  \BibitemOpen
  In the presently studied tight-binding 1D arrays, the connectivity is always twofold except at end sites. 
  As was seen, the NLCs under LS involving end sites are simply evaluated by setting the corresponding hopping element (on nonexisting links) to zero \BibitemShut {NoStop}%
\bibitem [{\citenamefont {Stockhofe}\ and\ \citenamefont
  {Schmelcher}(2015{\natexlab{a}})}]{Stockhofe2015_PRA_92_023605_Sub-Supercritical}%
  \BibitemOpen
  \bibfield  {title} {  }\bibfield  {author} {\bibinfo {author} {\bibfnamefont
  {J.}~\bibnamefont {Stockhofe}}\ and\ \bibinfo {author} {\bibfnamefont
  {P.}~\bibnamefont {Schmelcher}},\ }\bibfield  {title} {\enquote {\bibinfo
  {title} {Sub- and supercritical defect scattering in Schr\"odinger chains
  with higher-order hopping},}\ }\href {\doibase 10.1103/PhysRevA.92.023605}
  {\bibfield  {journal} {\bibinfo  {journal} {Phys. Rev. A}\ }\textbf
  {\bibinfo {volume} {92}},\ \bibinfo {pages} {023605} (\bibinfo {year}
  {2015}{\natexlab{a}})}\BibitemShut {NoStop}%
\bibitem [{\citenamefont {Stockhofe}\ and\ \citenamefont
  {Schmelcher}(2015{\natexlab{b}})}]{Stockhofe2015_PRA_91_023606_BlochDynamics}%
  \BibitemOpen
  \bibfield  {author} {\bibinfo {author} {\bibfnamefont {J.}~\bibnamefont
  {Stockhofe}}\ and\ \bibinfo {author} {\bibfnamefont {P.}~\bibnamefont
  {Schmelcher}},\ }\bibfield  {title} {\enquote {\bibinfo {title} {Bloch
  dynamics in lattices with long-range hopping},}\ }\href {\doibase
  10.1103/PhysRevA.91.023606} {\bibfield  {journal} {\bibinfo  {journal}
  {Phys. Rev. A}\ }\textbf {\bibinfo {volume} {91}},\ \bibinfo {pages}
  {023606} (\bibinfo {year} {2015}{\natexlab{b}})}\BibitemShut {NoStop}%
\bibitem [{\citenamefont {Haake}(2001)}]{Haake2001_QSC__15_TimeReversal}%
  \BibitemOpen
  \bibfield  {author} {\bibinfo {author} {\bibfnamefont {F.}\
  \bibnamefont {Haake}},\ }\bibfield  {title} {\enquote {\bibinfo {title} {Time
  reversal and unitary symmetries},}\ }in\ \href
  {http://link.springer.com/chapter/10.1007/978-3-662-04506-0_2} {\emph
  {\bibinfo {booktitle} {Quantum Signatures of Chaos}}}\ (\bibinfo  {publisher} {Springer Berlin Heidelberg},\
  \bibinfo {year} {2001}),\ pp.\ \bibinfo {pages} {15}\BibitemShut {NoStop}%
\bibitem [{\citenamefont {Kelley}\ and\ \citenamefont
  {Peterson}(2000)}]{Kelley2000____DifferenceEquations}%
  \BibitemOpen
  \bibfield  {author} {\bibinfo {author} {\bibfnamefont {W.~G.}\
  \bibnamefont {Kelley}}\ and\ \bibinfo {author} {\bibfnamefont {A.~C.}\
  \bibnamefont {Peterson}},\ }\href@noop {} {\emph {\bibinfo {title}
  {Difference Equations: An Introduction with Applications}}}\
  (\bibinfo  {publisher} {Academic Press},\
  \bibinfo {address} {San Diego},\ \bibinfo {year} {2000})\BibitemShut
  {NoStop}%
\bibitem [{\citenamefont {Lieb}\ and\ \citenamefont
  {Mattis}(1966)}]{Lieb1966____MathematicalPhysics}%
  \BibitemOpen
  \bibfield  {author} {\bibinfo {author} {\bibfnamefont {E.~H.}\
  \bibnamefont {Lieb}}\ and\ \bibinfo {author} {\bibfnamefont {D.~C.}\
  \bibnamefont {Mattis}},\ }\href@noop {} {\emph {\bibinfo {title}
  {Mathematical physics in one dimension: exactly soluble models of interacting
  particles}}}\ (\bibinfo  {publisher} {Academic Press},\ \bibinfo {year}
  {1966})\BibitemShut {NoStop}%
\end{thebibliography}
\end{document}